\documentclass[%
amsmath,amssymb,
twocolumn,
prb,
superscriptaddress,
floatfix,
longbibliography,
aps, 10pt
]{revtex4-2}

\usepackage{graphicx}
\usepackage{mathtools}
\usepackage{dcolumn}
\usepackage{bm}
\usepackage{xcolor}
\usepackage{physics}

\begin{document}
	
\title{Pulsed magnetophononics in gapped quantum magnets}

\author{B. Demazure}
\email{baptiste.demazure@psi.ch}
\affiliation{PSI Center for Scientific Computing, Theory and Data, CH-5232 Villigen-PSI, Switzerland}
\affiliation{Institute of Physics, Ecole Polytechnique F\'ed\'erale de Lausanne (EPFL), CH-1015 Lausanne, Switzerland}

\author{M. Krebs}
\affiliation{Condensed Matter Theory, Department of Physics, TU Dortmund University,  Otto-Hahn-Stra\ss{}e 4, 44227 Dortmund, Germany}

\author{G. S. Uhrig}
\affiliation{Condensed Matter Theory, Department of Physics, TU Dortmund University,  Otto-Hahn-Stra\ss{}e 4, 44227 Dortmund, Germany}

\author{B. Normand}
\affiliation{PSI Center for Scientific Computing, Theory and Data, CH-5232 Villigen-PSI, Switzerland}

\begin{abstract}
One route to the control of quantum magnetism at ultrafast timescales is magnetophononics, the modulation of magnetic interactions by coherently driven lattice excitations. Theoretical studies of a gapped quantum magnet subject to continuous, single-frequency driving of one strongly coupled phonon mode find intriguing phenomena including mutually repelling phonon-bitriplon excitations and global renormalization of the spin excitation spectrum. Because experiments are performed with ultrashort pulses that contain a wide range of driving frequencies, we investigate phonon-bitriplon physics under pulsed laser driving. We use the equations of motion to compute the transient response of the driven and dissipative spin-phonon system, which we characterize using the phonon displacement, phonon number, and triplon occupations. In the Fourier transforms of each quantity we discover a low-frequency energetic oscillation between the lattice and spin sectors, which is an intrinsically nonequilibrium collective mode, and demonstrate its origin as a beating between mutually repelling composite excitations. We introduce a phonon-bitriplon approximation that captures all the physics of hybridization, collective mode formation, and difference-frequency excitation, and show that sum-frequency phenomena also leave clear signatures in the repsonse. We model the appearance of such magnetophononic phenomena in the strongly-coupled spin-chain compound CuGeO$_3$, whose overlapping phonon and spin excitation spectra are well characterized, to deduce the criteria for their possible observation in quantum magnetic materials.
\end{abstract}

\maketitle
	

\section{Introduction}
\label{sintro}

Ultrafast laser driving has opened a new chapter in the study of nonequilibrium phenomena in quantum matter, with scientific and technological developments proceeding in synergy \cite{kampf13,zhang17b,salen19}. Ultrafast pump-probe techniques have been used to effect the transient alteration of materials properties including superconductivity \cite{caval18}, metallicity \cite{murak23}, quasiparticle dispersions (Floquet engineering) \cite{oka19}, lattice dynamics \cite{disa21}, and other degrees of freedom \cite{torre21}. 

One of the directions most attractive for both its intrinsic complexity and its technological relevance is the ultrafast engineering of magnetic properties. Multiple studies have effected the ultrafast manipulation of ordered ferromagnetic (FM) and antiferromagnetic (AFM) systems at frequencies up to the THz regime \cite{kampf11,kubac14,baier16,lu17,mashk19}, and one current research direction is towards applications in data storage \cite{behov23,khudo24}, spintronics \cite{walow16,nemec18} and magnonics \cite{pirro21}. From a physics perspective, quantum magnets provide a far wider spectrum of complex phenomena, ranging from strongly renormalized semiclassical behavior to systems that lack any kind of dipolar order \cite{broho20}, and here we restrict our considerations to the latter. Because these phenomena are delicate, and often difficult to couple directly to light, the energy scales involved have to date precluded detailed studies of ultrafast quantum magnetism. 

One of the most promising approaches to ultrafast quantum magnetism is magnetophononics, a term we use for the coherent driving of infrared-active lattice vibrations to modulate the magnetic interactions. Focusing first on the driving of lattice excitations, one may distinguish between ``linear phononics,'' which is the straightforward pumping of harmonic phonon modes, and``nonlinear phononics,'' which concerns the combination of anharmonic phonons to produce sum- and difference-frequency effects, including static distortions \cite{nova17,kozin19}. Although the term magnetophononics has been applied to describe a range of processes by which magnetism is affected by the lattice, here we consider only dynamical effects. The authors of Ref.~\cite{fechn18} applied the term to cover both cases of driven harmonic phonons modulating magnetic interactions at the phonon frequency and nonlinear phononic effects on static magnetic properties, whose analogy with magneto-optic effects has been discussed theoretically \cite{juras20}. 

Here we focus on the first case, of driven harmonic phonons whose frequency matches the spin excitation spectrum. Although we previously referred to this situation as ``linear magnetophononics'' \cite{yarmo23,giorg23}, in the present work we will find regimes where the nonequilibrium response is linear or nonlinear in the spin-phonon coupling, in the electric-field strength, and in the  expansion order to which we compute magnetic interactions from lattice distortions, as well as observables linear or bilinear in the phonon and spin operators. We therefore adopt the more specific terminology ``one-phonon magnetophononics'' for this type of driving. In practice, however, there is no guarantee that the characteristic frequencies of the magnetic and lattice excitations will be similar. Although initial studies of lattice-driven spin dynamics found a slow response at frequencies far below that of the THz drive \cite{disa20,afana21}, more recent work has demonstrated a magnetic response at the driving frequency \cite{mashk21,bossi21}. In the clearest experiment performed to date on a system with no magnetic order parameter, the characteristic frequencies did not match, but gapped quantum spin excitations were populated by a difference frequency between driven phonons \cite{giorg23}, which in our adopted nomenclature is ``two-phonon magnetophononics.'' 

Continuous, single-frequency driving of a single optic phonon coupled to an alternating (gapped) quantum spin chain provides an example of a nonequilibrium steady state (NESS) of both the spin and phonon sectors of the system \cite{yarmo21}. When the spin-phonon coupling is strong, these NESS display mutually repelling hybrid excitations, phonon self-blocking (a departure from resonance that obstructs its absorption of the driving energy), and the ability to engineer the entire band of spin excitations by selective phonon driving \cite{yarmo23}. However, estimates based on real materials parameters indicate that bulk driving of this type is accompanied by a massive heating problem, such that the system would be driven beyond the temperatures required for quantum coherence by a double-digit number of electric-field cycles \cite{yarmo21}. 

Real ultrafast experiments use ultrashort laser pulses, consisting of perhaps 1-5 significant electric-field maxima. Such short pulses contain a wide spectrum of frequencies, which can in principle drive most of the lattice and spin excitations in the system simultaneously. To investigate the effects of these pulses on a gapped quantum magnet in the framework of one-phonon magnetophononics, we adopt the same minimal model as in earlier studies \cite{yarmo21,yarmo23} and consider a single, zero-momentum phonon mode coupled to an alternating spin chain. We solve the equations of motion of this open quantum system to follow the time evolution of quantities including the phonon and triplon (spin excitation) occupations, and calculate the Fourier transforms of each quantity to obtain complete insight into the transient phenomena following excitation by an ultrashort pulse. We stress here that, while all pulsed physics is necessarily transient, some forms of nonequilibrium physics can arise in systems with steady driving or dissipation, whereas other forms arise only due to transient conditions. 

We will show that the NESS physics of strongly hybridized phonon-bitriplon states and band-engineering effects can be created and measured in a single pulse, whose broad range of driving frequencies makes self-blocking irrelevant. In addition to populating composite modes already present in the spectrum of the strongly coupled system, the driving pulse causes the phonon and triplon numbers to display a phenomenon in which the initial phonon energy flows into the spin system and then back for several cycles at an emergent low frequency. In the terminology of the previous paragraph, this is an intrinsically transient nonequilibrium excitation, which appears only due to the pulsed nature of the drive. 

We introduce a ``phonon-bitriplon'' approximation, in which the system is treated as a bulk phonon coupled to an ensemble of quantum harmonic oscillators representing the independent bitriplons. By computing the resolvant, which is equivalent to the linear response function, we obtain an analytical understanding of the entire phonon-bitriplon spectrum and deduce that the low-energy excitation is a beating whose characteristic frequency is the difference between the energies of the two states resulting from phonon-bitriplon hybridization. We then go beyond the features intrinsic to the material, namely the characteristic excitation frequencies and spin-phonon coupling, to compute the properties of the system that can be controlled by the pulse length and the peak electric-field strength. We conclude by discussing the possible observation of these phenomena in the strongly coupled spin-chain material CuGeO$_3$, which becomes dimerized below a spin-Peierls transition at 14.2 K \cite{hase93} and whose phonon and spin excitation spectra are known \cite{popov95,brade98a,brade02,regna96a,uhrig97a} to lie in the regime suitable for one-phonon magnetophononics.

The structure of this article is as follows. In Sec.~\ref{sec:mm} we summarize the minimal magnetophononic model, the equation-of-motion method, and our Fourier transform protocol. In Sec.~\ref{sec:res} we present results illustrating the role of the different intrinsic and extrinsic (pulse) model parameters in the transient phenomena we observe. In Sec.~\ref{sec:pba} we present a phonon-bitriplon approximation by which to interpret the spectrum in the limit of weak driving. In Sec.~\ref{sec:phen} we focus on the primary qualitative physics in our calculations, namely strong hybrid-state formation, {\it {in situ}} spin-band engineering, and the near-universal emergence of a low-frequency mode. In Sec.~\ref{sec:CuGeO3} we consider the spin-chain material CuGeO$_3$ in order to illustrate where these phenomena may be found in real ultrafast experiments. In Sec.~\ref{sec:dc} we discuss the wider implications of our results and conclude. 

\section{Model and methods}
\label{sec:mm}

\subsection{Spin chain Hamiltonian}

\begin{figure}[t]
	\includegraphics[width=\linewidth]{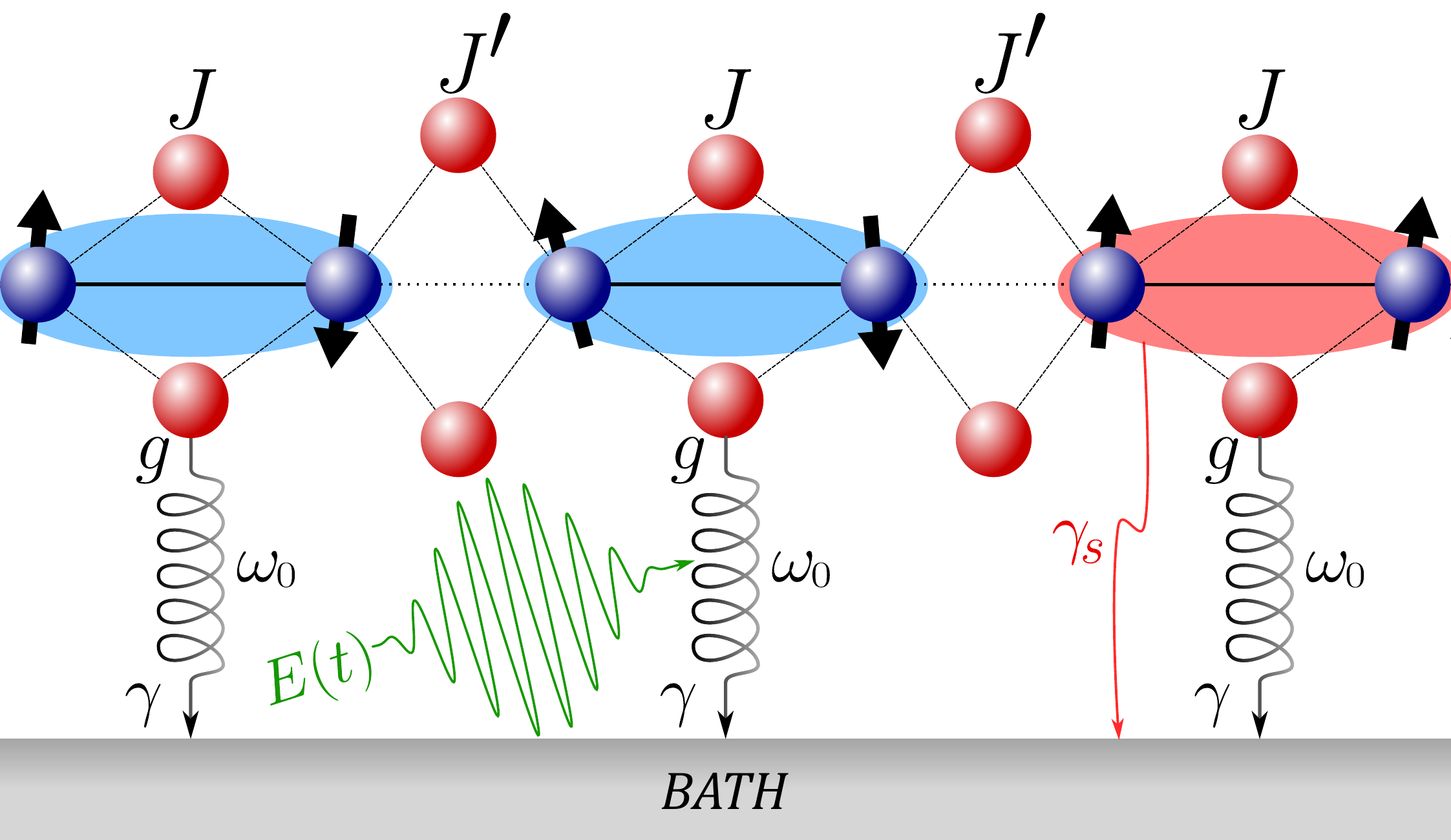}
	\caption{Schematic representation of an alternating spin chain with interaction parameters $J$ and $J'$, spin damping $\gamma_{\rm s}$, and a magnetophononic coupling $g$ to the strong bond ($J$). Blue ellipses denote dimer singlets and the red ellipse a triplon excitation. The optic phonon is nondispersive with frequency $\omega_0$ and damping $\gamma$. The incident laser pulse, characterized by its peak electric field ($E_0$), width ($T_p$), and central frequency ($\omega_d$, which we set equal to $\omega_0$), causes direct driving only of the phonon.}
	\label{fig:scheme}
\end{figure}

We adopt the minimal model of Refs.~\cite{yarmo21,yarmo23} for our study of pulsed magnetophononics. As a representative non-ordered quantum magnetic system, we consider an \(S = 1/2\) spin chain with alternating strong (\(J\)) and weak (\(J'\)) superexchange interactions, such that the magnetic excitations are gapped triplons moving in a background of singlets, as represented in Fig.~\ref{fig:scheme}. Quasi-one-dimensional (1D) magnetic materials based on \(S = 1/2\) spins are rather common in nature, with their dimerization appearing either directly in the crystal structure [as in \(\text{Cu}(\text{NO}_3)_{2}\) and \((\text{VO})_2\text{P}_{2}\text{O}_{7}\)] or as a consequence of a spin-Peierls transition, where the magnetic energy of dimer formation assists the structural distortion of a uniform spin chain to create two different superexchange paths, as in CuGeO$_3$. 

To realize magnetophononic effects, we consider one hypothetical optic phonon mode that involves the atom or atoms on the superexchange path of the strong bond; it was shown in Ref.~\cite{yarmo23} that the more general case differs only in minor details. Because the superexchange interaction depends very sensitively on the geometry of the paths, generally expressed as bond lengths and angles, a phonon excitation can in principle couple strongly to the magnetic interactions. Quantitative estimates based on experiments achieving  the controlled driving of one or more infrared (IR)-active phonon modes into highly excited states indicated that some superexchange parameters could be modulated by many tens of percent \cite{giorg23}. Beyond the single phonon mode selected for driving, all the other acoustic and optic phonons of the system, which are not driven, act to dissipate the input energy by weak phonon-phonon interactions \cite{yarmo23}, thereby forming a system-sized ``bath'' that  is the same at all sites (i.e.~is translationally invariant, as depicted in Fig.~\ref{fig:scheme}) and is responsible for the relaxational dynamics we treat in Sec.~\ref{sec:mm}C.

We express the Hamiltonian of the alternating antiferromagnetic \(S = 1/2\) chain as 
\begin{align} \label{eq:ham_spin}
	H_{\text{s}} & = \sum_{i=1}^{N} \big[ J \vec{S}_{1,i} \cdot \vec{S}_{2,i} + J'\vec{S}_{2,i} \cdot \vec{S}_{1,i+1} \big],
\end{align}
where indices 1 and 2 label the two spins on a dimer and \(i\) the unit cell, \(N\) is the number of dimers, and periodic boundary conditions are assumed. We define the dimerization as \(\lambda = J'/J\). As stated above, we consider that the parameter $J$ is modulated by a single optic phonon with Hamiltonian
\begin{align}
	H_{\text{p}} & = \sum_{\bf q} \omega_{q} b_{\bf q}^{\dagger} b_{\bf q}. 
\end{align}
However, the very long wavelength of THz light (approximately 10$^{-5}$ of the Brillouin zone for a small unit-cell material at 1 THz) ensures that it is an excellent approximation to consider that only the mode $q = 0$ couples to the driving laser, and in this approximation we neglect the sum over {\bf q}, substituting $\omega_{q}$ by the frequency \(\omega_{0}\) throughout this study. The last term of the equilibrium Hamiltonian is the spin-phonon coupling, which with a view to describing magnetic ions in the 3$d$ transition-metal series we write as
\begin{align}
	H_{\text{sp}} = \sum_j g  \big( b_{j} + b_{j}^{\dagger} \big) \vec{S}_{1,j} \cdot \vec{S}_{2,j},
	\label{eq:hsp}
\end{align}
where \(g\) is the strength of the coupling. In contrast to most of the mechanisms by which the electric field of light can couple directly to magnetism, which require some degree of spin anisotropy, the magnetophononic route is not sensitive to anisotropy and we use it here for a case in which the coupling is fully spin-isotropic. The THz laser enters the Hamiltonian as the source term
\begin{align}
	H_{\text{L}} = \sqrt{N} \big( b_{0} + b_{0}^{\dagger}\big) E(t), 
\end{align}
where the electric-field profile $E(t)$ is no longer the continuous driving of Refs.~\cite{yarmo21,yarmo23} but will instead be modeled as an ultrashort pulse. 

\subsection{Bond-operator treatment}

To diagonalize the spin Hamiltonian, we rewrite the spin operators using the bond-operator representation \cite{sachd90}, which is particularly well suited to the treatment of gapped $S = 1/2$ systems with an explicit dimerization pattern \cite{gopal94}. In this complete representation, the two spins on each strong (dimer) bond are expressed as
\begin{align}
	S_{1,2}^{\alpha} & = \pm\tfrac{1}{2}\left(s^{\dagger}t_{\alpha} + t_{\alpha}^{\dagger}s\right)
	- \tfrac{i}{2}\sum_{\beta\zeta}\epsilon_{\alpha\beta\zeta}t_{\beta}^{\dagger}t_{\zeta},
\end{align}
where the indices $\alpha,\beta,\zeta$ represent the triplet flavors $x,y,z$ in the bond operators that act on the vacuum $\ket{0}$ to create the states
\begin{subequations}
	\begin{align}
		\ket{s} & = s^{\dagger} \ket{0} = \tfrac{1}{\sqrt{2}} (\ket{\uparrow\downarrow} - \ket{\downarrow\uparrow}), \\
		\ket{t_{x}} & = t_{x}^{\dagger} \ket{0} = \tfrac{-1}{\sqrt{2}} (\ket{\uparrow\uparrow} - \ket{\downarrow\downarrow}),\\
		\ket{t_{y}} & = t_{y}^{\dagger} \ket{0} = \tfrac{i}{\sqrt{2}} (\ket{\uparrow\uparrow} + \ket{\downarrow\downarrow}), \\
		\ket{t_{z}} & = t_{z}^{\dagger} \ket{0} = \tfrac{1}{\sqrt{2}} (\ket{\uparrow\downarrow} + \ket{\downarrow\uparrow}).
	\end{align}
\end{subequations}
The spin algebra mandates that the bond operators be bosonic, albeit with hard-core nature as each dimer admits only one such state.

In the regime $J \gg J'$ ($\lambda \ll 1$), a condensate of dimer singlets provides an excellent approximation to the ground state, with the lowest excitations being individual gapped triplon modes, as depicted in Fig.~\ref{fig:scheme}. The singlet condensate in Eq.~\eqref{eq:ham_spin} contributes only a constant energy shift that we neglect henceforth. At lowest order in $\lambda$, we neglect triplon-triplon interactions and hence retain terms only to bilinear order in the triplet operators. Quantitatively, this approximation is found to remain accurate up to \(\lambda \approx 1/2\) \cite{norma11}. The spin Hamiltonian in reciprocal space then has the form
\begin{align}
	H_{s} = & \sum_{k,\alpha} \big[ J t_{k,\alpha}^{\dagger}t_{k,\alpha}  \\ 
	& - \tfrac{1}{4} J' \cos k \big( 2 t_{k,\alpha}^{\dagger}t_{k,\alpha}
	+ t_{k,\alpha}^{\dagger}t_{-k,\alpha}^{\dagger} + t_{k,\alpha}t_{-k,\alpha} \big) \big],  \nonumber
\end{align}
which we diagonalize by a Bogoliubov transformation to obtain 
\begin{align}
	H_{s} & = \sum_{k,\alpha}\omega_{k}\tilde{t}_{k,\alpha}^{\dagger}\tilde{t}_{k,\alpha}
\end{align}
with quasiparticle (triplon) dispersion relation 
\begin{align} 
	\omega_{k} & = J\sqrt{1 - \lambda \cos k}. \label{eq:disp_1d}
\end{align}

In the bond-operator representation, the spin-phonon coupling term is
\begin{align}
	\label{eq:hspbo}
	H_{\text{sp}} = \frac{1}{\sqrt{N}} g \big( b_{0} + b_{0}^{\dagger} \big) \sum_{k,\alpha} \big[ 
	& y_{k} \tilde{t}_{k,\alpha}^{\dagger} \tilde{t}_{k,\alpha} \\ & + \tfrac12 y'_{k} 
	\big( \tilde{t}_{k,\alpha}^{\dagger} \tilde{t}_{-k,\alpha}^{\dagger} + \text{H.c.} \big) \big], \nonumber
\end{align}
with coefficients $y_k  =  J(1 - \lambda \cos k /2)/\omega_k$ and $y'_k =  J' \cos k /(2 \omega_k)$. We adopt a mean-field decoupling of the phonon and spin parts \cite{yarmo21} by introducing the average phonon displacement 
\begin{align} \label{eq:def_q}
	\expval{q} & = \big\langle \tfrac{1}{\sqrt{N}} \big( b_{0} + b_{0}^{\dagger}\big) \big\rangle
\end{align}
and the summed spin expectation values 
\begin{subequations}
	\begin{align}
		U & = \frac{1}{N} \sum_{k,\alpha} y_{k} \expval{\tilde{t}_{k,\alpha}^{\dagger}\tilde{t}_{k,\alpha}} \! , \label{eq:U}\\
		V & = \frac{1}{N} \sum_{k,\alpha} y'_{k} \Re \expval{\tilde{t}_{k,\alpha}^{\dagger}\tilde{t}_{-k,\alpha}^{\dagger}} \! .
	\end{align}
\end{subequations}
The full Hamiltonian at the quadratic and mean-field level is then  
\begin{align} \label{eq:Ham_decoupled}
	H & = \omega_{0} b_{0}^{\dagger} b_{0}
	+ \sum_{k,\alpha} \omega_{k} \tilde{t}_{k,\alpha}^{\dagger} \tilde{t}_{k,\alpha} \nonumber\\
	& \quad + g \expval{q} \sum_{k,\alpha} \big[ y_{k} \tilde{t}_{k,\alpha}^{\dagger} \tilde{t}_{k,\alpha}
	+ \tfrac12 y'_{k} \big( \tilde{t}_{k,\alpha}^{\dagger} \tilde{t}_{-k,\alpha}^{\dagger}
	+ \text{H.c.} \big) \big] \nonumber\\
	& \quad + \sqrt{N} \big( b_{0} + b_{0}^{\dagger} \big) \big[ E(t) + g \big( U + V \big) \big],
\end{align}
by which is meant that we retain bilinear terms in the phonon and triplon operators but decouple the trilinear (phonon-bitriplon) term. The mean-field decoupling follows the standard form with time-dependent mean fields [$\expval{q}$ in the second line of Eq.~\eqref{eq:Ham_decoupled} and the square bracket in the third] also known as density-matrix formalism.

We observe that a single phonon interacts only with triplon pairs of equal and opposite momenta and net zero spin. These bitriplons, with energy \(2\omega_{k}\), are then the relevant magnetic excitations for the magnetophononic system with spin-phonon coupling of the type specified in Eq.~\eqref{eq:hsp}. We observe also that phonon creation and annihilation are driven both by the laser and by the ``back action'' of the spin sector, in the form of both the weighted triplon number ($U$) and the weighted bitriplon excitations ($V$). The scale of this effective quasiparticle exchange between the phonon and spin sectors is controlled by the coupling parameter, \(g\).

\begin{figure}[t]
	\includegraphics[width=\linewidth]{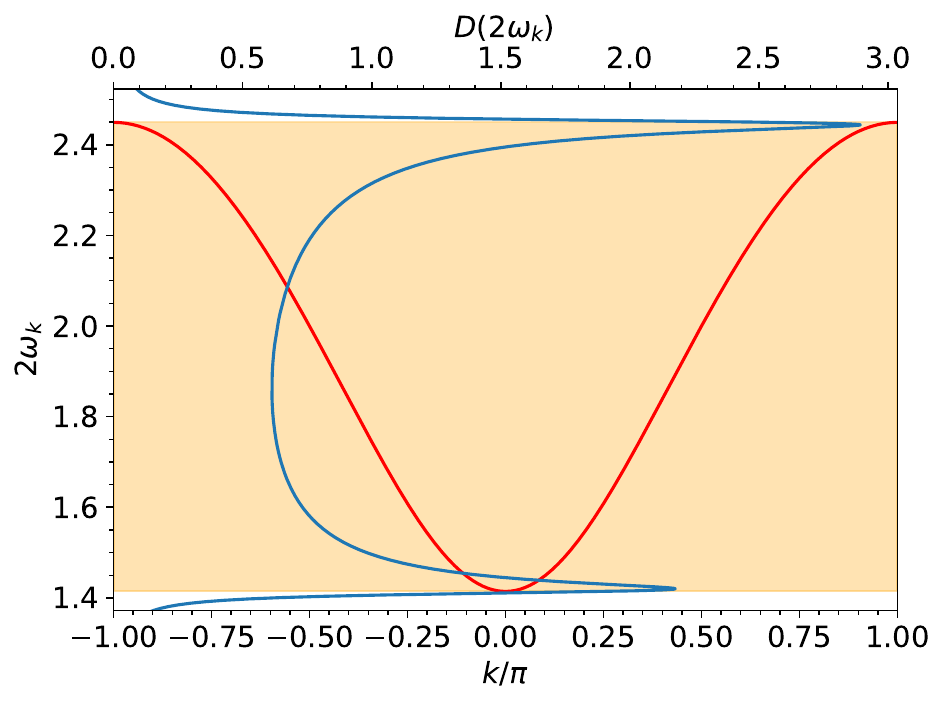}
	\caption{Bitriplon dispersion (\(2\omega_{k}\), red) and density of states [\(D(2\omega_{k})\), blue] for a spin chain with coupling ratio \(\lambda = 0.5\), computed for \(N =1000\) dimers using as the triplon broadening a spin damping \(\gamma_s = 0.01 J\). Beige shading represents the energy window of the bitriplon band for the infinite system of ideal (undamped) triplons. }
	\label{fig:bidisp}
\end{figure}

Returning to the bitriplons, their dispersion (\(2\omega_{k}\)) for our parameter choice $\lambda = 0.5$ is shown in Fig.~\ref{fig:bidisp}. The density of states reflecting the energetic distribution of bitriplons is given by 
\begin{equation}
\label{eq:dosgen}
D(E) = \frac{1}{N} \! \sum_{k} \delta(E - 2\omega_{k}) \approx \frac{1}{N} \! \sum_{k} \frac{1}{\pi}\frac{\gamma_{s}}{(E - 2\omega_{k})^{2} + \gamma_{s}^{2}}, \nonumber
\end{equation}
where $N$ is the number of dimers in the chain and we use a Lorentzian with the spin damping, $\gamma_{\rm s}$, to approximate the Dirac $\delta$-function. For the alternating chain with \(\gamma_{s} = 0\), this can be evaluated algebraically as
\begin{equation}
\label{eq:dos1d}
D(E) = \frac{\abs{E}/(2J)}{2\pi J\lambda\sqrt{1 - [(4J^{2} - E^{2})/(4 \lambda J^{2})]^{2}}},
\end{equation}
which diverges at the two bitriplon band edges ($E^2/J^2 = 2$ and 6). We will revisit this property of a 1D system when generalizing our results to real materials, but in practice $\gamma_{\rm s}$ produces a function that is finite, although still peaked very sharply at the band edges, as shown in Fig.~\ref{fig:bidisp}. 

\subsection{Driven dissipative dynamics}

The time evolution of the driven magnetophononic system is determined by its dissipative terms. The dynamics of this open quantum system are described most generally by the adjoint quantum master equation, or Gorini-Kossakowski-Sudarshan-Lindblad (GKSL) equation,
\begin{align}
	\frac{d}{dt} \! \expval{O} \! (t) & = i\expval{\comm{H}{O(t)}} \! \nonumber \\ 
	& \quad + \! \sum_{i} \! \gamma_{i} \Big( \! A_{i}^{\dagger}  		
	O A_{i} \! - \! \tfrac{1}{2} \acomm{A_{i}^{\dagger}A_{i}}{O} \! \Big) \!,
	\label{eq:gksl}
\end{align}
where \(O\) is any system observable, $H$ is the system Hamiltonian, and \(\{A_{i}\}\) are quantum jump operators that describe the dissipative effect of the environment, or bath, through associated damping coefficients \(\{\gamma_{i}\}\) \cite{breue07,weiss21}. The insight of Lindblad \cite{lindb76} was that the jump operators \(\{A_{i}\}\) can be constructed using only combinations of the system operators, and crucially need not include all possible bath operators. The nature of the system-bath interactions and the properties of the bath dictate the required operator combinations \(\{A_{i}\}\) and their accompanying decay rates \(\{\gamma_{i}\}\). In the magnetophononic model, the bath is composed of all the other phonons that are not driven by the laser (Sec.~\ref{sec:mm}A), but the jump operators governing how the driving energy proceeds from the system to the bath are constructed only from combinations of $b_0$, $b_0^\dag$, $\tilde{t}_k$, and $\tilde{t}_k^\dag$. 

For the phonon we take the generic choice \cite{yarmo21} \(A_{1} = b_{0}^{\dagger}\) and \(A_{2} = b_{0}\), with corresponding damping coefficients \(\gamma_{1} = \gamma n(\omega_{0})\) and \(\gamma_{2} = \gamma [1 + n(\omega_{0})]\), where $n(\omega)$ is the Bose occupation function. Without loss of generality we take the system to be at zero temperature, where \(n(\omega) = 0\) and hence the only allowed process is to remove phonons from the system. The time-dependence is contained in the expectation values of the physical observables, and hence the phononic observables of interest are
\begin{subequations}
	\begin{align}
	& q(t) = \tfrac{1}{\sqrt{N}} \big\langle \hat{b}_{0} + \hat{b}_{0}^{\dagger} \big\rangle, \\
	& p(t) = \tfrac{i}{\sqrt{N}} \big\langle \hat{b}_{0}^{\dagger} - \hat{b}_{0} \big\rangle, \\
	& n_{\text{ph}}(t) = \tfrac{1}{N} \big\langle \hat{b}_{0}^{\dagger}\hat{b}_{0} \big\rangle, 
	\end{align}
\end{subequations}
describing respectively the time-dependent phonon displacement [evaluating the quantity $\expval{q}$ defined in Eq.~\eqref{eq:def_q}], the conjugate momentum, and the number of phonons per dimer. The equations of motion in this textbook case \cite{breue07,weiss21} are those of the damped quantum harmonic oscillator. 

We caution that the Lindblad approach is very simple, if also very effective, and the derivation of the $\{\gamma_i\}$ coefficients is largely phenomenological. A more sophisticated treatment of the bath may be required under a range of circumstances, one of the most relevant for our considerations being that phonon modes at two or more different frequencies are driven simultaneously by the broad frequency range of the ultrashort driving pulse (Sec.~\ref{sec:mm}D). It has been shown in the context of molecular excited states that two such modes can develop correlations in a treatment where they are coupled to the same bath \cite{giava22}, and that this needs to be distinguished from the case of independent damping channels. In the system we consider, the very large bath ensures very weak occupation of the bath phonons responsible for energy dissipation, which are exclusively in their harmonic regime and hence remain independent. Thus we expect that a Lindblad treatment remains appropriate in this situation, even up to very high driven and dissipated pulse energies.

For the triplon dissipation we follow the same logic as for the phonon, taking \(A_{1} = \tilde{t}_{k,\alpha}^{\dagger}\) and \(A_{2} = \tilde{t}_{k,\alpha}\), with \(\gamma_{1} = 0\) and \(\gamma_{2} = \gamma_{s}\). As discussed in Refs.~\cite{yarmo21,yarmo23}, this choice allows dissipative processes that do not conserve spin, whereas the spin-phonon Hamiltonian of Eq.~\eqref{eq:hsp} is spin-isotropic and hence specifies that the phonons cannot change the spin state. In a more materials-specific treatment, spin-flipping dissipation terms would be appropriate for describing spin-anisotropic 4$d$ or 5$d$ magnetic ions, while more complex (two-body) jump operators would be required to model the spin bath for a 3$d$ ion. However, we continue to follow the simplified  treatment of Refs.~\cite{yarmo21,yarmo23} in order to preserve the minimal ($k$-diagonal) equations of motion for a meaningful comparison of pulsed-driving phenomena with these studies. The observables of the spin sector are
\begin{subequations}
	\begin{align}
		u_{k}(t) & = \sum_{\alpha} \expval{ \tilde{t}_{k,\alpha}^{\dagger} \tilde{t}_{k,\alpha}} \! , \\
		v_{k}(t) & = \Re \sum_{\alpha} \expval{ \tilde{t}_{k,\alpha}^{\dagger} \tilde{t}_{-k,\alpha}^{\dagger}} \! , \\
		w_{k}(t) & = \Im \sum_{\alpha} \expval{ \tilde{t}_{k,\alpha}^{\dagger} \tilde{t}_{-k,\alpha}^{\dagger}} \! ,
	\end{align}
\end{subequations}
where \(u_{k}(t)\) tracks the triplon number for each momentum \(k\) while \(v_{k}\) and \(w_{k}\) describe the dynamics of bitriplon creation. Applying the GKSL equation~\eqref{eq:gksl} with $H$ from Eq.~\eqref{eq:Ham_decoupled} yields the \(3N+3\) equations of motion 
\begin{subequations} \label{eq:eom}
	\begin{align} 
		& \tfrac{d}{dt} q(t) = \omega_{0} p(t) - \tfrac{1}{2} \gamma q(t), \\		
		& \tfrac{d}{dt} p(t) = -\omega_{0} q(t) - \tfrac{1}{2}\gamma p(t) \nonumber \\ 
		& \hspace{1.31cm} - 2 \left( E(t) + g \left[ U(t) + V(t) \right] \right), \label{eq:dqdt} \\
		& \tfrac{d}{dt} n_{\text{ph}} (t) = - \left( E(t) + g \left[ U(t) + V(t) \right] \right) p(t) \nonumber \\
		& \hspace{1.69cm} - \gamma [n_{\text{ph}}(t) - n(\omega_{0}) ], \label{eq:dnphdt} \\
		& \tfrac{d}{dt} u_{k} (t) = 2 g y'_{k} q(t) w_{k}(t) - \gamma_{s} \left(u_{k} (t) - 3 n(\omega_{k}) \right), \label{eq:dukdt}\\
		& \tfrac{d}{dt} v_{k} (t) = - 2 \left( \omega_{k} + g y_{k} q(t) \right) w_{k}(t) - \gamma_{s} v_{k}(t), \\
		& \tfrac{d}{dt} w_{k} (t) = 2 \left( \omega_{k} + g y_{k} q(t) \right) v_{k}(t) \nonumber \\
		& \hspace{1.55cm} + 2 g y'_{k} q(t) \left( u_{k} + \tfrac{3}{2} \right) - \gamma_{s} w_{k}(t).
	\end{align}
\end{subequations}

\subsection{Driving electric field and calculation parameters}

To parameterize the laser pulse driving the selected phonon, we use a cosine-shaped electric-field envelope centered around \(t = 0\),  
\begin{subequations}
	\begin{align}\label{eq:pulse} 
		E(t) & = \tfrac{1}{2} E_{0} \Big[ 1 + \cos (\tfrac{2\pi t}{T_{p}}) \Big] \cos(\omega_{d} t) B(t) \quad {\rm {with}} \\
		B(t) & = \Theta (t + T_{p}/2)  - \Theta(t - T_{p}/2).
	\end{align} 
\end{subequations}

\begin{figure}[t]
	\includegraphics[width=0.96\linewidth]{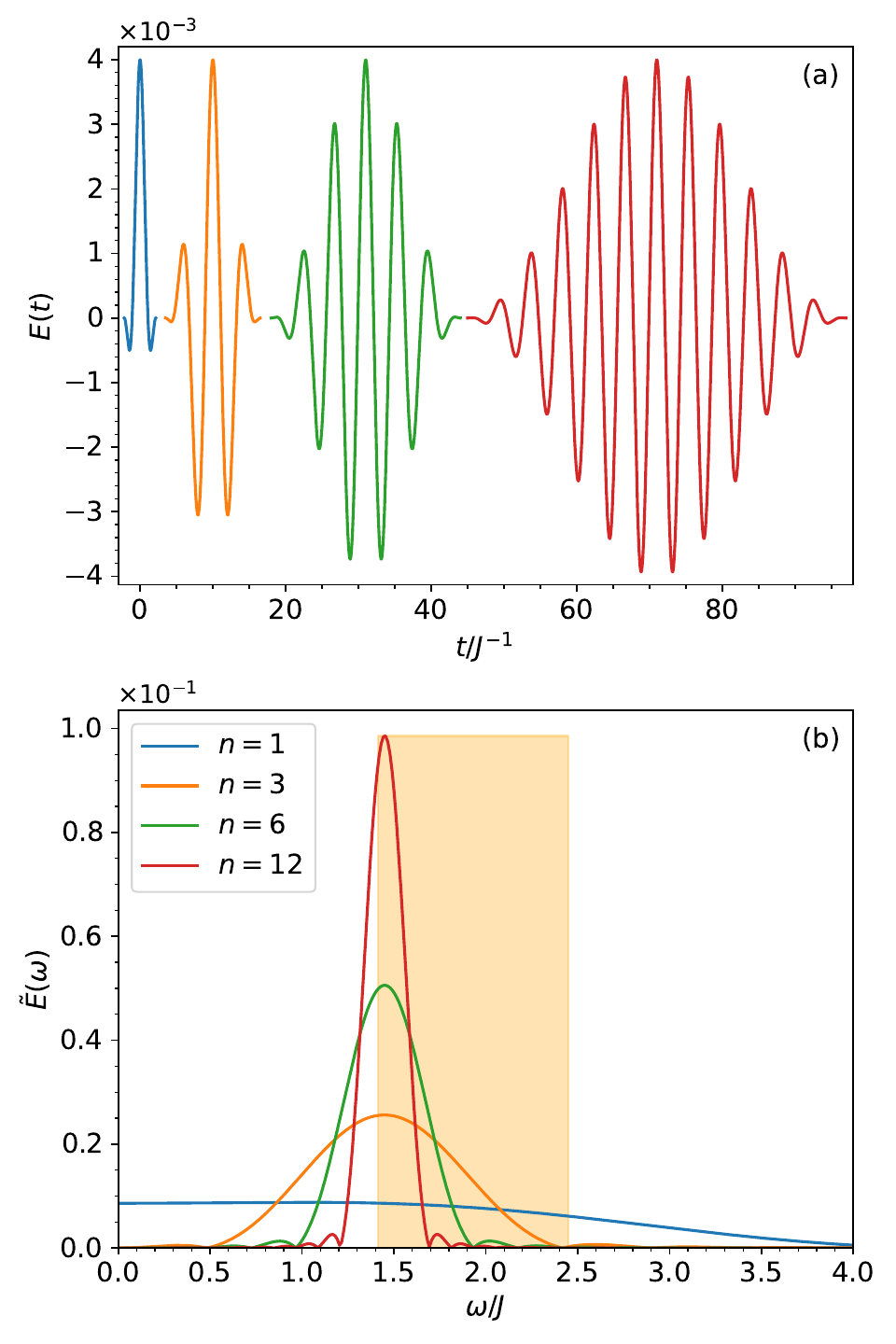}
	\caption{(a) Time traces showing the electric fields of four light pulses with increasing widths, $T_{p} = 2 \pi n / \omega_{d}$ for \(n = 1\), 3, 6, and 12. The drive frequency is \(\omega_{d} = 1.45 J\) and the peak electric field is \(E_{0} = 4\times 10^{-3} \omega_{d}\). We set $\hbar = 1$ and measure the time in units of $J^{-1}$. (b) Moduli of the Fourier transforms of these four pulses, illustrating the increase in spectral content with decreasing pulse width.}
	\label{fig:pulses}
\end{figure}

To isolate a single pulse, we multiply the cosine function by two Heaviside functions. We choose the expression in the square brackets rather than a Gaussian envelope to avoid any issues with long-time tails. The key advantage of this envelope is that it is continuously differentiable even at the end points of the interval set by \(B(t)\), so that only little ringing is induced. The resulting pulse is characterized by three parameters, its amplitude \(E_{0}\), its central frequency \(\omega_{d}\), and its width \(T_{p}\). To control the number of peaks inside the cosine envelope, we set \(T_{p} = 2 \pi n / \omega_{0}\) and in Fig.~\ref{fig:pulses}(a) show examples of pulses with four different \(n\) values. We remark that pulses with odd \(n\) have \(n\) positive and \(n+1\) negative peaks, while those with even \(n\) have the converse. The Fourier transforms of these pulses, shown in Fig.~\ref{fig:pulses}(b), display a rapid broadening of spectral content as the pulse length is reduced. Modern ultrafast and ultra-intense THz light sources produce very short pulses, approaching the single-cycle limit (\(n = 1\)), and we will focus most of our efforts on the representative case \(n = 3\). We defer a discussion of the maximum pulse intensity set by our theoretical approach to Sec.~\ref{sec:phen}.

Suitable parameters for investigating the magnetophononic driving effect of these pulses are as follows. We compute the time-evolution of Eqs.~\eqref{eq:eom} from \(t = -T_{p}/2-1\) to \(t=2000\) with a time step of \(\delta t = 0.01\) using the 4th-order explicit Runge-Kutta method, which we implemented using the \textit{Julia} language \cite{bezanson17}. We perform these calculations on systems of \(N = 500\) dimers, a size that yields accurate results for low computation times. To compare our results with previous studies using a continuous driving \cite{yarmo23}, we adopt an alternating chain with the same physical parameters, i.e.~intradimer coupling \(J = 1.0\), interdimer coupling \(J' = \lambda J = 0.5\), phonon damping \(\gamma = 0.02 \omega_{0}\), and spin damping \(\gamma_{s} = 0.01J\). We study only resonant phonon driving and hence set \(\omega_{d} = \omega_{0}\), such that the four variable parameters to investigate are \(g\), \(\omega_{0}\), \(E_{0}\), and \(n\).

The three diagnostic quantities we compute are the phonon displacement \(q(t)\), the phonon number \(n_{\text{ph}}(t)\), and the triplon number \(n_{\text{t}}(t)\). In contrast to the case with continuous driving at a single frequency \cite{yarmo23}, it is the Fourier transforms of these quantities that contain the most valuable information. These we obtain by applying the numerical FFT scheme provided by the \textit{SciPy} libraries \cite{virtanen20} over the full time range of the calculation. Because the transient phenomena at very short times are of primary interest, and because the pulse effects decay almost to zero at our final times, unless otherwise stated we do not apply a window function during the FFT. Although this comes at the cost of a finite spectral leakage, the frequency positions of the dominant features are such that this has no influence on our results. 

\begin{figure}[t]
	\includegraphics[width=\linewidth]{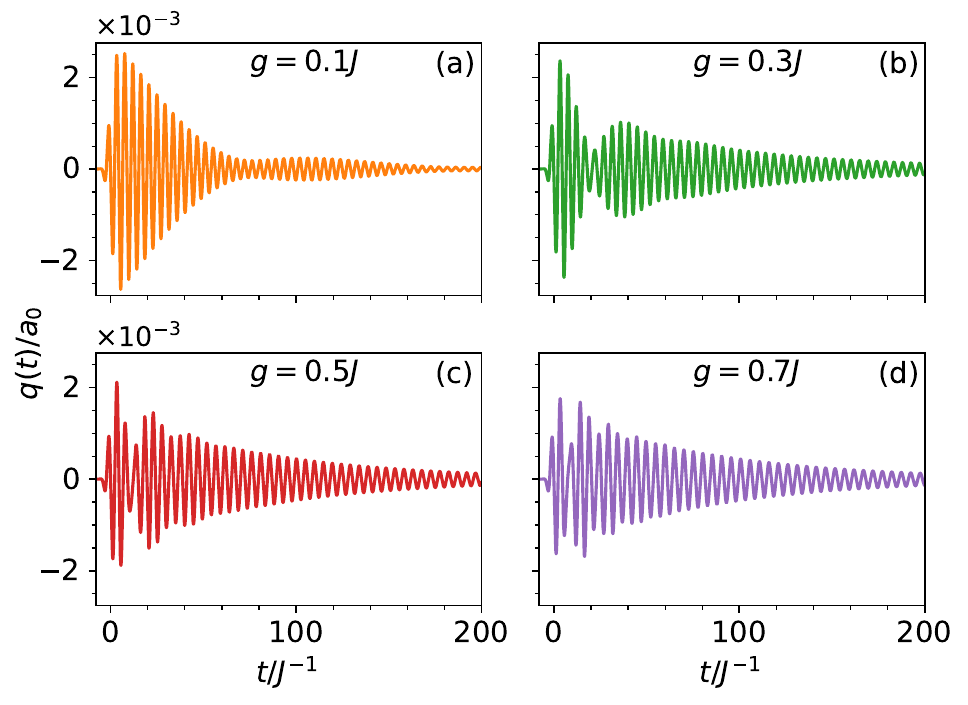}
	\caption{Time evolution of the phonon displacement, \(q(t)\), obtained with a standard driving pulse of width \(T_p = 3 (2\pi / \omega_{0})\) and peak electric field \(E_{0} = 4 \times 10^{-3} \omega_{0}\) incident on a spin chain coupled to a single phonon mode with frequency \(\omega_0 = 1.45 J\) with four different coupling strengths, \(g/J = 0.1\), 0.3, 0.5, and 0.7.}
	\label{fig:q_q0_w0_145}
\end{figure}

\section{Roles of model and pulse parameters}
\label{sec:res}
	
\subsection{Spin-phonon coupling \(g\)} 
\label{ssec:g}
	
We begin by scanning the dependence of the driven dynamics emerging from the solution of Eqs.~\eqref{eq:eom} on the spin-phonon coupling strength, \(g\). Although \(g\) is in general fixed for a given material and phonon, it is not impossible that cavity magnetophononic experiments could offer the possibility of changing the strength of this coupling. To fix the other parameters to standard values, we investigate a phonon frequency close to the lower band edge, \(\omega_{0} = 1.45J\), with pulse width \(T_{p} = 3 (2\pi / \omega_{0})\) (i.e.~ \(n = 3\)) and peak driving electric field \(E_{0} = 4 \times 10^{-3}\omega_{0}\) \cite{yarmo23}. From Fig.~\ref{fig:pulses}(b), the frequency content of this pulse covers most of the two-triplon excitation band, with a full width at half maximum height from approximately $J$ to $2J$.

In Fig.~\ref{fig:q_q0_w0_145} we show the phonon displacement, \(q(t)\). While the natural unit of $q$ is the oscillator length, \(q_{\rm osc} = \sqrt{\hbar / (M \omega_{0})}\), for our purposes it is more appropriate to measure the phonon displacement in units of the lattice constant, $a_0$. Following Ref.~\cite{yarmo21}, we equate $a_0$ very loosely to the value of \(q_{\rm osc}\) for an electron and thereby obtain \(q(t)/a_0\) by scaling with the factor \(\left(m_{e}/M\right)^{1/4}\); taking $M$ as the mass of an O atom yields a factor of 1/13. The dynamics of \(q(t)/a_0\) reflect two timescales. The first is a rapid oscillation set by \(\omega_{0}\) and the second is a slower oscillation of the shape of the envelope that appears at and above \(g = 0.3J\) [Fig.~\ref{fig:q_q0_w0_145}(b)]. The frequency characterizing the slow oscillations clearly rises with \(g\), as one may read from the location of the first minimum of the envelope. At very short times, the rate at which \(q(t)\) rises as the pulse first acts on the system becomes lower as \(g\) increases. At long times, beyond \(t \approx 100/J^{-1}\), \(q(t)\) shows only the exponential decay of an underdamped oscillator on a timescale set by the inverse phonon damping, \(\gamma^{-1}\). 
	
\begin{figure}[t]
	\includegraphics[width=\linewidth]{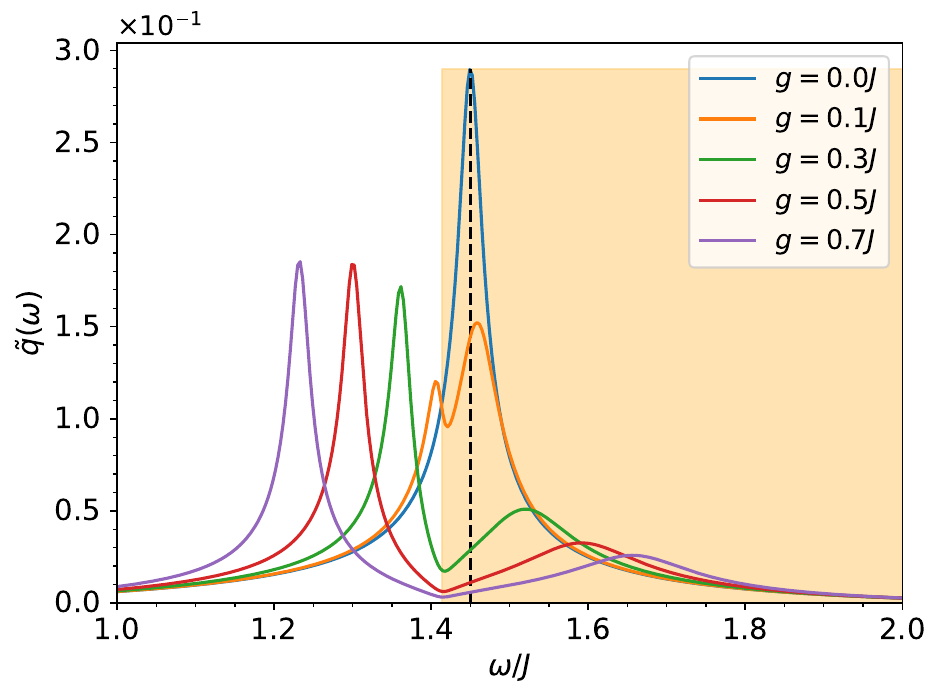}
	\caption{Fourier transform, \(\tilde{q} (\omega)\), of the four \(q(t)\) signals in Fig.~\ref{fig:q_q0_w0_145}, compared with the situation for a chain with \(g = 0\).}
	\label{fig:ft_q_w0_145}
\end{figure}

\begin{figure}[b]
	\includegraphics[width=\linewidth]{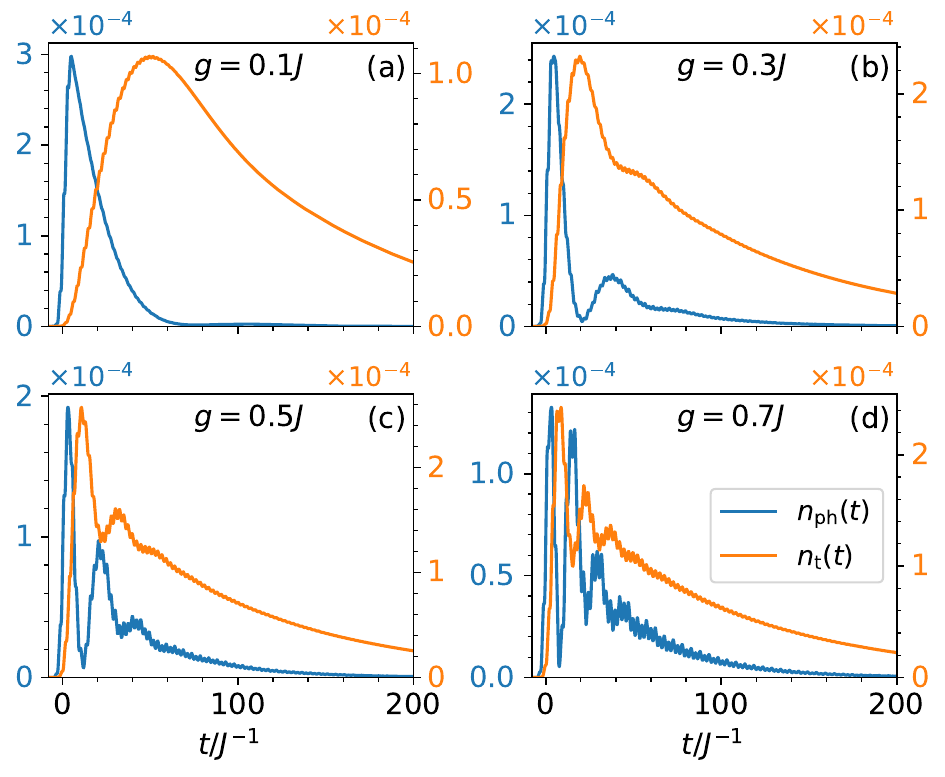}
	\caption{Nonequilibrium phonon occupation, \(n_{\text{ph}} (t)\), and triplon occupation, \(n_{\text{t}} (t)\), shown for an alternating chain with four different \(g\) values subject to driving by a standard pulse; the corresponding evolution of \(q(t)\) is shown in Fig.~\ref{fig:q_q0_w0_145}.}
	\label{fig:nph_nt_w0_145}
\end{figure}

\begin{figure*}[t]
	\includegraphics[width=0.96\linewidth]{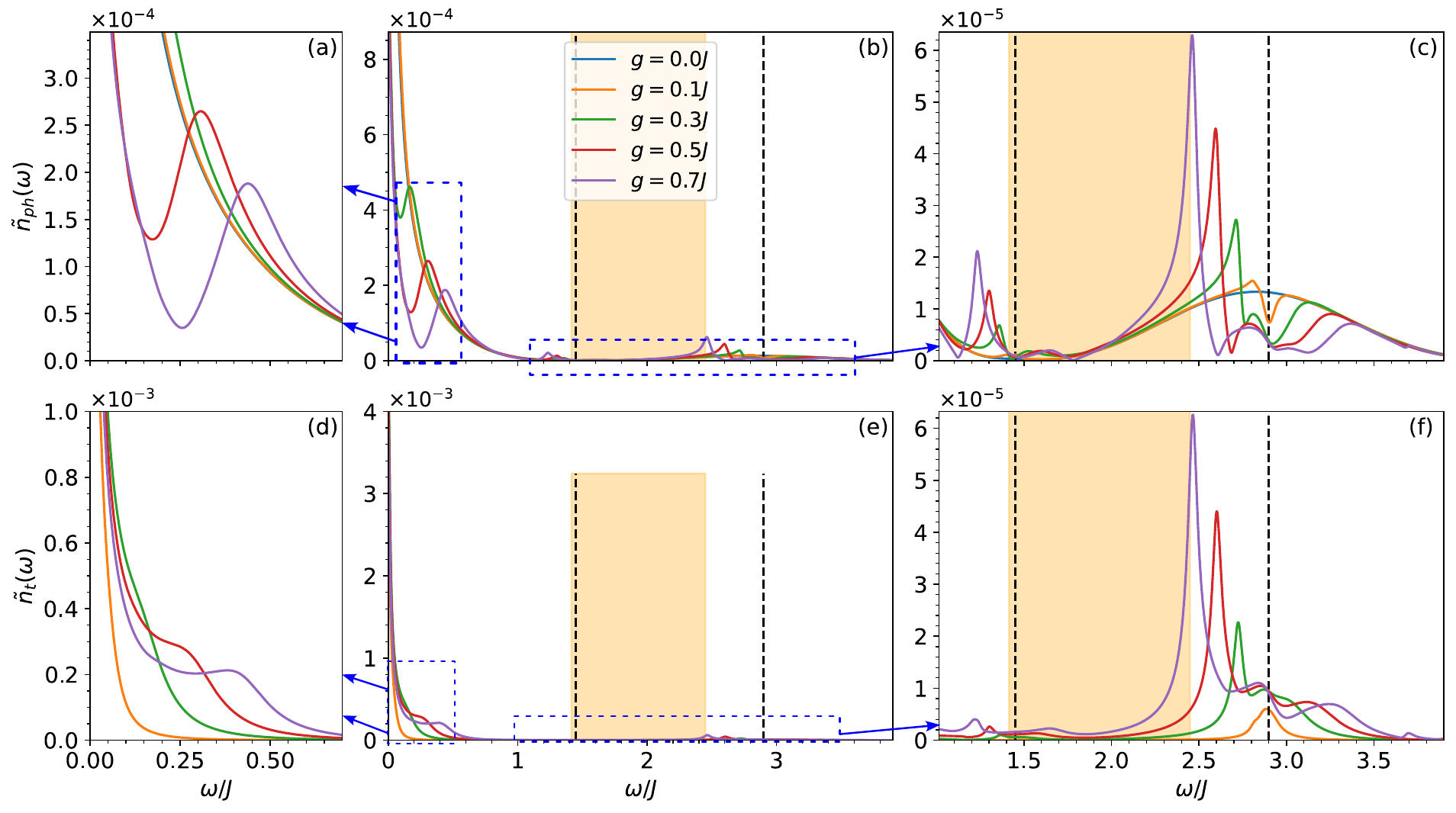}
	\caption{(a-c) Fourier transform, \(\tilde{n}_{\text{ph}} (\omega)\), of \(n_{\text{ph}} (t)\) shown in Fig.~\ref{fig:nph_nt_w0_145}. Dashed lines mark the phonon frequency, \(\omega_0\), and \(2\omega_0\), while beige shading marks the bitriplon bandwidth. The full response, shown in panel (b), has strong contributions at low frequencies, shown on an expanded frequency scale in panel (a), and a complex response around \(\omega_0\) and \(2\omega_0\), shown on an expanded intensity scale in panel (c). (d-f) Corresponding Fourier transform, \(\tilde{n}_{\text{t}} (\omega)\), of \(n_{\text{t}} (t)\) from Fig.~\ref{fig:nph_nt_w0_145}.}
	\label{fig:ft_nph_nt_w0_145}
\end{figure*}

In Fig.~\ref{fig:ft_q_w0_145} we show the Fourier transforms, \({\tilde q} (\omega)\), corresponding to each of the cases in Fig.~\ref{fig:q_q0_w0_145}. At \(g = 0\) we recover the Fourier transform of the damped harmonic oscillator, a Lorentzian centered around \(\omega_{0}\) with a width set by $\gamma$. In the presence of spin-phonon coupling, the spectrum splits into  two characteristic frequencies whose separation increases with \(g\), as expected from standard considerations of hybridization and level repulsion \cite{yarmo23}. We name the hybrid state that is repelled outside the two-triplon band, in this case below the lower band edge (\(2 \omega_{k=0} = \sqrt{2}J\)), \(\omega_{\text{out}}\) and the state that is pushed deeper into the band \(\omega_{\text{in}}\). While the \(\omega_{\text{out}}\) excitation is based on one state from the hybridized phonon-bitriplon spectrum and hence is well described as a single mode, the \(\omega_{\text{in}}\) excitation involves the hybridization of the phonon with multiple bitriplon states and is better characterized as a ``resonance'' \cite{yarmo23}.

The presence of these two nearby but mutually repelling modes suggests that the slow (envelope) dynamics in Fig.~\ref{fig:q_q0_w0_145} should be the consequence of a beating process with characteristic frequency \(\omega_{\text{df}} = \omega_{\text{in}} - \omega_{\text{out}}\). This difference frequency, and the corresponding sum frequencies, reflect nonlinearities in the response that arise due to $g$ and are also present in \(\tilde{q}(\omega)\), as we will see later, but are very weak. By contrast, observables which are bilinear in the phonon and triplon operators naturally display sum- and difference-frequency excitations, which are therefore expected to be the primary features in the phonon number, \(n_{\text{ph}}(t) = \tfrac{1}{2}[q^{2}(t) + p^{2}(t) - 1]\), and the triplon number, \(n_{\text{t}}(t)\). On this basis we study \(n_{\text{ph}}(t)\) and \(n_{\text{t}}(t)\) as two further diagnostic measures. 

In Fig.~\ref{fig:nph_nt_w0_145} we find that the rapid oscillations in \(n_{\text{ph}}(t)\) and \(n_{\text{t}}(t)\) form a small part of the total signals, which also display a slow oscillation and an exponential decay. The uptake of energy by the phonon from the field is not instantaneous, but takes 2-3 cycles to reach its maximum, which in turn drops from \(3\times 10^{-4}\) to \(1.5 \times 10^{-4}\) as \(g\) is increased. The triplon number starts to increase one cycle after the phonon number, because the phonon acts as the drive for triplon creation, and increases both more rapidly and to a higher maximum value as $g$ is raised. Again the slow oscillation appears in both quantities at and above \(g = 0.3J\). While it becomes more evident in \(n_{\text{t}}(t)\) with increasing $g$, \(n_{\text{ph}}(t)\) shows the striking feature that it decreases almost to zero at the end of its first cycle. At the same time, the triplon number reaches a maximum, indicating almost ideally efficient energy exchange between the phonon and the spin sector, at least for \(\omega_{0} = 1.45J\), and the establishment of an antiphase oscillation that persists at the beat frequency for another 2-3 weakening cycles. In this regard the magnetophononic system shows the same physics as two coupled classical pendula, where one or other can reach zero oscillation amplitude on the slow timescale. 

Figure \ref{fig:ft_nph_nt_w0_145} shows that the Fourier transforms, \(\tilde{n}_{\text{ph}} (\omega)\) and \(\tilde{n}_{\text{t}} (\omega)\), contain a number of noteworthy spectral features. The strongest is the zero-frequency component [Figs.~\ref{fig:ft_nph_nt_w0_145}(a,d)], which is due to the non-zero means of \(n_{\text{ph}}(t)\) and \(n_{\text{t}}(t)\) and are common to all $g$ values. The next strongest, lying between \(\omega \approx 0.1J\) and \(0.3J\) is a peak corresponding to \(\omega_{\text{df}}\) [Figs.~\ref{fig:ft_nph_nt_w0_145}(a,d)], which is one of our most important indicators of magnetophononic dynamics and will have a key role in the remainder of this study. Around \(\omega =\omega_{0}\) we recognize relatively weak peaks at the frequencies \(\omega_{\text{out}}\) and \(\omega_{\text{in}}\) [Figs.~\ref{fig:ft_nph_nt_w0_145}(c,f)]; these grow in amplitude with \(g\), indicating their origin in \(\tilde{n}_{\text{ph}} (\omega)\) as a consequence of spin feedback through the term \(U(t)\) in Eq.~\eqref{eq:dnphdt}, which then imprints them on \(\tilde{n}_{\text{t}} (\omega)\). Finally, the spin-phonon hybridization produces not only a clear difference-frequency component at \(\omega_{\text{df}}\) but also the sum frequencies \(2\omega_{\text{out}}\), \(\omega_{\text{out}} + \omega_{\text{in}}\), and \(2\omega_{\text{in}}\), all of which are clearly visible around \(\omega = 2\omega_{0}\) in \(\tilde{n}_{\text{ph}} (\omega)\) [Fig.~\ref{fig:ft_nph_nt_w0_145}(c,f)]. 

The work of Ref.~\cite{yarmo23} investigated the same magnetophononic model subject to continuous driving at a single frequency to establish a NESS. Because all transient excitations have decayed to zero in the long-time limit, the frequency content of a NESS consists only of the driving frequency and its higher harmonics. The additional composite collective excitations appearing at the sum and difference frequencies are therefore not present in a NESS, and hence constitute one of the primary pieces of intrinsically transient nonequilibrium phenomenology to explore by pulsed driving. We remark that the processes contributing to \(n_{\text{ph}}(t)\) and \(n_{\text{t}}(t)\) have different matrix elements in the equations of motion \eqref{eq:eom}, and hence the peaks in their Fourier spectra differ slightly by position \cite{yarmo23} and considerably in amplitude, with the latter also affecting their visibility. In general, the envelope oscillations in \(n_{\text{t}}(t)\) are less pronounced than in \(n_{\text{ph}}(t)\) [Fig.~\ref{fig:nph_nt_w0_145}], making \(\omega_{\text{df}}\) less a peak than a kink in the zero-frequency component of \(\tilde{n}_{\text{t}} (\omega)\), while the contrast in the \(\omega_{0}\) and \(2 \omega_{0}\) peaks is also weaker at this choice of phonon frequency.  

\begin{figure}[t]
	\includegraphics[width=\linewidth]{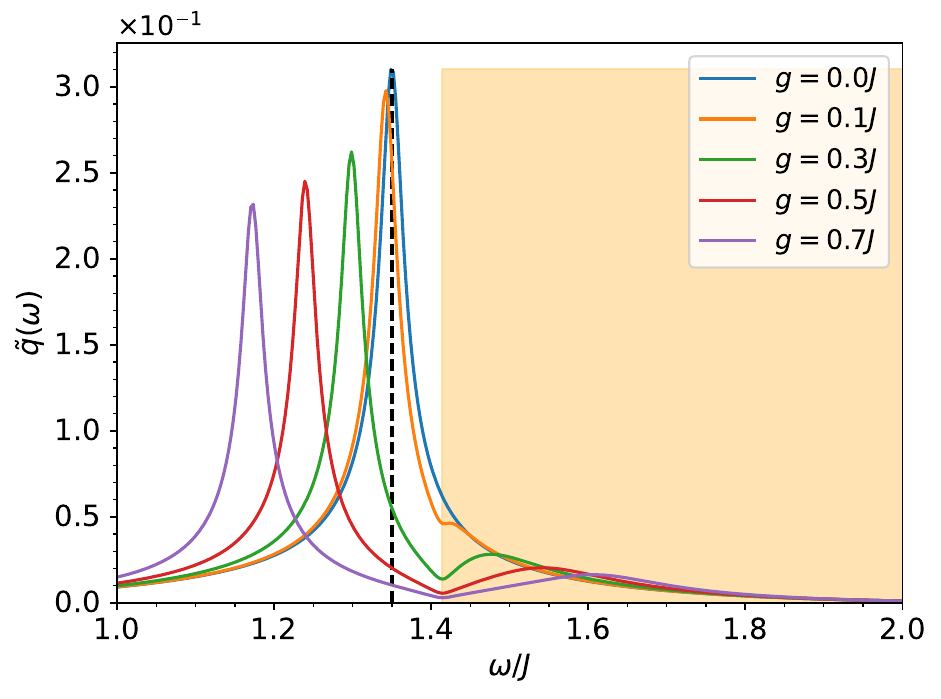}
	\caption{Fourier transform, \(\tilde{q} (\omega)\), obtained with a standard driving pulse incident on a spin chain for five different couplings, \(g\), to a phonon of frequency \(\omega_{0} = 1.35J\).}
	\label{fig:ft_q_w0_135}
\end{figure}

Studying the Fourier transforms of the three diagnostic observables as functions of \(g\) leads us to observe that the response of the system to a single, ultrashort pulse that excites a wide range of frequencies allows one to recover all of the results shown in  Fig.~4 of Ref.~\cite{yarmo23}. We recall that these were obtained by many long-time runs on a dense frequency grid to reach a NESS, in which the only remaining frequency is that of the drive and the observable probed is the amplitude of the response. Borrowing from the physical explanations of these authors, at large $g$ \(\omega_{\text{out}}\) and \(\omega_{\text{in}}\) are the frequencies of two mutually repelling magnetoelastic excitations that both have strongly mixed phonon-bitriplon character. The nature of the composite excitations is that the driven phonons are strongly dressed by the bitriplons they create through Eq.~\eqref{eq:hsp}, with their hybridization controlled by $g$. The beating phenomenon is a consequence of the fact that this hybridization establishes pairs of modes at relatively close frequencies, and the excitation appearing at \(\omega_{\text{df}}\) is explicitly a transient nonequilibrium feature that does not appear in a NESS. We defer to Sec.~\ref{sec:phen} a systematic analysis of the parameter-dependence of \(\omega_{\text{df}}\).
	
\subsection{Resonant frequency \(\omega_{0}\)}

In Sec.~\ref{sec:res}A we tested the effect of \(g\) at one phonon frequency, \(\omega_{0}=1.45J\), which lies just inside the bitriplon band in a region with high density of bitriplon states. Here we illustrate the differences arising when the phonon frequency lies just outside the bitriplon band, at  \(\omega_{0} = 1.35J\), and when the phonon lies well inside the band, at \(\omega_{0} = 1.7J\). We do not show results for frequencies around the upper bitriplon band edge (\(2\omega_{\rm max}=\sqrt{6}J\)) because the phenomena are qualitatively the same.

\begin{figure}[t]
	\includegraphics[width=\linewidth]{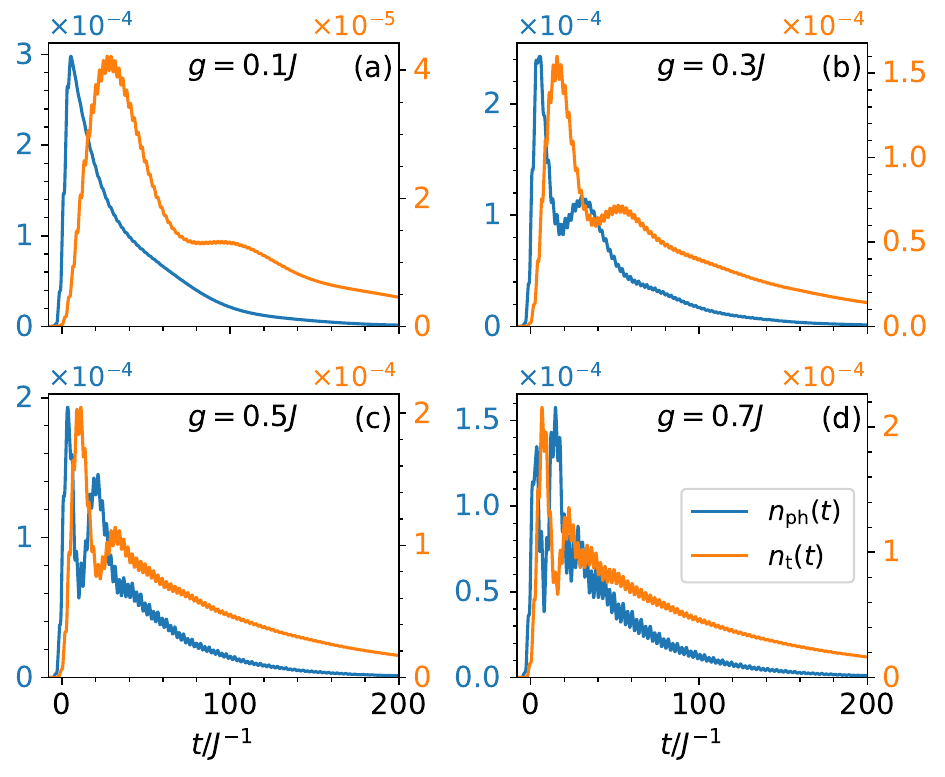}
	\caption{Nonequilibrium phonon occupation, \(n_{\text{ph}} (t)\), and triplon occupation, \(n_{\text{t}} (t)\), shown for an alternating chain coupled to a phonon of frequency \(\omega_{0} = 1.35J\) by four different \(g\) values and subject to driving by a standard pulse.}
	\label{fig:nph_nt_135}
\end{figure}

\subsubsection{\(\omega_{0} = 1.35J\)}

In principle choosing \(\omega_{0}=1.35J\) instead of \(1.45J\) exchanges the character of the ``phononic'' and ``triplonic'' hybrids, with the former now the mode at \(\omega_{\text{out}}\) and the latter the resonance at \(\omega_{\text{in}}\). Nevertheless, \({\tilde q} (\omega)\) in Fig.~\ref{fig:ft_q_w0_135} is very similar to the situation in Fig.~\ref{fig:ft_q_w0_145} as regards the effect of \(g\) on the mode repulsion. The most noticeable difference in the \(\omega_{\text{out}}\) mode is that its intensity is larger at small \(g\) and falls with increasing \(g\), as phonon driving energy inserted outside the band is transferred into the band by the coupling. The \(\omega_{\text{in}}\) resonance is very small at small \(g\) and grows from the lower band edge, becoming broader in energy rather than higher in intensity as \(g\) becomes large. 

Turning to the observables \(n_{\text{ph}} (t)\) and \(n_{\text{t}} (t)\), the presence of two mutually repelling hybrid modes ensures that the phenomenology of the slow envelope oscillation (beat) remains equally robust, as shown in Fig.~\ref{fig:nph_nt_135}. The most important contrast with Fig.~\ref{fig:nph_nt_w0_145} is that the return of \(n_{\text{ph}} (t)\) to precisely zero at the end of the first cycle is no longer assured. We state without illustration that this has only a minor effect on the visibility of the $\omega_{\rm df}$ peak in the corresponding \(\tilde{n}_{\text{ph}} (\omega)\) and \(\tilde{n}_{\text{t}} (\omega)\) [cf.~Figs.~\ref{fig:ft_nph_nt_w0_145}(a,d)].

\begin{figure}[t]
	\includegraphics[width=\linewidth]{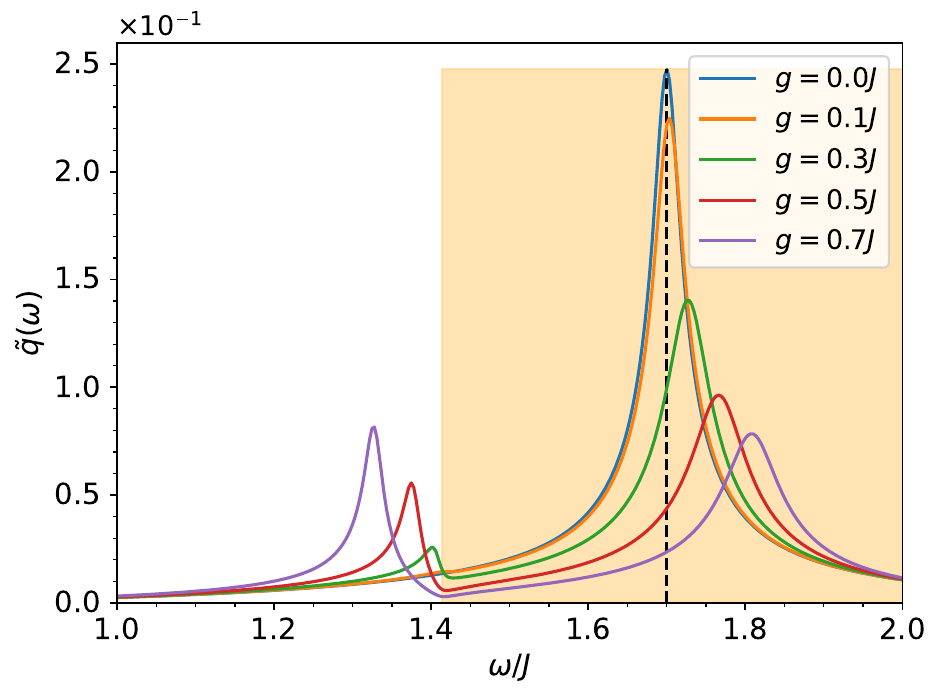}
	\caption{Fourier transform, \(\tilde{q} (\omega)\), obtained with a standard driving pulse incident on a spin chain for five different couplings, \(g\), to a phonon of frequency \(\omega_{0} = 1.7J\).}
	\label{fig:ft_q_w0_170}
\end{figure}

\subsubsection{\(\omega_{0} = 1.7J\)}

When \(\omega_0\) lies well inside the bitriplon band, the phononic hybrid again falls in amplitude as \(g\) increases and its triplon dressing becomes stronger, as shown in Fig.~\ref{fig:ft_q_w0_170}, but this resonance remains far sharper at large \(g\) than the analogous situation at \(\omega_{0} = 1.45J\) (Fig.~\ref{fig:ft_q_w0_145}). This result can be traced to the significantly smaller density of bitriplon states at \(\omega_{0} = 1.7J\) (Fig.~\ref{fig:bidisp}). Despite the large separation in frequency between the phonon and the lower band edge, a mutual level repulsion remains clear as a triplonic hybrid mode is generated and pushed below the bitriplon band by increasing $g$. We remark that this situation represents a qualitatively different origin for the difference (beat) frequency $\omega_{\rm df}$, in that it is dominated by $|\omega_0 - 2\omega_{\rm min}|$ rather than by $g^2$, as was the case at \(\omega_{0} = 1.35J\) and \(1.45J\), and we will revisit this difference in Sec.~\ref{sec:phen}.

\begin{figure}[t]
	\includegraphics[width=\linewidth]{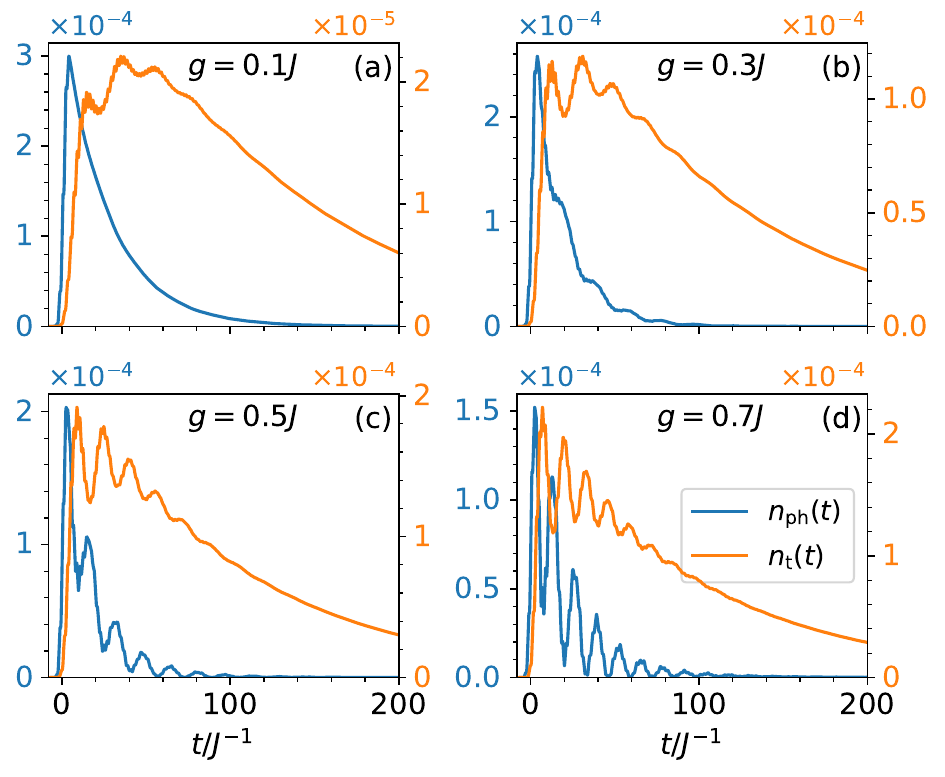}
	\caption{Nonequilibrium phonon occupation, \(n_{\text{ph}} (t)\), and triplon occupation, \(n_{\text{t}} (t)\), shown for an alternating chain coupled to a phonon of frequency \(\omega_{0} = 1.7 J\) by four different \(g\) values and subject to driving by a standard pulse.}
	\label{fig:nph_nt_w0_170}
\end{figure}

In this case \(n_{\text{ph}} (t)\) and \(n_{\text{t}} (t)\) show very rapid antiphase envelope oscillations (Fig.~\ref{fig:nph_nt_w0_170}), reflecting the large $|\omega_0 - 2\omega_{\rm min}|$, which survive to a significantly larger number of repetitions (6-7 at large $g$) than was the case at \(\omega_{0} = 1.35J\) and \(1.45J\). This more underdamped behavior can also be taken as a consequence of the lower triplon DOS at this value of $\omega_0$, which restricts the rate at which complete phonon-bitriplon mixing can be achieved. Despite the fidelity of the intersectoral energy oscillations at \(\omega_{0} = 1.7J\), the amplitude of the  beating envelope is not as large in \(n_{\text{ph}}(t)\) as was the case at \(\omega_{0} = 1.45J\), and certainly there is no reduction to zero at the end of the first cycle. By contrast, the envelope oscillations in \(n_{\text{t}}(t)\) are in fact more pronounced than those in Fig.~\ref{fig:nph_nt_w0_145}. 
	
\begin{figure*}[t]
	\includegraphics[width=\linewidth]{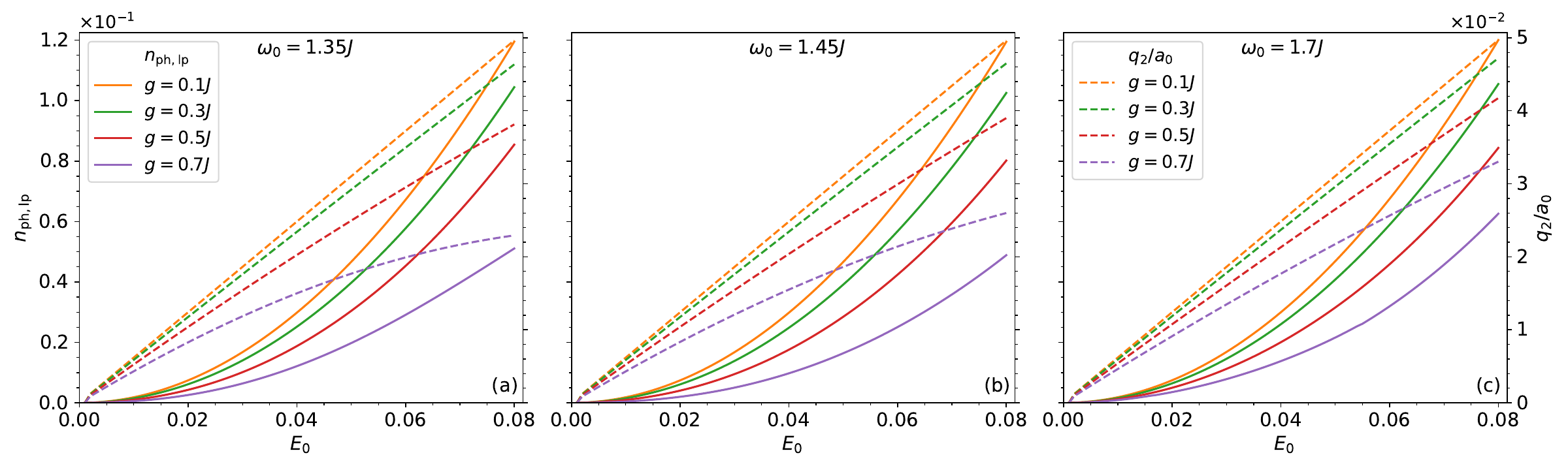}
	\caption{Evolution of the phonon displacement and phonon number with the peak value of the electric field, \(E_{0}\), in the pulse, shown for four values of $g$. To characterize the displacement, we consider the second peak in the time trace of $q(t)/a_{0}$, denoted as \(q_{2}(t)/a_{0}\) (dashed lines). To characterize the number, we consider the last peak in the time trace of $n_{\text{ph}}(t)$ that falls within the pulse envelope, denoted as \(n_{\rm ph,lp}(t)\) (solid lines). These results illustrate the departure from linearity of $q_{2}(t)/a_{0}$ with increasing $g$ and the quadratic dependence of $n_{\text{ph,lp}}(t)$, albeit with a $g$-dependent prefactor.}
	\label{fig:max_q_nph_E0}
\end{figure*}

\begin{figure*}[t]
	\includegraphics[width=0.96\linewidth]{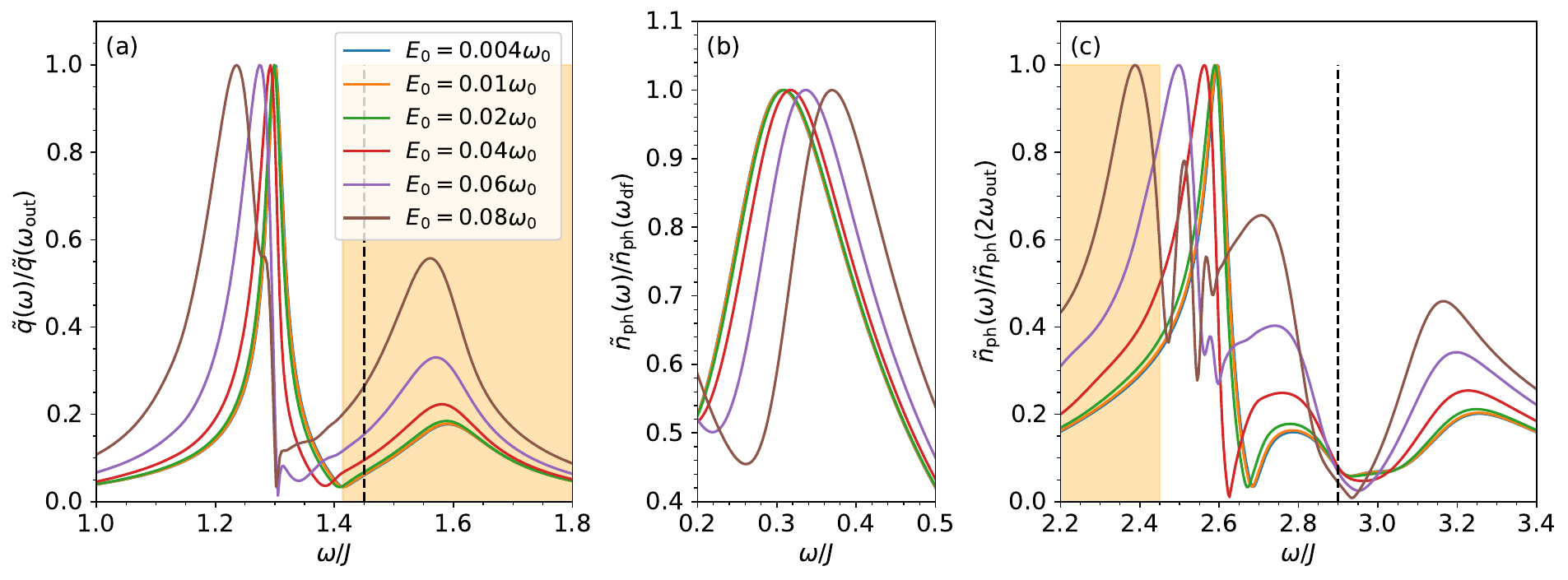}
	\caption{Relative absolute values of the Fourier transforms of the standard diagnostic quantities computed with \(\omega_{0} = 1.45J\) and \(g = 0.5J\) for a range of peak electric-field values, \(E_{0}\), of the incident pulse.
	(a) Phonon displacement, \(\tilde{q}(\omega)\).
	(b) Phonon number, \(\tilde{n}_{\rm ph}(\omega)\), in a window around the difference frequency, \(\omega_{\rm df}\). 
	(c) Phonon number, \(\tilde{n}_{\rm ph}(\omega)\), in a window covering the three sum frequencies, which appear as the three primary peaks.}
	\label{fig:ft_nph_E0}
\end{figure*}

We conclude this subsection by summarizing again the structure of the Fourier transforms, \({\tilde q} (\omega)\), shown in the spectra of Figs.~\ref{fig:ft_q_w0_145}, \ref{fig:ft_q_w0_135}, and \ref{fig:ft_q_w0_170}. The sharp modes appearing outside the bitriplon band are single Lorentzian functions whose width is determined by the damping parameters $\gamma$ and $\gamma_s$. The origin of these peaks is the phonon in Fig.~\ref{fig:ft_q_w0_135}, whereas in Figs.~\ref{fig:ft_q_w0_145} and \ref{fig:ft_q_w0_170} a single mode of triplonic origin is pushed progressively further below the bitriplon band with increasing $g$. The broad resonances inside the band can be considered as superpositions of multiple Lorentzians appearing as many bitriplons are excited by the driving process. If the phonon is inside the bitriplon band then the peak appears at a renormalized value of $\omega_0$, and otherwise the peak position is determined by the contributions from excited bitriplon modes in a broad distribution whose width is dictated by $g$ (i.e.~by hybridization). The precise peak positions $\omega_{\text{in}}$ and $\omega_{\text{out}}$ then determine the beat frequency \(\omega_{\text{df}}\) appearing as a peak in Fig.~\ref{fig:ft_nph_nt_w0_145}, and analogously at the other driving phonon frequencies. 

\subsection{Laser amplitude \(E_{0}\)} \label{ssec:re0}

Increasing the electric field, \(E_{0}\), raises the input energy per pulse and hence the number of excited phonon quanta, which are then transferred to triplon excitations. In a real material, very strong $E_0$ will produce heating out of the quantum regime, a situation considered for continuous driving in Ref.~\cite{yarmo21}. The equations of motion we solve [Eqs.~\eqref{eq:eom}] do allow the possibility of including finite temperatures, which enter in the Bose occupation functions appearing in the dissipative terms in the discussion below Eq.~\eqref{eq:gksl}. For our present analysis, however, we use the fact that the driving pulse is ultrashort to consider only the nonthermal dynamical regime soon after pulse application, and do not consider the system-wide thermal equilibration that is expected at much longer times.

In our treatment a large $E_0$ could have two possible nonphysical consequences. One would be a violation of the condition for the validity of the quadratic triplon approximation, which we set \cite{yarmo21} as being that the number of triplons per dimer should remain below 0.2. The other would be an instability in our numerical (Runge-Kutta) solution of the equations of motion, leading to a divergence of the expectation values we compute. We find that in the physical range of $E_0$ values it is not difficult to avoid both. Borrowing approximate values from Sec.~\ref{sec:CuGeO3} and working to one significant figure, $E_0  = 0.1\omega_0$ with $\omega_0 \approx 10$ meV, dropped over a unit cell size of 1 nm, is an electric field of 10 MV cm$^{-1}$, which is close to the limit of modern ultrafast laser technology. We therefore take a maximum electric field of \(E_{0} = 8 \times 10^{-2}\omega_{0}\), which because our standard field amplitude is \(4 \times 10^{-3}\omega_{0}\) gives us a factor-20 range over which to analyze $E_0$ effects.
  
\begin{figure}[t]
	\includegraphics[width=\linewidth]{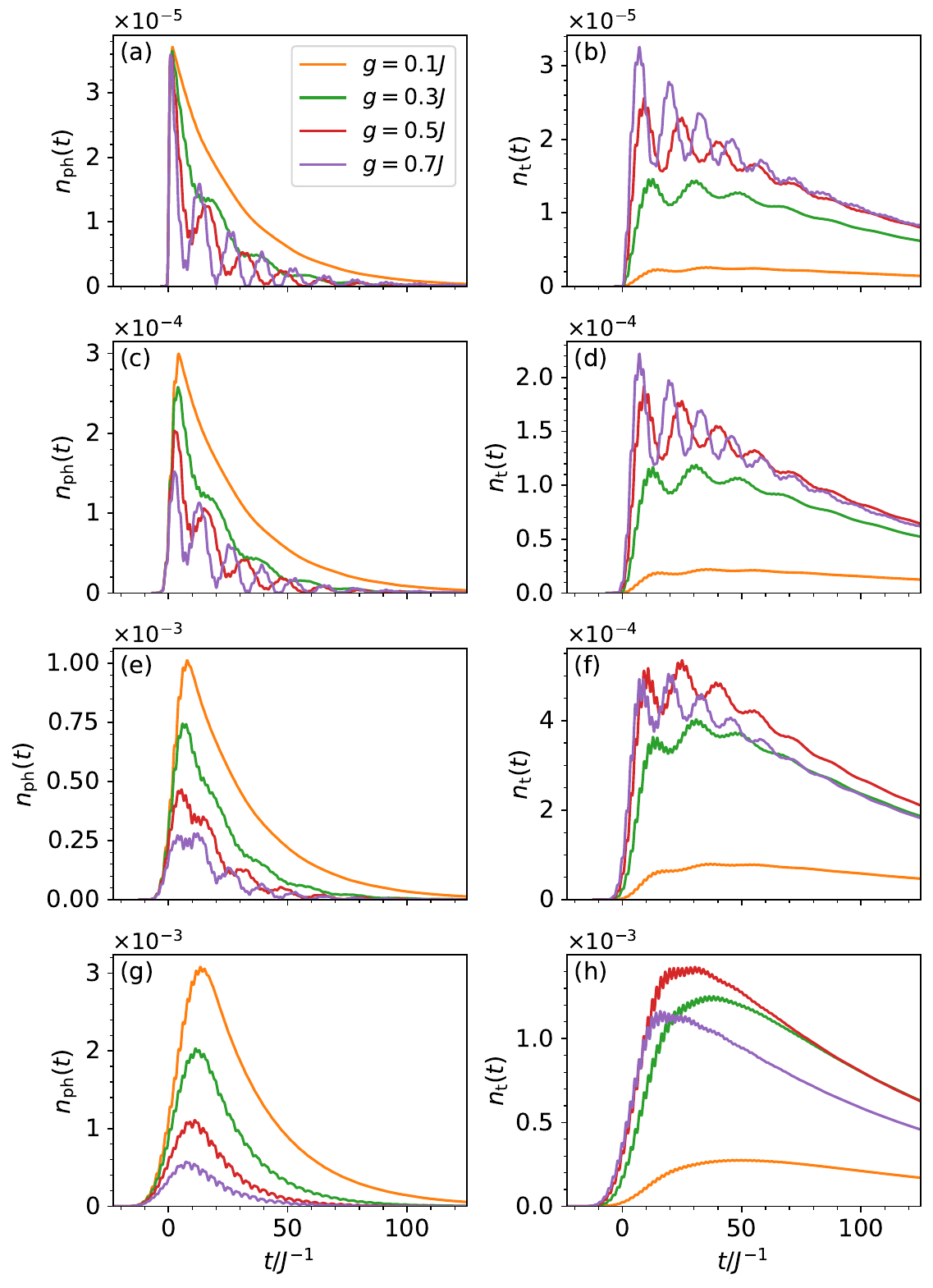}
	\caption{Nonequilibrium phonon and triplon occupations, \(n_{\text{ph}}(t)\) (a,c,e,g) and \(n_{\text{t}}(t)\) (b,d,f,h), shown for an alternating chain coupled with different strengths \(g\) to a phonon of frequency \(\omega_{0} = 1.7J\), subject to driving by pulses with the four different lengths shown in Fig.~\ref{fig:pulses}(a), i.e.~$T_p = 2 \pi n/\omega_0$ with \(n = 1\) (a,b), 3 (c,d), 6 (e,f), and 12 (g,h).}
	\label{fig:nph_pulselength}
\end{figure}

The equations of motion [Eqs.~\eqref{eq:eom}] are deterministic expressions for the expectation values, and hence quantities such as the maximum or average of \(q(t)\) should increase linearly with \(E_{0}\), while \(n_{\text{ph}}(t)\) and therefore \(n_{\text{t}}\) should increase quadratically. This behavior is clearly visible at low $g$ in Fig.~\ref{fig:max_q_nph_E0}, where the peak values of $q(t)$ and $n_{\text{ph}}(t)$ at equivalent points in their respective time traces are shown as functions of the peak electric field. The departure from ideal linear or quadratic dependence, which reflects clearly the nonlinear feedback in the magnetophononic equations of motion \cite{yarmo21}, grows systematically with $g$ and is strongest for phonon frequencies close to the lower band edge (i.e.~to energy regimes with high densities of bitriplon states).

On the energy axis, increasing the electric field at low $E_0$ values does not change the position, the shape, or the number of oscillations. At larger $E_0$, however, we observe in Fig.~\ref{fig:ft_nph_E0} that the characteristic frequencies appearing in the spectral function do begin to shift. Here we recall that the previous analysis with continuous driving \cite{yarmo23} discovered finite field-induced $q$ values and hence investigated shifts of the interaction parameters, leading to ``spin-band engineering'' by the electric field. In Sec.~\ref{sec:phen}C we will investigate more quantitatively how such band engineering can be recovered in an experiment using ultrashort pulses. For the parameters of Fig.~\ref{fig:ft_nph_E0}, it is clear that the spectral peaks around $2\omega_0$ in ${\tilde n}_{\rm ph}(\omega)$ are renormalized downwards [Fig.~\ref{fig:ft_nph_E0}(c)], while $\omega_{\rm df}$ moves slightly upwards [Fig.~\ref{fig:ft_nph_E0}(b)], indicating that \(\omega_{\rm out}\) and \(\omega_{\rm in}\) are shifted downward by differing amounts [Fig.~\ref{fig:ft_nph_E0}(a)].

\subsection{Pulse width \(T_{p}\)}

Finally we illustrate the effect of changing the pulse width, \(T_{p} = 2 \pi n/ \omega_{0}\). Generically, driving for a longer time adds more energy to the system, leading to higher values of \(n_{\text{ph}} (t)\) in Fig.~\ref{fig:nph_pulselength}. We observe that the single-cycle $n = 1$ pulse causes the same driving effects as our standard $n = 3$ pulse, with rapid oscillations at $\omega_0$ and envelope oscillations at $\omega_{\rm df}$. However, driving with a long $n = 12$ pulse destroys the antiphase oscillation effect and leaves only minimal slow oscillations of the envelope. We assume that this is a consequence of the long pulse continuing to drive the phonon while its natural tendency in the absence of driving would be to return energy to the spin system, thereby cancelling this effect. We stress that this does not mean the resonance frequencies of the system ($\omega_{\rm df}$, \(\omega_{\rm out}\), \(\omega_{\rm in}\), and the sum frequencies) have changed, only that the initial conditions have altered the visibility of the antiphase oscillation in the time domain. Finally, the intermediate case of an $n = 6$ pulse shows that the clarity of the antiphase oscillation is already affected by pulses of this length. Given that conventional experiments in condensed matter with coherent THz light use ultrashort pulses, varying from single-cycle to few-cycle ($n = 1$ and $n = 3$ in our formulation), we focus henceforth only on $n = 3$ as a representative short pulse. 

\section{Phonon-bitriplon approximation}
\label{sec:pba}

\subsection{Phonon-Bitriplon Hamiltonian}

To achieve an analytical description of the magnetophononic problem, we introduce an approximation that is based on taking bitriplons as the quasiparticles of the system. As noted in Sec.~\ref{sec:mm}, the phonon in the spin-phonon coupling expressed by Eq.~\eqref{eq:hsp} does not couple to single triplons, but only to pairs of triplons with total momentum zero, meaning with opposite momenta of the two participating triplons. Because a single triplon has spin \(S = 1\), a bitriplon can have total spin \(S = 0\), 1, or 2. However, the phonons we consider have zero spin and thus couple only to the $S = 0$ bitriplon, which is the symmetric superposition of all three flavors \(\alpha = x\), \(y\), and \(z\) in \(\sum_{k,\alpha} \tilde{t}^{\dagger}_{k,\alpha} \tilde{t}^{\dagger}_{-k,\alpha}\). 

We introduce the bitriplon annihilation and creation operators, respectively
\begin{equation}
\label{eq:bto}
\hat{B}_{k,\alpha} = \tilde{t}_{k,\alpha} \tilde{t}_{-k,\alpha} \;\; {\rm and} \;\; \hat{B}_{k,\alpha}^{\dagger} = \tilde{t}_{k,\alpha}^{\dagger} \tilde{t}_{-k,\alpha}^{\dagger}. 
\end{equation}
To reexpress the Hamiltonian of Eq.~\eqref{eq:Ham_decoupled}, we consider first the case with no driving and at zero or low temperature. The approximation makes use of the fact that the diagonal triplon terms are very small, \(\tilde{t}_{k,\alpha}^{\dagger} \tilde{t}_{k,\alpha} \approx 0\) for all \(k\), to neglect these in the coupling to the phonon. This can be justified by the argument that the influence of the diagonal terms is limited at low energies because they do not change the triplon occupation, and in particular if the system is in its triplon vacuum then it would stay there if only the diagonal terms were present.

Proceeding from Sec.~\ref{sec:mm}, in the absence of the laser we consider
\begin{equation}
	H = H_{\rm s} + H_{\rm {sp}} + H_{\rm p}
\end{equation}
with
\begin{equation}
	\label{eq:phonon}
	H_{\rm p} = \omega_{0} \hat{b}_{0}^{\dagger} \hat{b}_{0} 
\end{equation}
up to constant terms. For use in describing the phonon dynamics, we define $\hat x = \tfrac{1}{\sqrt{2}} (\hat{b}_{0}^{\dagger}+ \hat{b}_{0})$ and $\hat p = \tfrac{1}{\sqrt{2}} i(\hat{b}_{0}^{\dagger} - \hat{b}_{0})$. For the properties of the bitriplon operators [Eq.~\eqref{eq:bto}], we distinguish the cases \(k \neq 0, \pi\) from \(k = 0,\pi\) and express the commutators as
\begin{subequations}
	\begin{align}
		\comm{\hat{B}_{k,\alpha}}{\hat{B}_{k,\alpha}} & = \comm{\hat{B}_{k,\alpha}^{\dagger}}{\hat{B}_{k,\alpha}^{\dagger}} = 0 
		\text{, }\forall \,\, k. \\
		\comm{\hat{B}_{k,\alpha}}{\hat{B}_{k,\alpha}^{\dagger}} & = 1 + \tilde{t}_{k,\alpha}^{\dagger}\tilde{t}_{k,\alpha} + \tilde{t}_{-k,\alpha}^{\dagger}\tilde{t}_{-k,\alpha}\\
		& \approx 1 \text{, for } 0<k<\pi. \\
		\comm{\hat{B}_{k,\alpha}}{\hat{B}_{k,\alpha}^{\dagger}} 
		& = 2 + 4\tilde{t}_{k,\alpha}^{\dagger}\tilde{t}_{k,\alpha} \approx 2 \text{, for } k \in \{0,\!\pi\},
	\end{align}
\end{subequations}
where we again neglect the diagonal terms. Although these terms will acquire finite values when the system is driven, neglecting them is justified as an approximate starting point that is borne in mind when interpreting the subsequent results. The crucial advantage of this approximation is that the bitriplon operators become standard bosonic operators except at the momenta $k \in \{0,\!\pi\}$.

To formulate a proper bosonic operator also for the cases $k \in \{0,\!\pi\}$, we define the renormalized operators \(B_{0,\alpha} = \tfrac{1}{\sqrt{2}} \tilde{t}_{0,\alpha} \tilde{t}_{0,\alpha}\) and \(B_{\pi,\alpha} = \tfrac{1}{\sqrt{2}} \tilde{t}_{\pi,\alpha} \tilde{t}_{-\pi,\alpha}\). Now we observe that the commutator of the bitriplon creation operator with the diagonal Hamiltonian, 
\begin{equation}
\comm{H_{s}}{\hat{B}_{k,\alpha}^{\dagger}} = 2\omega_{k}\hat{B}_{k,\alpha}^{\dagger} \text{, }\forall \,\,\, k,
\end{equation}
is identical to its commutator with a diagonal bitriplon term
\begin{equation}
	\comm{\hat{B}_{k,\alpha}^{\dagger}}{2\omega_{k}\hat{B}^{\dagger}_{k,\alpha}\hat{B}_{k,\alpha}} 
		= 2\omega_{k}\hat{B}_{k,\alpha}^{\dagger}\text{, }\forall \,\,\, k,
\end{equation}
which justifies replacing the diagonal triplon Hamiltonian by a diagonal bitriplon Hamiltonian,
\begin{equation}
	\label{eq:spin}
	H_{\rm s} = \sum_{0\le k\le\pi,\alpha}2\omega_{k}\hat{B}_{k,\alpha}^{\dagger}\hat{B}_{k,\alpha}. 
\end{equation}

Finally, neglecting the diagonal triplon term in the spin-phonon coupling [Eq.~\eqref{eq:hspbo}] yields
\begin{align}
	\label{eq:spin-phonon}
		H_{\rm sp} & = \frac{\sqrt{2} g \hat{x}}{\sqrt{N}} \bigg[ \sum_{0<k<\pi,\alpha} y'_{k} \left( \hat{B}_{k,\alpha}^{\dagger} + \hat{B}_{k,\alpha} \right) \nonumber \\ 
		& \hspace{2.0cm} + \sum_{k \in \{0,\!\pi\},\alpha} \tfrac{1}{\sqrt{2}} y'_{k} \left( \hat{B}_{k,\alpha}^{\dagger} + \hat{B}_{k,\alpha}\right) \bigg],
	\end{align}
where the factor \(\tfrac{1}{2}\) in Eq.~\eqref{eq:hspbo} is compensated by using the interval \(0 < k < \pi\) and the symmetry $k \leftrightarrow -k$ instead of the full Brillouin zone. Because the approximate Hamiltonian, $H$, consisting of $H_\text{p}$ in Eq.~\eqref{eq:phonon}, $H_\text{s}$ in Eq.~\eqref{eq:spin}, and $H_\text{sp}$ in Eq.~\eqref{eq:spin-phonon} is bilinear in bosonic operators, it can be diagonalized with only moderate effort. 

One way to determine the eigenvalues of $H$ is to define the position and momentum operators 
\begin{subequations}
	\begin{align}
		\hat{X}_{k,\alpha} &= \tfrac{1}{\sqrt{2}}\left( \hat{B}_{k,\alpha}^{\dagger} + \hat{B}_{k,\alpha}\right)
		\\
		\hat{P}_{k,\alpha} &= \tfrac{i}{\sqrt{2}}\left( \hat{B}_{k,\alpha}^{\dagger} - \hat{B}_{k,\alpha} \right)
	\end{align}
\end{subequations}
that reexpress the bitriplons [Eq.~\eqref{eq:bto}] in a form with the canonical commutation relations of harmonic oscillators. This transformation leads to 
\begin{equation}
	H_{\rm s} = \sum_{0\le k\le\pi,\alpha}\omega_{k}
	\left(\hat{X}_{k,\alpha}^{2} + \hat{P}_{k,\alpha}^{2}\right)	,
\end{equation}
where again we have removed constant terms (energy shifts). The spin-phonon term,
\begin{align}
	H_{\rm sp} & \! = \! \frac{\sqrt{2}g\hat{x}}{\sqrt{N}} \! \bigg[ \! \sum_{0<k<\pi,\alpha} \!\! \sqrt{2}y'_{k} \hat{X}_{k,\alpha} \! + \! \sum_{k\in\{0,\pi\},\alpha} \!\! y'_{k} \hat{X}_{k,\alpha} \! \bigg],
\end{align} 
is reduced to a coupling between the coordinate of a ``phononic'' harmonic oscillator and the \(3(N+1)/2\) coordinates of ``bitriplonic'' harmonic oscillators. To express the phonon-bitriplon Hamiltonian in a compact and convenient final form, we renormalize the operators again to the forms \(\hat{x} \rightarrow \hat{x} \sqrt{\omega_{0}}\), \(\hat{p} \rightarrow \hat{p} /\sqrt{\omega_{0}}\), \(\hat{X}_{k,\alpha} \rightarrow \hat{X}_{k,\alpha} \sqrt{2\omega_{k}}\), and \(\hat{P}_{k,\alpha} \rightarrow \hat{P}_{k,\alpha}/\sqrt{2\omega_{k}}\), and define the coefficients
\begin{subequations}
	\begin{align}
		f_k &= 2g\sqrt{2\omega_0\omega_k} y'_k & &\text{for} \quad 0<k<\pi
		\\
		f_k &= 2g\sqrt{\omega_0\omega_k} y'_k &  &\text{for} \quad k\in\{0,\pi\}.
	\end{align}
\end{subequations}
These measures then allow us to write the Hamiltonian as 
\begin{align} 
	H & = \frac{1}{2} \Big( \hat{p}^{2}+ \!\!\mathop{\sum_{0\le k\le\pi}}_\alpha \hat{P}_{k,\alpha}^{2}\Big) 
	+ \frac{1}{2}\Big(\omega_{0}^{2} \hat{x}^{2} + \!\!\mathop{\sum_{0\le k\le\pi}}_\alpha  
	4\omega_{k}^{2}\hat{X}_{k,\alpha}^{2}\Big) \nonumber \\
	& \hspace{2.0cm} + \frac{\hat{x}}{\sqrt{N}}\mathop{\sum_{0\le k\le\pi}}_\alpha  f_k \hat{X}_{k,\alpha} . 
	\label{eq:PBA_Ham_XP}
\end{align}

Because the momentum part (the first bracket) of Eq.~\eqref{eq:PBA_Ham_XP} is diagonal, one need only diagonalize the position part, which can be represented by a $1 + \tfrac{3}{2}(N+1)$-dimensional matrix, $M$. The term in the second bracket is also purely diagonal, while the last term represents matrix elements that fill only the first column and the first row. It is sufficient to diagonalize $M$ to obtain the squares of the eigenvalues of the bilinear bosonic Hamiltonian. For our purposes, however, the eigenvalues are less of a focus than the linear response function, $\chi(\omega)$, of the primary variables to the driving of the phonon coordinate. To reintroduce this driving, we transform the coupling to the external electric field into the new variables as 
\begin{subequations}
	\label{eq:laser}
	\begin{align}
		\label{eq:laser-a}
		H_\text{L} &= E(t) \sqrt{N} (b_0 + b_0^\dag) 
		\\
		\label{eq:laser-b}
		& = E(t) \sqrt{2\omega_0N} \hat x = E(t) N \hat q.
	\end{align}
\end{subequations}

\subsection{Linear response function}

Although the Kubo formula presents a standard method for obtaining linear response functions, here we determine the linear response directly because the inclusion of dissipation is more transparent in this approach. We first establish the equations of motion obtained from the adjoint master equation [Eq.~\eqref{eq:gksl}] by using as the Lindblad operators $b_0^\dag$ and its Hermitian conjugate with damping coefficient $\gamma$ and $\tilde t_{k,\alpha}^\dag$ and its Hermitian conjugate with damping $\gamma_s$. We stress that here we use the triplonic creation operator again, not the bitriplonic one, which implies a factor-2 difference in the effect of the damping, as we will see below. The adjoint master equation describes the behavior of expectation values of the operators, which we denote by the same symbols as the operators but without the hats (schematically $\langle \hat{o} \rangle = o$). For the phonon dynamics we obtain
\begin{subequations}
	\begin{align}
		& \dot{x}  = p - \tfrac12 {\gamma} x, \\
		& \dot{p} = - \omega_{0}^{2} x - \tfrac12 {\gamma} p \nonumber \\
		& \qquad - \frac{1}{\sqrt{N}}\mathop{\sum_{0\le k\le\pi}}_\alpha f_k {X}_{k,\alpha} - E \sqrt{2\omega_0 N},
	\end{align}
\end{subequations}	
and for the bitriplons
\begin{subequations}
	\begin{align}		
		& \dot{X}_{k,\alpha} = {P}_{k,\alpha} - \gamma_s {X}_{k,\alpha} \\
		& \dot{{P}}_{k,\alpha} = - 4 \omega_{k}^{2} {X}_{k,\alpha} - \gamma_s {P}_{k,\alpha} - \tfrac{1}{\sqrt{N}} f_k x.
	\end{align}
\end{subequations}

\begin{figure}[t]
	\includegraphics[width=\linewidth]{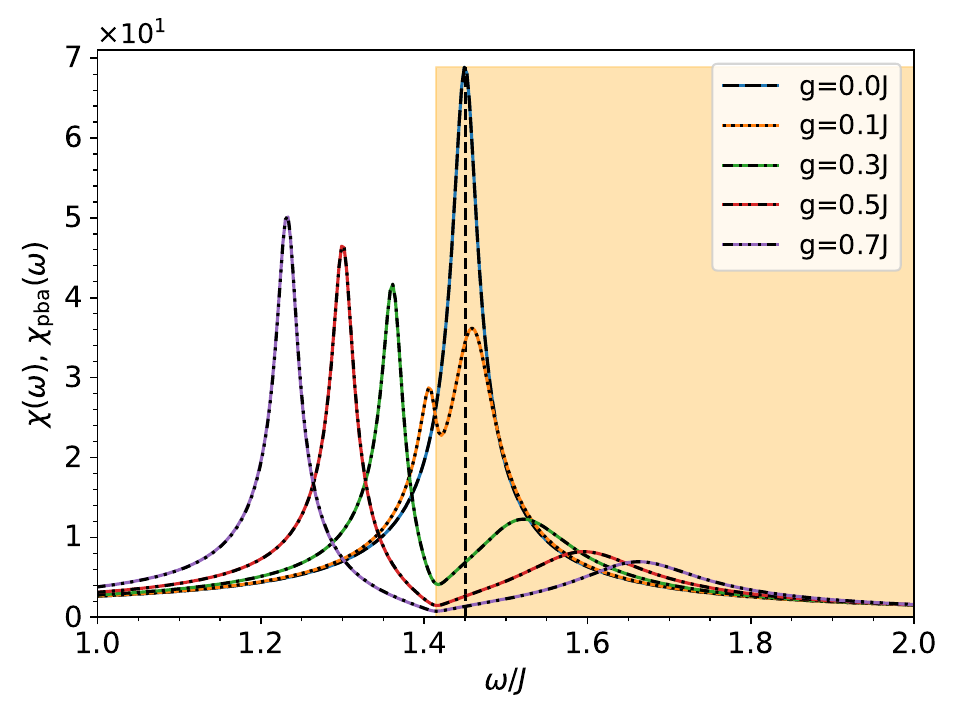}
	\caption{Linear response functions \(\chi_{\rm pba}(\omega)\), computed within the phonon-bitriplon approximation (PBA) of Eq.~\eqref{eq:resolvant2} (dashed lines), and \(\chi (\omega)\), computed from Eqs.~\eqref{eq:eom} (solid lines). Both sets of calculations are performed for five different \(g\) values at phonon frequency \(\omega_{0} = 1.45J\) with \(\gamma = 0.02\omega_{0}\) and \(\gamma_{\rm s} = 0.01J\). The solid lines are precisely the data of Fig.~\ref{fig:ft_q_w0_145} for $\tilde q(\omega)$ divided by the peak electric field, $E_0 = 4 \times 10^{-3} \omega_0$.}
	\label{fig:chi_bitrip}
\end{figure}

Next we transform all of the time-dependent quantities into frequency space and solve for $\tilde x(\omega)$. In our standard Fourier-transform convention, the time derivatives are replaced with multiplication by $-i\omega$, bringing the four equations of motion to the forms
\begin{subequations}
	\begin{align}
		\label{eq:lin-res-x}
		- i (\omega + i \tfrac12 \gamma) \tilde x  & = \tilde p, \\
		\label{eq:lin-res-p}
		- i (\omega + i \tfrac12 \gamma) \tilde p  & = - \omega_0^2 \tilde x - \frac{1}{\sqrt{N}} \mathop{\sum_{0 \le k \le \pi}}_\alpha f_k {\tilde X}_{k,\alpha}, \\ \nonumber
		& \qquad\qquad - \tilde E \sqrt{2\omega_0 N} \\
		- i (\omega + i \gamma_s) {\tilde X}_{k,\alpha} & = {\tilde P}_{k,\alpha}, \\
		- i (\omega + i \gamma_s) {\tilde P}_{k,\alpha} & = - 4 \omega_{k}^{2}{\tilde X}_{k,\alpha} - \tfrac{1}{\sqrt{N}} f_k \tilde x.
	\end{align}
\end{subequations}
Solving the latter two equations for ${\tilde X}_{k,\alpha}$ yields
\begin{equation}
	\tilde{X}_{k,\alpha}  = \frac{1}{\sqrt{N}}\frac{f_k \tilde x}{(\omega + i\gamma_{s})^{2} - 4\omega_{k}^{2}},
\end{equation}
which inserted into Eq.~\eqref{eq:lin-res-p} and combined with Eq.~\eqref{eq:lin-res-x} leads to 
\begin{equation}
	\left[(\omega + i \tfrac12 \gamma)^2-\omega_0^2 \right] \tilde x = \sigma(\omega) \tilde x + \sqrt{2\omega_0 N} \tilde{E},
\end{equation}
where we have defined the self-energy
\begin{align} 
	\label{eq:selfenergy}
	\sigma(\omega) & = \frac{3}{N}
	\sum_{0\le k\le\pi}\frac{f_k^2}{(\omega + i\gamma_{s})^{2} - 4\omega_{k}^{2}}.
\end{align}
In this way the dynamics of the phononic coordinate are expressed by the equation 
\begin{align} 
	\label{eq:resolvant1}
	\tilde x &= \frac{\sqrt{2\omega_0 N}}{(\omega + \tfrac12 i \gamma)^2 - \omega_0^2 - \sigma(\omega)} \tilde{E},
\end{align}
which from Eq.~\eqref{eq:laser-b} is equivalent to
\begin{align} 
	\label{eq:resolvant2}
	\tilde q &= \frac{{2\omega_0}}{(\omega + \tfrac12 i \gamma)^2 - \omega_0^2 - \sigma(\omega)} \tilde{E}.
\end{align}
The prefactor of $\tilde E$ on the right-hand side of Eq.~\eqref{eq:resolvant2} is the susceptibility, $\chi_\text{pba}(\omega)$, obtained within the phonon-bitriplon approximation.

$\chi_\text{pba}(\omega)$ is readily evaluated and is shown by the dashed lines in Fig.~\ref{fig:chi_bitrip} for the standard parameters of Sec.~\ref{sec:res}A. The solid lines in Fig.~\ref{fig:chi_bitrip} show the analogous susceptibility, $\chi(\omega) = \tilde q(\omega)/\tilde E(\omega)$, obtained numerically from the full equations of motion [Eqs.~\eqref{eq:eom}]. The excellent quantitative agreement expected at small $g$ in fact extends to all values of $g$ we consider, at least for the linear response and in the limit of weak $E_0$. In the next section we will observe where feedback processes of higher order in $E_0$ result in the phonon-bitriplon approximation failing to capture the full response. 

At the qualitative level, Eqs.~\eqref{eq:selfenergy} and \eqref{eq:resolvant2} reveal the structure of the spectral function. The bitriplonic degrees of freedom behave as damped harmonic oscillators with eigenfrequency \(2\omega_{k}\) and damping coefficient $\gamma_s$. The self-energy, $\sigma(\omega)$, is a sum over the responses of all the bitriplons that arises physically from the phonon-bitriplon coupling. The modification of the phononic oscillator due to its interactions with the bitriplons is captured well by this self-energy. The feedback of the phonon to the bitriplonic oscillators is, however, missing from the analytic approximation, which as a consequence cannot be expected to account for spin-band engineering. Nevertheless, the result that bitriplons form the basis for a reliable description of the spin sector in terms of damped harmonic oscillators builds a solid foundation for the discussion of mode energies and the physcal interpretation of hybridization and mutual repulsion invoked in our previous analysis and applied in the next section.

\section{Pulsed driving phenomenology} 
\label{sec:phen}

In Sec.~\ref{sec:res} we illustrated how, for a short driving pulse, the phonon-bitriplon excitation spectrum varies with the frequency of the coupled phonon (\(\omega_{0}\)), the strength of this coupling (\(g\)), and the electric-field strength of the pulse (\(E_{0}\)). Here we focus on the primary phenomena of near-resonant pulsed magnetophononics, using the insight gained in Sec.~\ref{sec:pba} to obtain a systematic understanding and comparing or contrasting the results with the situation obtained on continuous driving \cite{yarmo23}.

\subsection{Composite hybrid states} 

\begin{figure}[t]
	\includegraphics[width=\linewidth]{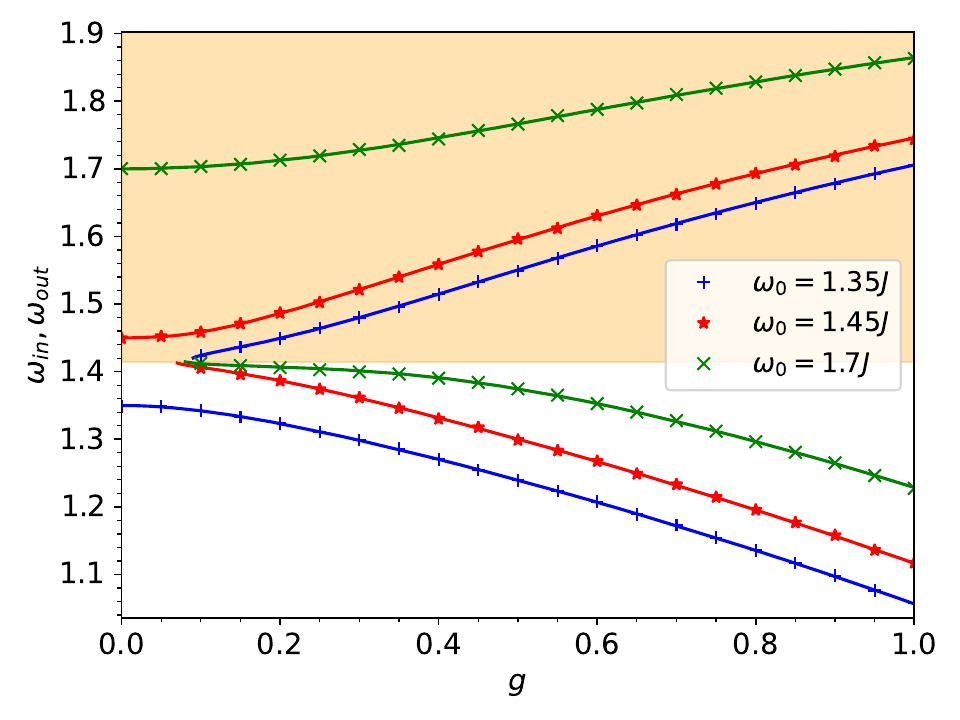}
	\caption{Peak frequencies (symbols) of the two mutually repelling hybrid excitations, \(\omega_{\rm in}\) lying within the bitriplon band and \(\omega_{\rm out}\) lying outside (below) it, obtained by standard driving of single phonons with three different frequencies around the lower bitriplon band edge and shown as functions of the spin-phonon coupling, \(g\). Solid lines show results obtained from the PBA [Eq.~\eqref{eq:resolvant2}].}
	\label{fig:hybg}
\end{figure}

Figures \ref{fig:ft_q_w0_145}, \ref{fig:ft_q_w0_135}, and \ref{fig:ft_q_w0_170} demonstrated the fundamental phenomenon of magnetophononics, that two composite modes are created by the hybridization of the single phonon with the bitriplons and their separation in energy rises with $g$. In Fig.~\ref{fig:hybg} we show how this mutual repulsion varies over a very wide range of $g$ values, and for each of our standard phonon frequencies. For \(\omega_{0} = 1.35J\), the lower-lying hybrid \(\omega_{\rm out}\) starts at $\omega_0$, making it a ``phononic'' mode at weak $g$, while \(\omega_{\rm in}\) lies initially at the lower band edge, making it a ``triplonic'' combination of high-density bitriplons. When $\omega_0$ lies within the bitriplon band, as for the other two choices, \(\omega_{\rm in}\) is necessarily phononic while \(\omega_{\rm out}\) lies initially at the lower band edge (triplonic) and is repelled downwards. 

The linear response function of Eq.~\eqref{eq:resolvant2}, shown by the solid lines in Fig.~\ref{fig:hybg}, captures the $g$-dependent repulsion exactly for all three phonon frequencies. This is a consequence of the fact that the repulsion is a linear phenomenon, with even our strongest $g$ values remaining in the linear regime for weak $E_0$ (Fig.~\ref{fig:ft_nph_E0}). One may also compare Fig.~\ref{fig:hybg} with Fig.~4(e) of Ref.~\cite{yarmo23}, where very similar results were found for \(g \leq 0.5\). In that work it was stated that both frequencies (\(\omega_{\rm in}\) and \(\omega_{\rm out}\)) follow a \(g^{2}\) form, which indeed is largely the case for small \(g\) also with pulsed driving, but for stronger $g$ values we find the behavior to be increasingly linear. 

\begin{figure}[t]
	\includegraphics[width=0.96\linewidth]{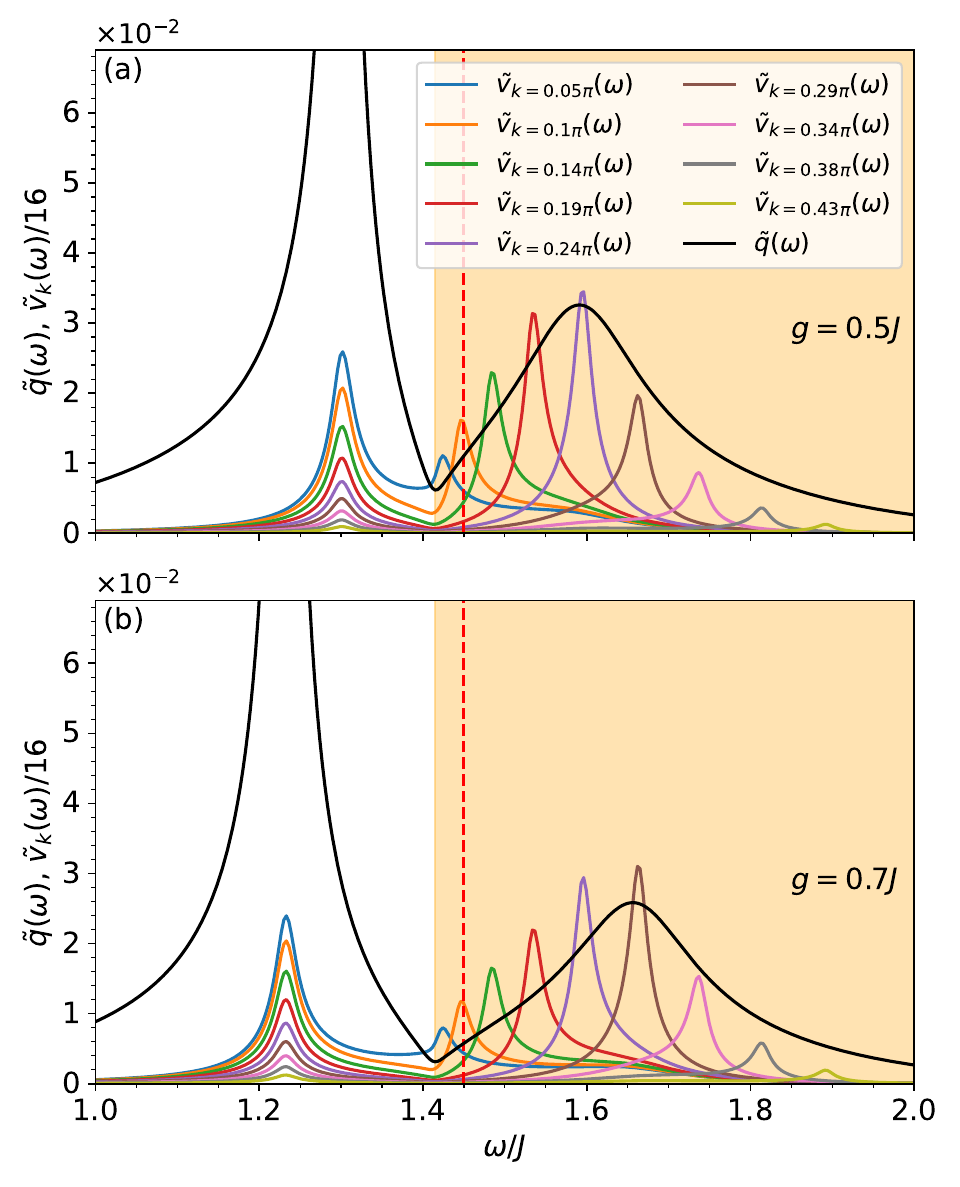}
	\caption{Comparison of the phonon displacement, $\tilde{q} (\omega)$, at two values of $g$ with series of functions $\tilde{v}_k (\omega)$ represented for selected $k$ values to illustrate the nature of the mutually repelling composite excitations. While the out-of-band peak is due to a single phonon-bitriplon state with a single, $g$-dependent value of $\omega$, the in-band resonance is composed of many contributions with different $k$ and $\omega$ values, causing its broad nature and a $g$-dependence dictated by that of the individual components.}
	\label{fig:vk}
\end{figure}

\begin{figure}[t]
	\includegraphics[width=\linewidth]{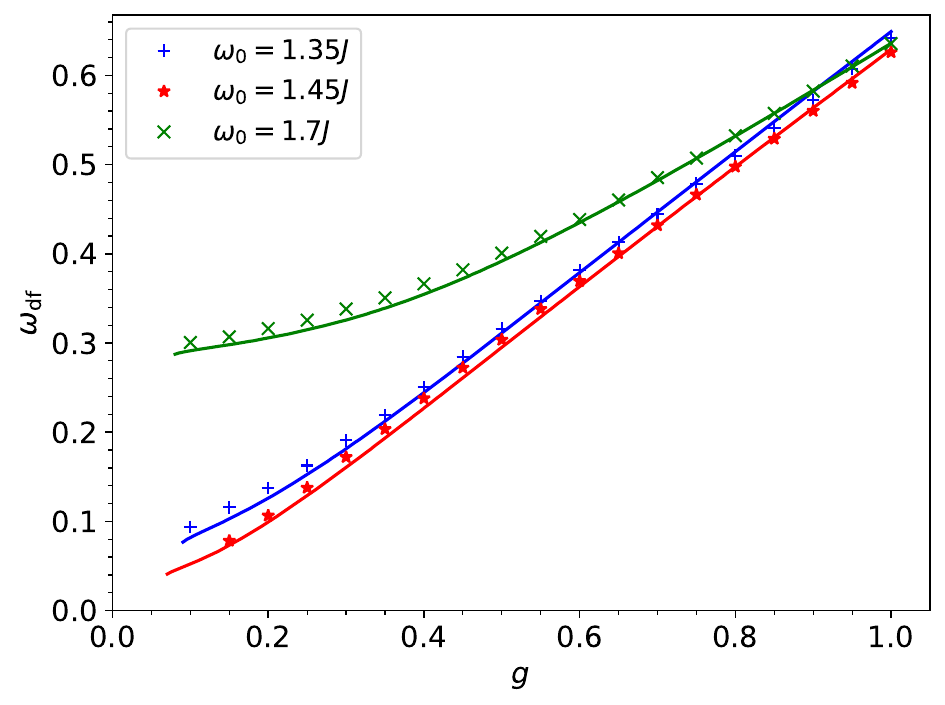}
	\caption{Difference frequency, \(\omega_{\rm df}\), shown as a function of \(g\). Solid lines are obtained from the PBA by taking the separation (\(\omega_{\rm in} - \omega_{\rm out}\)) of the peaks in Fig.~\ref{fig:hybg}, which match almost exactly the frequencies obtained from the equations of motion [Eqs.~\eqref{eq:eom}]. Symbols show \(\omega_{\rm df}\)  extracted directly from the low-energy peak in $\tilde{n}_{\text{ph}} (\omega)$.}
	\label{fig:wsfb}
\end{figure}

\begin{figure*}[t]
	\includegraphics[width=0.96\linewidth]{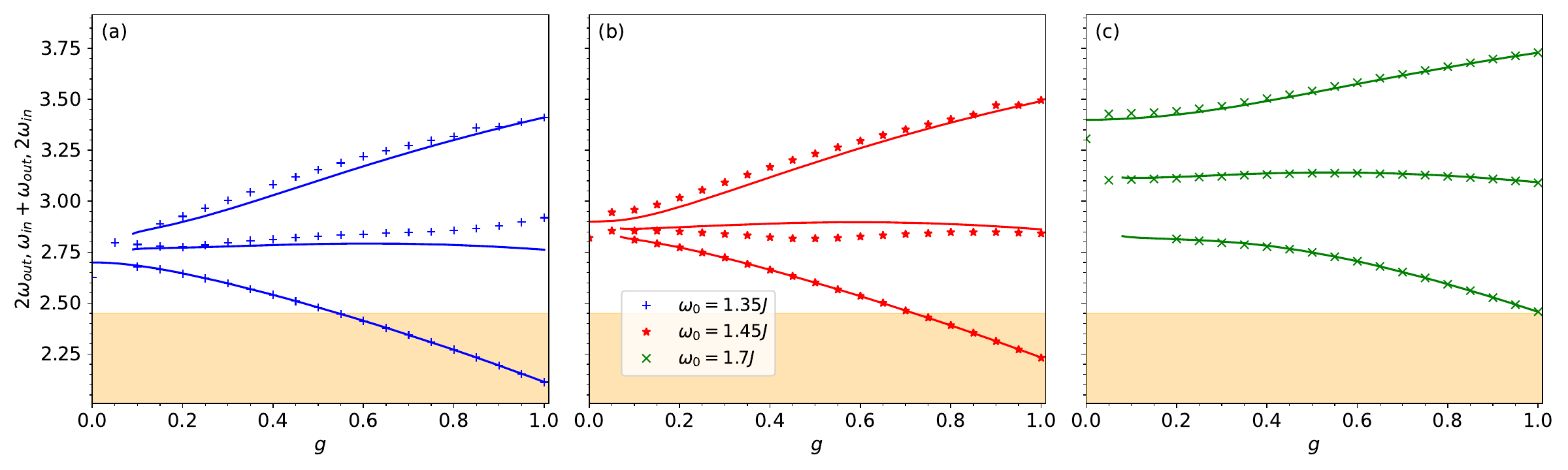}
	\caption{Sum frequencies \(2\omega_{\rm in}\) (top), \(\omega_{\rm in} + \omega_{\rm out}\) (center), and \(2\omega_{\rm out}\) (bottom), shown as functions of \(g\) for three selected phonon frequencies. Symbols were extracted from \(\tilde{n}_{\rm ph}\) and solid lines calculated from the linear response function [Eq.~\eqref{eq:resolvant2}].}
	\label{fig:sumfreq}
\end{figure*}

With regard to the structure of the excitation spectrum, even at large $g$ we find that the rather narrow \(\omega_{\rm out}\) peak consists of a single excited hybrid mode, either based on the phonon for $\omega_0 < \sqrt{2} J$ or based on a single, strongly repelled bitriplon mode for $\omega_0 > \sqrt{2} J$. By contrast, and as noted in Sec.~\ref{sec:res}, the \(\omega_{\rm in}\) peak is a resonance formed by the excitation of multiple phonon-bitriplon modes, with its peak frequency given by that with the strongest contribution; this we illustrate in Fig.~\ref{fig:vk} by comparing $\tilde{q} (\omega)$, which is a measure of the phononic component, with some of the $\tilde{v}_k (\omega)$ functions, which represent the bitriplon component. The width of the \(\omega_{\rm in}\) resonance is an intrinsic quantity dependent on $g$, which also dictates the widths of the peaks arising from higher-order processes, of the types we discuss next.

\subsection{Sum- and difference-frequency excitations}

These excitations are processes bilinear in the phonon-bitriplon operators that are intrinsic to the transient nonequilibrium situation, but not readily available with continuous driving (i.e.~in a NESS). In Sec.~\ref{sec:res}A we stated that the low-frequency feature in the Fourier transforms [Fig.~\ref{fig:ft_nph_nt_w0_145}(a,d)] of the phonon number, \(\tilde{n}_{\rm ph}(\omega)\), and the triplon number, \(\tilde{n}_{\rm t}(\omega)\), is a beat between the two mutually repelling hybrid states, and hence its frequency should match the quantity \(\omega_{\rm in} - \omega_{\rm out}\). To verify this, in Fig.~\ref{fig:wsfb} we compare the PBA prediction of the difference \(\omega_{\rm in} - \omega_{\rm out}\), calculated using the linear reponse function of Eq.~\eqref{eq:resolvant2}, with \(\omega_{\rm df}\) extracted directly from the absolute value of the Fourier transform of \(\tilde{n}_{\rm nph}\), where the corresponding peak is well defined. Clearly the results have qualitatively the same behavior and quantitatively are very close, which confirms our statement. The small discrepancies arise due to the uncertainties in defining the precise positions of the \(\omega_{\rm in}\), \(\omega_{\rm out}\), and indeed \(\omega_{\rm df}\) peaks due to their intrinsic widths (the identity \(\omega_{\rm df} = \omega_{\rm in} - \omega_{\rm out}\) would hold rigorously for $\delta$-function peaks). 

We observe that \(\omega_{\rm df}\) increases with \(g\) in a near-linear fashion for \(\omega_{0} = 1.35J\) and \(\omega_{0} = 1.45J\), except at rather small \(g\), whereas with \(\omega_{0}=1.7J\) the evolution remains nonlinear up to \(g \approx 0.8J\). This behavior indicates the difference between driving phonons where the repulsion of the two is dominated by \(\abs{\omega_0 - 2\omega_{k,{\rm min}}}\) or directly by \(g\). For phonons close to resonance with the high densities of states at the band edges, $g$ comes to dominate \(\omega_{\rm df}\) beyond very small values. By contrast, when the initial separation \(\abs{\omega_0 - 2\omega_{k,{\rm min}}}\) is large, this continues to dictate the value of \(\omega_{\rm df}\) for in-band phonons until $g$ is large (perhaps \(g = 0.7J\) for the \(\omega_{0} = 1.7J\) phonon in Fig.~\ref{fig:wsfb}), while phonons a similar distance outside the bitriplon band would require even stronger $g$ to show a similar repulsion.

We remark that the difference-frequency feature in the triplon number can be hard to discern, primarily because \(\omega_{\rm df}\) is small enough to match the width of the zero-frequency Lorentzian. In Fig.~\ref{fig:ft_nph_nt_w0_145}(d), the \(\omega_{\rm df}\) peak is completely masked for \(g = 0.1J\) and \(0.3J\), and appears only as a kink at \(g = 0.5J\). To improve the visibility of these features in both \(\tilde{n}_{\rm ph}(\omega)\) and \(\tilde{n}_{\rm t}(\omega)\), here we apply a Lanczos window (sinc window) before taking the Fourier transform, but nevertheless some low-$g$ features cannot be retrieved (an example being the missing data point for \(\omega_{0} = 1.45J\) with \(g = 0.1J\) in Fig.~\ref{fig:wsfb}). This visibility issue of the \(\omega_{\rm df}\) peak will arise again in Sec.~\ref{sec:CuGeO3}.

Next we apply the same procedure for the other features at bilinear order, which are the three sum frequencies located around \(2\omega_{0}\). Given the visibility of the \(\omega_{\rm df}\) features, particularly at the smaller \(g\) values relevant in most materials, the sum-frequency features also stand as potential diagnostic markers for intrinsically nonequilibrium magnetophononic processes. In Fig.~\ref{fig:sumfreq} we compute the sums \(2\omega_{\rm out}\), \(\omega_{\rm out} + \omega_{\rm in}\), and \(2\omega_{\rm in}\) using Eq.~\eqref{eq:resolvant2}, which match the results obtained from reading \(\omega_{\rm out}\) and  \(\omega_{\rm in}\) from \(\tilde{q} (\omega)\), and compare these with the positions of the peaks in \(\tilde{n}_{\rm ph} (\omega)\). Other than at small $g$, where some of the sum-frequency peaks (notably \(2\omega_{\rm in}\) at \(\omega_{0} = 1.35J\) and \(2\omega_{\rm out}\) at \(\omega_{0} = 1.7J\)) can be indiscernible or obscure each other, we find three clear features as expected from Figs.~\ref{fig:ft_nph_nt_w0_145}(c) and \ref{fig:ft_nph_E0}(c). 

Qualitatively, the overall $g$-dependence of each feature is as expected from the behavior of \(\omega_{\rm in}\) and \(\omega_{\rm out}\) in Fig.~\ref{fig:hybg}, with \(2\omega_{\rm out}\) evolving to lower frequencies, \(2\omega_{\rm in}\) shifting higher, and \(\omega_{\rm out} + \omega_{\rm in}\) shifting little, because it is the sum of these two competing but asymmetric trends. Quantitatively, Fig.~\ref{fig:sumfreq} makes clear that the comparison with the PBA prediction for \(2\omega_{\rm out}\), which is a single mode pushed outside the bitriplon band, is excellent for all three choices of \(\omega_{0}\). The PBA results for \(\omega_{0} = 1.7J\), where the mode separation is controlled by \(\abs{\omega_0 - 2\omega_{k,{\rm min}}}\), are also quantitatively accurate for all three sums [Fig.~\ref{fig:sumfreq}(c)]. These are the cases where the peaks of the hybrid states ($\omega_{\rm in}$ from Fig.~\ref{fig:ft_q_w0_170} and all the $\omega_{\rm out}$ states) are rather sharply defined. By contrast, for \(\omega_{0} = 1.35J\) and \(1.45J\), the PBA predictions are less accurate for the two sum-frequency excitations involving \(\omega_{\rm in}\) [Figs.~\ref{fig:sumfreq}(a,b)] due to the much greater peak widths in these cases (Figs.~\ref{fig:ft_q_w0_145} and \ref{fig:ft_q_w0_135}). 

In general terms, insight from the PBA of Sec.~\ref{sec:pba} allows us to comment on the underlying structure of the difference- and sum-frequency peaks in the excitation spectrum. As noted above, the \(\omega_{\rm out}\) peak is in essence a single Lorentzian whereas the \(\omega_{\rm in}\) peak is a superposition of multiple Lorentzians whose energies span a range controlled by $g$, this extended width corresponding to a short quasiparticle lifetime. The bilinear features at \(\omega_{\rm df}\), \(\omega_{\rm out} + \omega_{\rm in}\), and \(2\omega_{\rm in}\) can then also be interpreted as superpositions of multiple phonon-bitriplon modes and also inherit their width (lifetime) from \(\omega_{\rm in}\).

\begin{figure*}[t]
	\includegraphics[width=0.96\linewidth]{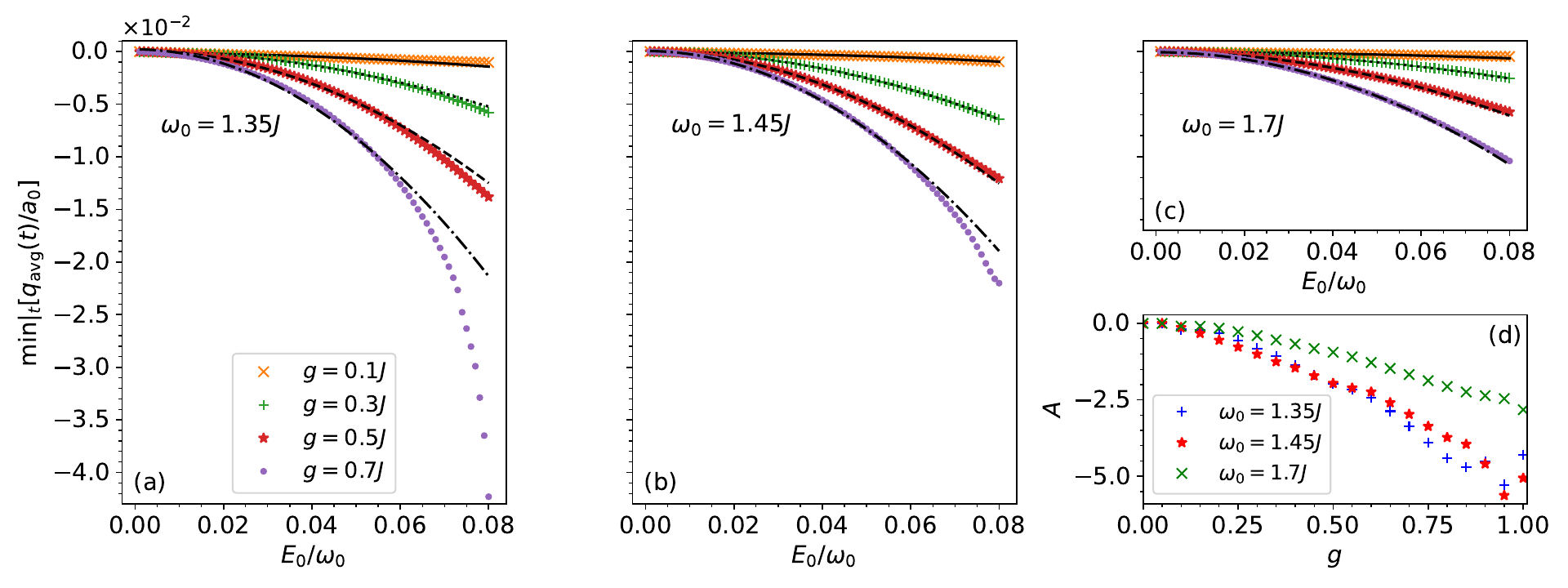}
	\caption{(a-c) Minimum value, \(\min|_{t}[q_{\rm avg}(t)]\), of the sliding average [Eq.~\eqref{eq:sliding}] of the phonon displacement, \(q(t)\), shown for four choices of the coupling strength, \(g\), at the usual three phonon frequencies, $\omega_0$. Black lines are fits to the form \(AE_{0}^{2}\), where the prefactor \(A\) is a free parameter. (d) Prefactor $A$ obtained by fitting to the electric-field range $0 \le E_0 \le 0.06$, shown as a function of $g$ for the three phonon frequencies.}
	\label{fig:sliding}
\end{figure*}

\subsection{Transient band engineering with \(E_{0}\)} \label{ssec:bee0}

\subsubsection{Phonon displacement}

Figure \ref{fig:ft_nph_E0} demonstrated that the electric field, \(E_{0}\), of the driving pulse is also an important parameter that influences the frequency content of the diagnostic quantities we study. It showed that our standard driving field can be regarded as a regime of weak pumping, and that increasing the pulse amplitude outside this regime creates further nonlinear effects. The leading effect of stronger $E_0$ was an overall downward shift of \(\omega_{\rm in}\) and \(\omega_{\rm out}\) [Fig.~\ref{fig:ft_nph_E0}(a)], although the shifts were not identical and thus $\omega_{\rm df}$ was found  to increase [Fig.~\ref{fig:ft_nph_E0}(b)]. Such a band shift is familiar from the case of continuously driven magnetophononics studied in Ref.~\cite{yarmo23}, where it was shown that the driving creates a static displacement, $q_0$, of the phonon coordinate from its equilibrium position, and this was translated to a lowest-order renormalization of the dimer coupling constant (Fig.~\ref{fig:scheme}) from $J$ to $ J(1 + g q_{0})$, with $q_0 < 0$. These authors then proposed a lowest-order expression for the field-induced engineering of the triplon band dispersion, 
\begin{equation} \label{eq:be_disp}
	\tilde{\omega}_{k} = J (1 + gq_{0}) \sqrt{1 - \lambda \cos(k)/(1 + gq_{0})}.
\end{equation} 

\begin{figure*}[t]
	\includegraphics[width=0.96\linewidth]{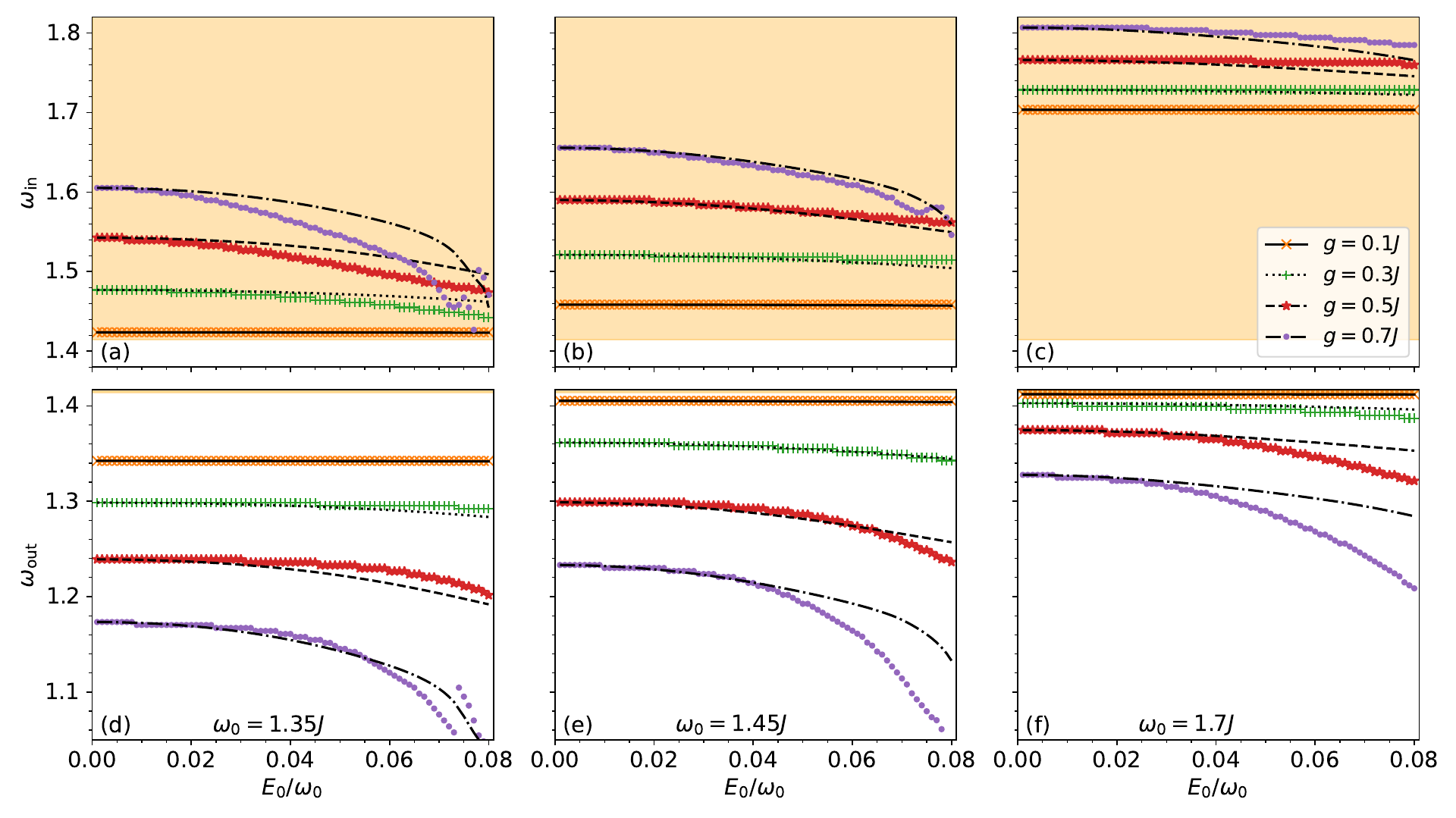}
	\caption{Frequencies of the two mutually repelling hybrid excitations, \(\omega_{\rm in}\) (a-c) and \(\omega_{\rm out}\) (d-f), shown as functions of the peak driving electric field, $E_0$, for phonon frequencies \(\omega_{0} = 1.35J\) (a,d), \(1.45J\) (b,e), and \(1.7J\) (c,f) with four different spin-phonon couplings, \(g\). Black lines are estimates based on applying  Eqs.~\eqref{eq:be_disp} and \eqref{eq:sliding} to obtain the shifted bitriplon band for each \(g\).}
	\label{fig:hybE0}
\end{figure*}

The transient analog of $q_0$ in a pulsed experiment is the slow-moving component of \(q(t)\), meaning that averaged over one or more full pulse cycles, which we name \(q_{\rm avg}(t)\). There are several routes by which such a static offset could be obtained and one is to apply a low-pass filter in Fourier space, removing all frequencies in \(\tilde{q} (\omega)\) higher than some cut-off (for example \(\omega_{\rm df}\)) before back-transforming this signal. Remaining instead in the time domain, one may define the sliding average
\begin{align} \label{eq:sliding}
	q_{\rm avg}(t) = \frac{1}{T} \int_{t}^{t+T} dt' \, q(t'),
\end{align}
where \(T\) is a period for a full cycle. Both approaches have some degree of arbitrariness, in the frequency cut-off or the chosen cycle time, that leaves a certain amount of slow but nonzero oscillations in \(q_{\rm avg}(t)\), but all variations of both methods return estimators that vary only by some 10s of percent. Here we adopt an approach of applying the sliding average twice, using periods \(T_{\rm in} = 2 ({2\pi}/{\omega_{\rm in}})\) and \(T_{\rm out} = 2 ({2\pi}/{\omega_{\rm out}})\), to effect the complete removal of all fast oscillations, and for each $E_0$ take the minimum reached by \(q_{\rm avg}(t)\) (i.e.~the maximum offset in the time trace) as the analog of $q_0$. 

\begin{figure*}[t]
	\includegraphics[width=0.96\linewidth]{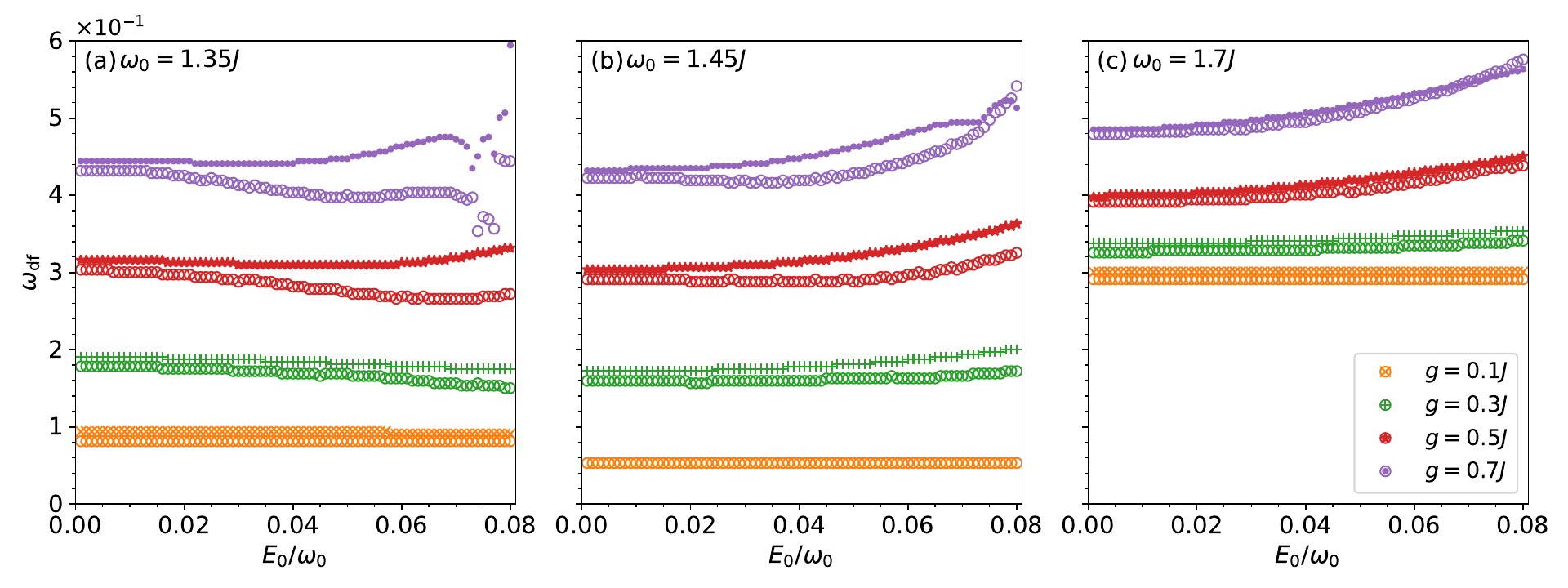}
	\caption{Difference frequency, \(\omega_{\rm df}\), shown as a function of \(E_{0}\) for our standard $g$ and $\omega_0$ values. Open circles are the difference between \(\omega_{\rm in}\) and \(\omega_{\rm out}\) taken from Fig.~\ref{fig:hybE0} and the other symbols are extracted from features visible in \(\tilde{n}_{\rm ph}(\omega)\); only open circles are present at $g = 0.1$ in panel (b) because the $\omega_{\rm df}$ peak in \(\tilde{n}_{\rm ph}(\omega)\) was overwhelmed by the zero-frequency peak in this case.}
	\label{fig:wsfbE0}
\end{figure*}

In Fig.~\ref{fig:sliding} we display the behavior of this quantity as a function of \(E_{0}\) for four choices of \(g\) and three of \(\omega_{0}\). In every case other than \(\omega_{0} = 1.35J\) and \(\omega_{0} = 1.45J\) with the highest \(g\) and $E_0$, we find a quadratic dependence of the transient static displacement on \(E_{0}\). This dependence can be traced to the spin-feedback term, \(U(t)\), which is a $y_k$-weighted sum over triplon occupations [Eq.~\eqref{eq:U}]. Following Ref.~\cite{yarmo23}, by averaging over Eq.~\eqref{eq:dqdt} with the slow-moving parts \(q(t)\) and \(U(t)\) as the only finite quantities, one obtains a relation of the form \(q_0 = - 2 g U_0/\omega_{0}\). Because $n_{\rm t}$, like $n_{\rm ph}$ [Fig.~\ref{fig:max_q_nph_E0}], varies quadratically with $E_0$ and the weights \(y_{k}\) are largely close to unity, the quadratic dependence of \(q_{\rm avg}(t)\) is evident. The coefficient $A$ of the quadratic fits, shown in Fig.~\ref{fig:sliding}(d), is less evident. At \(g = 0.1J\), feedback through the spin sector is minimal and the quadratic increase of \(U(t)\) with $E_0$ has only a very small effect on \(q_{\rm avg}(t)\). At the other $g$ values, the standard spin-feedback process operates and $A$ has an approximately linear increase with $g$. However, for \(g = 0.7J\) and when \(E_0 > 0.06 \omega_{0}\), we find a dramatic growth of \(q_{\rm avg}(t)\) with $E_0$ beyond the quadratic form for phonon frequencies close to the band edge, from which we deduce that the system has entered a different driving regime [Figs.~\ref{fig:sliding}(a,b)]. Because this regime is relevant only at extremely strong coupling, we do not investigate it further here. 

\subsubsection{Spin-band engineering}

Turning to pulse-strength effects on the characteristic magnetophononic excitations, in Fig.~\ref{fig:hybE0} we represent \(\omega_{\rm in}\) and \(\omega_{\rm out}\) as functions of \(E_{0}\). In the interpretation of Ref.~\cite{yarmo23}, shifts of the composite hybrid states at fixed $g$ should constitute direct evidence for field-induced modification of the bitriplon band dispersion, moving the energies of the peaks in the bitriplon density of states. The black lines in Fig.~\ref{fig:hybE0} show predictions obtained for \(\omega_{\rm out}\) and \(\omega_{\rm in}\) by using Eq.~\eqref{eq:be_disp}, with the results from Fig.~\ref{fig:sliding} as $q_0$, to determine the renormalization of $2\omega_k$ for the bitriplon mode $k$ at their low-$E_0$ peaks. Deviations of \(\omega_{\rm in}\) from these lines may therefore imply that the effect of $E_0$ is to change the most strongly contributing mode, whereas for \(\omega_{\rm out}\) they imply only a breakdown of the approximation.

At \(g = 0.1J\) we find that feedback effects are simply negligible; with no appreciable feedback channel, which is controlled by $g$, no amount of applied field can alter the triplon dispersion beyond $g$-linear hybridization. With \(g = 0.3J\) and \(0.5J\), feedback effects become relevant and we begin to observe weak field effects on both \(\omega_{\rm in}\) and \(\omega_{\rm out}\), which can be traced to the underlying triplon bands. The piecewise plateaus appearing over intervals of \(E_{0}\) are a consequence of the weak energy shifts and the frequency bins of the Fourier transform. The predictions of the lowest-order band-shifting approximation are qualitatively correct in direction and magnitude, but begin to show quantitative deviations at intermediate $E_0$ values that vary with both $\omega_0$ and $g$ (although generally setting in at lower $E_0$ for higher $g$). The nature of the discrepancy is not systematic, with the prediction tending to overestimate the renormalization in some cases ($\omega_0 = 1.7 J$) and to underestimate it in others ($\omega_0 = 1.35 J$). One may assume that these discrepancies arise primarily because the strong $E_0$ values we study in the pulsed framework can go far beyond the linear regime to which continuous driving, and the estimate of Eq.~\eqref{eq:be_disp}, are restricted. Nevertheless, the concept of a global triplon band renormalization arising only from the reduction of $J$ is also an approximation, and our results offer the possibility of performing a more detailed exploration of this phenomenon. 

At \(g = 0.7J\) we find not only the strongest renormalization but for \(\omega_{0} = 1.35J\) and \(1.45J\) also the change of regime above \(E_{0} = 0.06 \omega_0\) [Figs.~\ref{fig:hybE0}(a,b,d,e)]. Here the frequency content of \(q(t)\) contains multiple nonlinear features, which appear as new spectral peaks that impact the definitive extraction of \(\omega_{\rm in}\) and \(\omega_{\rm out}\). Quite generally, the shifts of \(\omega_{\rm in}\) and \(\omega_{\rm out}\) are not identical for any $\omega_0$, with weaker shifts for the phononic hybrid (the one matching the phonon at $g = 0$) and stronger ones for its triplonic counterpart. The stronger shifts for \(\omega_{0} = 1.35J\) and \(1.45J\) compared to \(\omega_{0} = 1.7J\) highlight the importance of resonance with the high density of bitriplon states to the strength of nonlinear feedback effects.

We conclude our systematic analysis of \(E_{0}\) effects by considering the difference and sum frequencies. Following Fig.~\ref{fig:ft_nph_E0}(b), we anticipate that \(\omega_{\rm df}\) should increase slowly at larger $E_0$, and this is confirmed in Fig.~\ref{fig:wsfbE0}(b). Also at \(\omega_{0} = 1.7J\) [Fig.~\ref{fig:wsfbE0}(c)], the difference frequency shows a monotonic and more clearly quadratic increase with $E_0$ for each choice of \(g\) (beyond the linear regime of $g = 0.1 J$). By contrast, and as expected from the differing trends in Fig.~\ref{fig:hybE0}, \(\omega_{\rm df}\) at \(\omega_{0} = 1.35J\) first decreases with $E_0$ [Fig.~\ref{fig:wsfbE0}(a)] before a plateau and possible rise for higher $g$ and \(E_{0}\). We also verify how closely the  difference \(\omega_{\rm in} - \omega_{\rm out}\) corresponds at large $E_0$ to the spectral feature(s) around \(\omega_{\rm df}\), given that the two are read from different diagnostic quantities. For \(\omega_{0} = 1.7J\) there is good agreement at all $g$ and $E_0$, but for the two phonons near resonance with the lower band edge this agreement is lost beyond small $E_0$ for most $g$ values. This reflects again that the difference calculation contains predominantly the linear response and does not account for the higher-order renormalization processes (strongest with a maximal density of phonon-bitriplon states) that are manifestly created by $E_0 \gtrsim 0.01 \omega_0$ once $g \gtrsim 0.2 J$. We remark that, given the relatively weak variation of \(\omega_{\rm df}\) with \(E_{0}\), its presence constitutes a good diagnostic in an experiment where the amplitude of the electric field is not known.

Finally, the evolution of the three sum frequencies with $E_0$, which we do not show, is largely as expected from \(\omega_{\rm in}\) and \(\omega_{\rm out}\) [Fig.~\ref{fig:hybE0}]. All three features shift downwards as \(E_{0}\) increases, but at different rates that lead to some features obscuring others. Computing the sum frequencies with data from Fig.~\ref{fig:hybE0} again yields excellent agreement for \(2\omega_{\rm out}\) and for all three frequencies at \(\omega_{0} = 1.7J\), but poorer agreement for the other four cases for the reasons already discussed in the analysis of \(\omega_{\rm df}\).

In summary, magnetophononic phenomena show a linear response regime in $E_0$, beyond which transient band-engineering effects are visible in every excitation of the system. Their precise form depends strongly on \(g\), which controls the amount and range of energy transfer processes between the spin and phonon sectors, and hence the strength of feedback phenomena.

\section{Alternating spin-chain material: C\lowercase{u}G\lowercase{e}O$_{\bf 3}$} 
\label{sec:CuGeO3}

\subsection{Introduction}

The experimental observation of magnetophononic phenomena requires a quantum magnetic material with strong spin-phonon coupling. Here we choose as an illustrative example the quasi-1D inorganic compound CuGeO\(_3\), whose strong $g$ is manifest in the fact that it undergoes a spin-Peierls transition at \(T_{\rm sp} = 14.2\) K \cite{hase93}. Below \(T_{\rm sp}\), the Cu-O-Cu chains become dimerized, but the fact that no single phonon is observed to soften suggests that the dimerization must be a more complex process involving multiple Peierls-active phonon modes \cite{gros98}. Extensive studies of the lattice dynamics in both the undimerized \cite{popov95,brade98a,brade02} and dimerized \cite{takeh00,spitz25} phases have characterized a wide spectrum of IR-active phonons that offer a range of possibilities for ultrafast coherent laser driving at frequencies in different parts of the bitriplon spectrum \cite{yarmo23}.

The triplon excitation spectrum of CuGeO\(_3\) \cite{regna96a} reveals an appreciable interchain coupling, and a straightforward 2D model was introduced \cite{uhrig97a} for an accurate description of this dispersion. Although this model was later generalized \cite{knett01} to provide an optimal account of multiple experiments, here we will simplify it to the minimal ingredients required to give an approximate description suitable for analyzing magnetophononic phenomena. Still, changing the dimensionality from one to two has fundamental repercussions for the density of states, which in 2D does not diverge at the band edges. The strong effects we have studied for the dimerized chain are undeniably consequences of the very strong peaks in the 1D density of states, and hence one may question their visibility, and indeed their nature, in higher-dimensional materials. 

\begin{figure}[t]
	\includegraphics[width=0.9\linewidth]{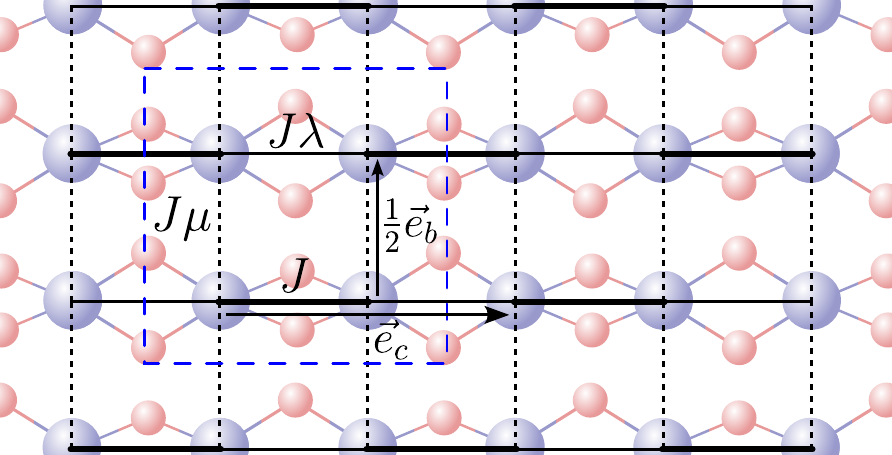}
	\caption{Simplified two-dimensional dimerized lattice of CuGeO\(_3\) inspired by Fig.~1 of Ref.~\cite{uhrig97a}. The spins form dimers on the strong bond, \(J\), and these pairs are coupled by the weak bond, \(J\lambda\). \(J\mu\) couples the spins chains in the perpendicular direction. The unit cell is indicated by the dashed blue rectangle.}
	\label{fig:lattice_2d}
\end{figure}

\subsection{2D Framework}

One essential property of the magnetophononic framework we apply (Sec.~\ref{sec:mm}) is the approximation of noninteracting triplons, meaning that triplon terms of wave vector $k$ mix only with other $k$ terms, and not with any $k'$ terms; the only other matrix element for each triplon is with the phonon. The bond-operator description is applicable in any dimension, and thus the considerations of Sec.~\ref{sec:mm} can be used to diagonalize the spin sector and obtain the triplon dispersion. Although in Secs.~\ref{sec:mm}-\ref{sec:phen} we stressed the structure of energy space, meaning the density of states per unit energy (Fig.~\ref{fig:bidisp}), over the structure of {\bf k}-space, the spin-phonon coupling does have a direct {\bf k}-dependence through the coefficients $y_k$ and $y'_k$, whose effects we will encounter below. We proceed by considering the 2D lattice of Fig.~\ref{fig:lattice_2d} \cite{uhrig97a}, where dimerized spin chains identical to those treated in Secs.~\ref{sec:mm}-\ref{sec:phen} (Fig.~\ref{fig:scheme}), aligned in the lattice \(\hat{c}\) direction with alternating dimer positions, are coupled in the \(\hat{b}\) direction by an interchain superexchange interaction \(\mu J\).  

\begin{figure}[t]
	\includegraphics[width=\linewidth]{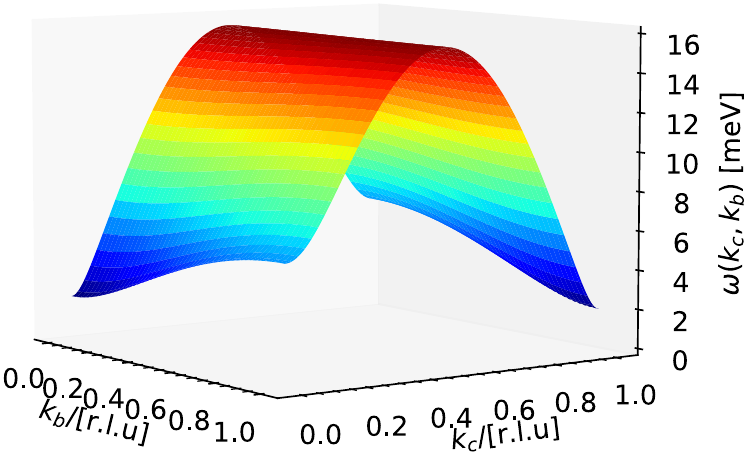}
	\caption{Triplon dispersion of CuGeO$_3$ approximated by Eq.~\eqref{eq:disp_2d} and represented in the space of $k_b$ and $k_c$. Using the parameters \(J = 11.52\) meV, \(\lambda = 0.858\), and \(\mu = 0.056\) yields a spin gap of 2.0 meV, a saddle-point energy of 5.8 meV and an upper band edge of 15.7 meV.}
	\label{fig:disp_2d}
\end{figure}

The triplon dispersion of this system is 
\begin{align} \label{eq:disp_2d}
	\omega(k_{c}, k_{b}) & \! = \! J\sqrt{1 \! - \! \lambda \cos(k_{c}) \! - \! 2 \mu \cos(\tfrac{1}{2}k_{c}) \cos(\tfrac{1}{2}k_{b})},
\end{align}
which without interchain coupling reduces to the alternating-chain model of Eq.~\eqref{eq:disp_1d}. With finite $\mu$, the dimer geometry (Fig.~\ref{fig:lattice_2d}) alters the periodicity and establishes a frustration in the chain direction, leading to an incommensurate position of the energy maximum. To mimic the dispersion of CuGeO$_3$, we choose the parameters \(J\), \(\lambda\), and \(\mu\) to match the known values of the spin gap, the band maximum, and the low-energy saddle point \cite{regna96a}. The results we obtain, a characteristic chain energy \(J = 11.52\) meV, a rather weak dimerization \(\lambda = 0.858\), and a very weak net interchain coupling \(\mu = 0.056\), are similar to those of Ref.~\cite{uhrig97a} and we draw the corresponding dispersion for the rectangular unit cell of two dimers (Fig.~\ref{fig:lattice_2d}) in Fig.~\ref{fig:disp_2d}. 

\begin{figure}[t]
	\includegraphics[width=\linewidth]{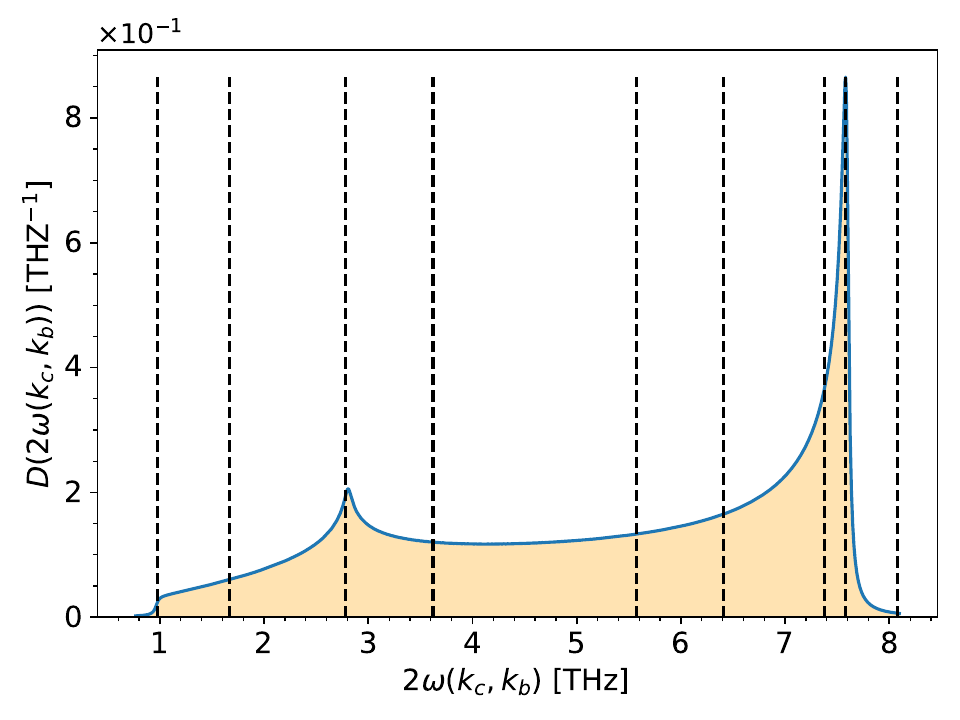}
	\caption{Density of states (DOS) computed from the bitriplon dispersion of Fig.~\ref{fig:disp_2d} with a broadening of \(\gamma_{\rm s} = 0.01J\). Black dotted lines indicate the 9 phonon frequencies at which data are shown in Figs.~\ref{fig:2d_t_q_g} to \ref{fig:2d_ft_up}.}
	\label{fig:dos_2d}
\end{figure}

As stated above, the key quantity entering the analysis of magnetophononic phenomena remains the bitriplon density of states, which we compute in the approximation of the minimal bond-operator dispersion and show in THz units in Fig.~\ref{fig:dos_2d}. Quite in contrast to the 1D result shown in Fig.~\ref{fig:bidisp}, there is no strong accumulation of states at the lower band edge, only a step up to a small, finite value, as expected for a gap at one point in 2D reciprocal space. The saddle point causes a logarithmic singularity in the density of states, whose broadening by $\gamma_{\rm s}$ induces a moderate peak appearing at 2.78 THz. Only the upper band edge has a large peak in the density of states, which appears because the energy of the band maximum in $k_c$ is almost identical for all $k_b$ (Fig.~\ref{fig:disp_2d}), forming a 1D manifold in the 2D space. 

Another contrast with the $\lambda = 0.5$ case studied in Secs.~\ref{sec:mm}-\ref{sec:phen} is the much smaller gap-to-bandwidth ratio in CuGeO$_3$, which leads to a factor-7 range in bitriplon energies. With an effective $\lambda$ of 0.86, CuGeO$_3$ represents a chain in which $J'$ approaches $J$, and as a result we will see quantitative but not qualitative changes to the spectral features, enriched by the interchain coupling. On the methodological side, this value of $\lambda$ lies beyond the validity of the noninteracting triplon approximation we apply in Sec.~\ref{sec:mm}, and we do not claim quantitative accuracy in our CuGeO$_3$ calculations. However, we remark that the density of states for our approximate triplon dispersion differs from the more precise results calculated in Ref.~\cite{uhrig97a} at the 5\% level on the vertical axis of Fig.~\ref{fig:dos_2d}, while the primary features are by design located correctly on the horizontal (energy) axis. 

\begin{figure*}[t]
	\includegraphics[width=0.98\linewidth]{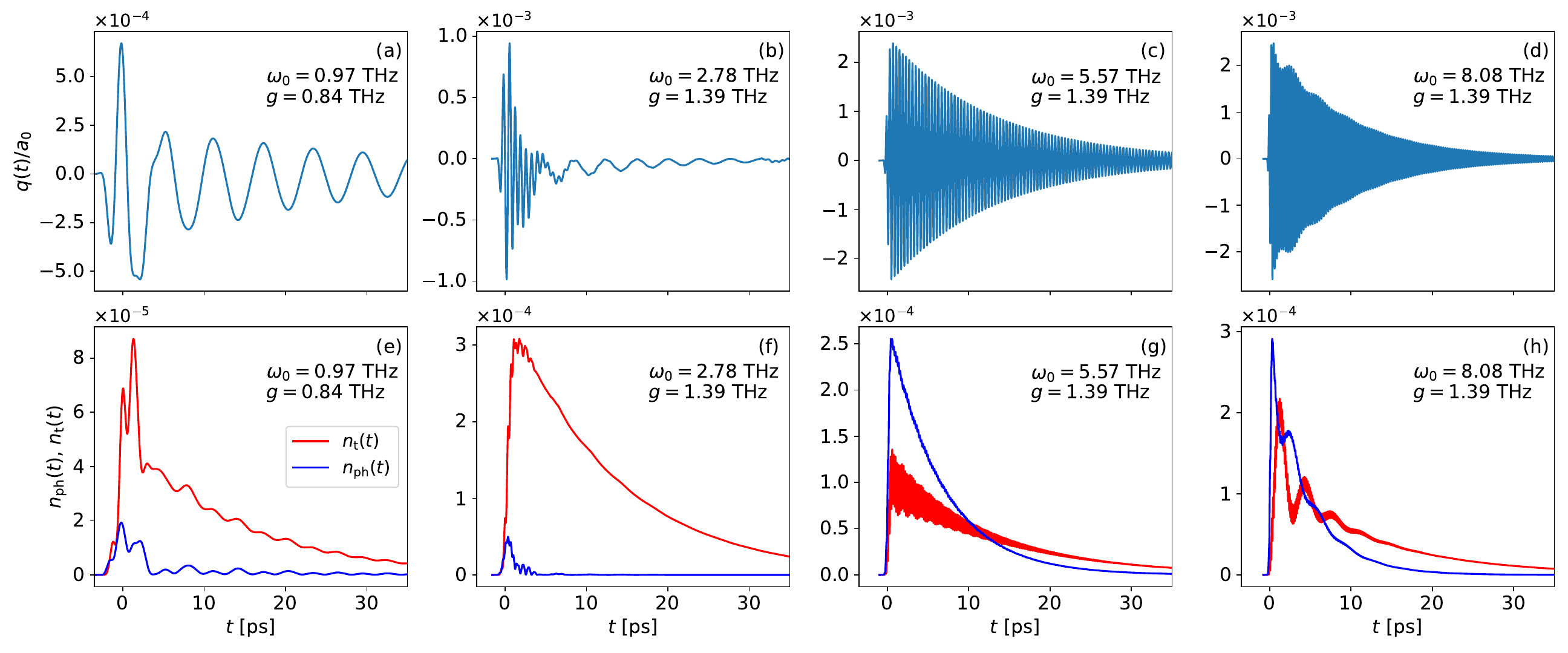}
	\caption{Phonon displacement (\(q(t)/a_0\)), phonon number [\(n_{\rm ph}(t)\), dark blue], and triplon number [\(n_{\rm t}(t)\), red] calculated for one example phonon near the lower band edge (a,e), near the saddle point (b,f), near the band center (c,g), and near the upper band edge (d,h). Only the strongest spin-phonon coupling is shown in each case [$g = 0.84$ THz in panels (a,e), otherwise \(g = 1.39\) THz].}
	\label{fig:2d_t_q_g}
\end{figure*}

For the illustrative calculation of magnetophononic effects in this system, we do not attempt to model the precise lattice dynamics of CuGeO$_3$, or even to account for the precise structure of O and Ge atoms. Instead we assume only that we are able to achieve coherent driving of phonons located at the known frequencies of the IR phonons in dimerized CuGeO$_3$, and that these will lead to the modulation of the superexchange interactions by an effective phonon coordinate, $q$, following the conjectures made in Sec.~\ref{sec:mm}. Our aim is to establish the visibility of magnetophononic effects in the time and frequency domains, in order to refine these with more detailed input information when experiments appear feasible. For this we require only minimal adaptations of the equations of motion [Eqs.~\eqref{eq:eom}]: $k$ becomes a label for the 2D vector ($k_c,k_b$) in the quantities \(\omega_{k}\), $y_k$, $y'_k$, \(u_{k}(t)\), \(v_{k}(t)\), and \(w_{k}(t)\). For reference we express the mixing coefficients in this general framework as 
\begin{equation}
\label{eq:ykykpgen}
y_k  = \frac{J (1 + \omega_k^2/J^2)}{2\omega_k} \;\, {\rm and} \;\, y'_k = \frac{J (1 - \omega_k^2/J^2)}{2 \omega_k},
\end{equation}
where the $k$-dependence is contained in $\omega_k$ and the expressions below Eq.~\eqref{eq:hspbo} are a simplification for the 1D case. The total system size becomes $N = N_c N_b$, but we still solve \(3 N + 3\) equations (with $N_c = 500$ and $N_b = 50$). 

\subsection{Exploring magnetophononics in CuGeO$_3$}

Mindful of our observation in Sec.~\ref{sec:phen} that the energetic separation of the two composite hybrid states can be dominated by $g$ or dominated by $|\omega_0 - \omega_{\rm nf}|$, where $\omega_{\rm nf}$ marks a notable feature in the bitriplon density of states, we begin by exploring the magnetophononic response obtained by placing a total of 24 different phonons (as defined in Sec..~\ref{sec:mm}) at different separations from the four most notable features in Fig.~\ref{fig:dos_2d}. These we take as (i) the lower band edge (0.95 THz), (ii) the saddle point ($\omega_{\rm sp} = 2.78$ THz), (iii) the very flat region above the saddle point, and (iv) the upper band edge ($\omega_{\rm ube} = 7.72$ THz). Generically, there are 60 phonon modes in the dimerized phase of CuGeO$_3$, all with different spin-phonon coupling constants, $g_i$ ($i = 1, \dots 60$), and we do not yet relate our chosen frequencies to specific phonon modes of CuGeO$_3$ \cite{popov95,brade98a,brade02}.

For this study we use a pulse with the same parameters as in Sec.~\ref{sec:res}, namely \(E_{0} = 4 \times 10^{-3} \omega_d\), \(\omega_{d} = \omega_{0}\), and \(T_{p} = 3 (2\pi/\omega_{0})\), and do not address the possibility of band engineering by varying $E_0$. The spin-phonon coupling constants, $g_i$, have been estimated for a small number of phonon modes in CuGeO$_3$ that were selected for their Peierls-active symmetry \cite{werne99,feldk02}. Even among these modes, the $g_i$ values can vary widely and the early calculations did not agree on quantitative values. While in principle one would know these fixed materials constants, and could use them as input for a calculation of nonequilibium mode frequencies and occupations, in practice our approach is necessarily more exploratory. To cover a reasonable range of possible $g_i$ values, we perform our calculations at each $\omega_0$ with $g = 0.1J$, $0.2J$, $0.3J$, $0.4J$, and $0.5J$, where $J = 2.78$ THz. We retain the damping coefficients used in Sec.~\ref{sec:res}, with \(\gamma = 0.01\omega_{0}\) scaled to the phonon frequency and \(\gamma_{\rm s} = 0.01J\) scaled to the magnetic energy, but we remark that these coefficients may enter an unphysical regime at very low \(\omega_{0}\), where the phonon may have a slower decay rate than the triplons (as well as an energy comparable to our stronger $g$ values). As noted above, we use a system of \(N_{b} = 50\) chains each with \(N_{c} = 500\) dimers, and we compute the time-evolution to a final time of 1000 ps with time step \(\delta t = 0.01\) ps to ensure accurate Fourier transforms. For an efficient solution of this high number of equations within a 4th-order explicit Runge-Kutta method, we employ the Julia package \textit{Differentialequations.jl} \cite{rackauckas17}. 

\begin{figure*}[t]
	\includegraphics[width=0.98\linewidth]{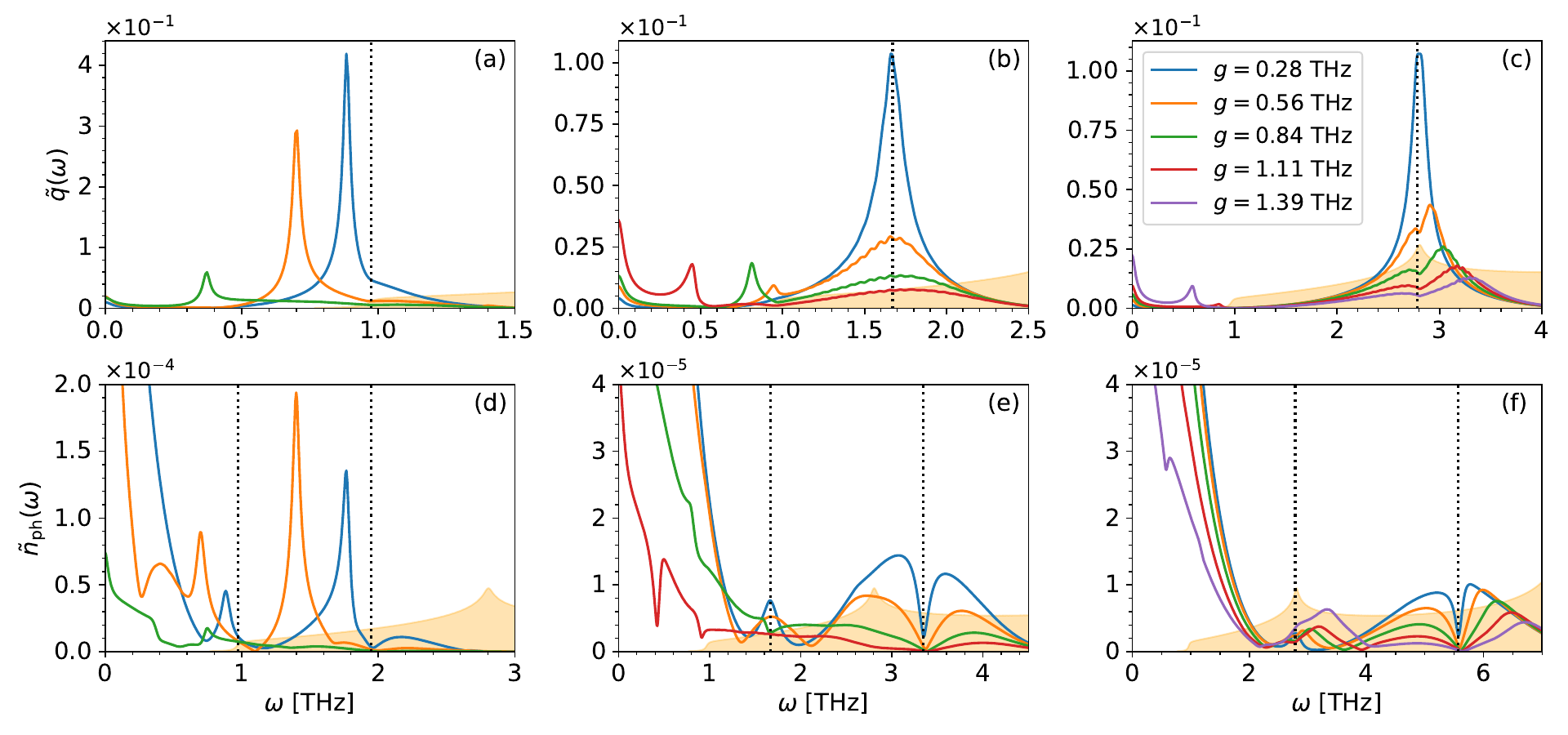}
	\caption{Absolute values of the Fourier transforms of the phonon displacement, \(\tilde{q} (\omega)\) (a-c), and phonon number, \(\tilde{n}_{\text{ph}}(\omega)\) (d-f), for three phonon frequencies selected near the lower band edge [0.97 THz (a,d)], between this edge and the saddle point [1.67 THz (b,e)], and near the saddle point [2.78 THz (c,f)]. Meaningful solutions were not obtained for our largest $g$ values at these low frequencies (a,b,d,e).}
	\label{fig:2d_ft_low}
\end{figure*}

In Fig.~\ref{fig:2d_t_q_g} we show four representative examples of the time evolution of the phonon displacement and number at frequencies selected across the entire range of the bitriplon band, all with strong spin-phonon coupling. Looking first at $q(t)$, at low frequencies [Fig.~\ref{fig:2d_t_q_g}(a)] we find an anomalously slow oscillation dominated by the lower hybrid phonon-bitriplon. At the saddle-point frequency [Fig.~\ref{fig:2d_t_q_g}(b)] we observe an extremely rapid decay indicative of very efficient energy transfer to the magnetic sector, whereas at the band center [Fig.~\ref{fig:2d_t_q_g}(c)] the decay time is simply that of the phonon ($1/\gamma$). These results reflect less the density of bitriplon states, which does not vary dramatically between the two frequencies (Fig.~\ref{fig:dos_2d}), than the value of the coupling coefficient $y'_k$ [Eq.~\ref{eq:ykykpgen}], which can be large for $k \equiv (k_b,k_c)$ around the gap and can vanish near the band center \cite{yarmo21}, although not for all $k$ values simultaneously [Eq.~\eqref{eq:ykykpgen}]. Only at phonon frequencies near the upper band edge [Fig.~\ref{fig:2d_t_q_g}(d)] do we see a slow oscillation of the $q(t)$ envelope. 

In \(n_{\rm ph}(t)\) and \(n_{\rm t}(t)\) we observe another consequence of the large $y'_k$ values at low frequencies [Figs.~\ref{fig:2d_t_q_g}(e,f)], which is that phonon driving becomes extremely efficient compared to the ratio of \(n_{\rm t}(t)\) to \(n_{\rm ph}(t)\) achieved in 1D [Figs.~\ref{fig:nph_nt_w0_145}, \ref{fig:nph_nt_135}, and \ref{fig:nph_nt_w0_170}]. At the mid-band frequencies this ratio has returned to a more conventional value [Fig.~\ref{fig:2d_t_q_g}(g)] and at high frequencies we observe multiple oscillations in both \(n_{\rm ph}(t)\) and \(n_{\rm t}(t)\) [Fig.~\ref{fig:2d_t_q_g}(h)]. This indicates that the effects of strong hybridization between the phonon and spin systems can indeed be observed in 2D when the driven phonon is close to a high density of bitriplon states.

Following the analysis of our 1D model in Sec.~\ref{sec:res}, we now inspect the Fourier transforms of the phonon displacement, \(\tilde{q}(\omega)\), and the phonon number, \(\tilde{n}_{\rm ph}(\omega)\). From the standpoint of observing unconventional features at frequencies around and below the gap in the bitriplon band, the physical parameters of CuGeO$_3$ present a significant problem. The small gap-to-bandwidth ratio places these features at frequencies considerably lower than the band-center energy scale, $2J$, while the widths of the zero-frequency peaks in the Fourier transforms (Fig.~\ref{fig:ft_nph_nt_w0_145}) are proportional to $J$, making it particularly difficult to discern weak signatures arising from difference-frequency phenomena.

In Fig.~\ref{fig:2d_ft_low} we can discern mutually repelling hybrids in $\tilde{q}(\omega)$ when the driven phonon is placed at [Fig.~\ref{fig:2d_ft_low}(a)] or above [Fig.~\ref{fig:2d_ft_low}(b)] the lower edge of the bitriplon band. As expected from the 1D model, the sharp $\omega_{\rm out}$ peak is repelled strongly downwards in frequency with increasing $g$, whereas a ``phononic'' $\omega_{\rm in}$ peak changes little in frequency but strongly in intensity. In \(\tilde{n}_{\rm ph}(\omega)\) we find a complex response with multiple low-frequency contributions [Figs.~\ref{fig:2d_ft_low}(d,e)]. As noted above, the phonon frequencies, damping rates, and $g$ itself all become comparable in this regime, and we find no physically meaningful solution at our larger $g$ values. Although the two well defined hybrid states give rise to a robust sum-frequency signal, there is no possibility of ascribing any low-frequency features to a difference-frequency effect. 

\begin{figure*}[t]
	\includegraphics[width=0.98\linewidth]{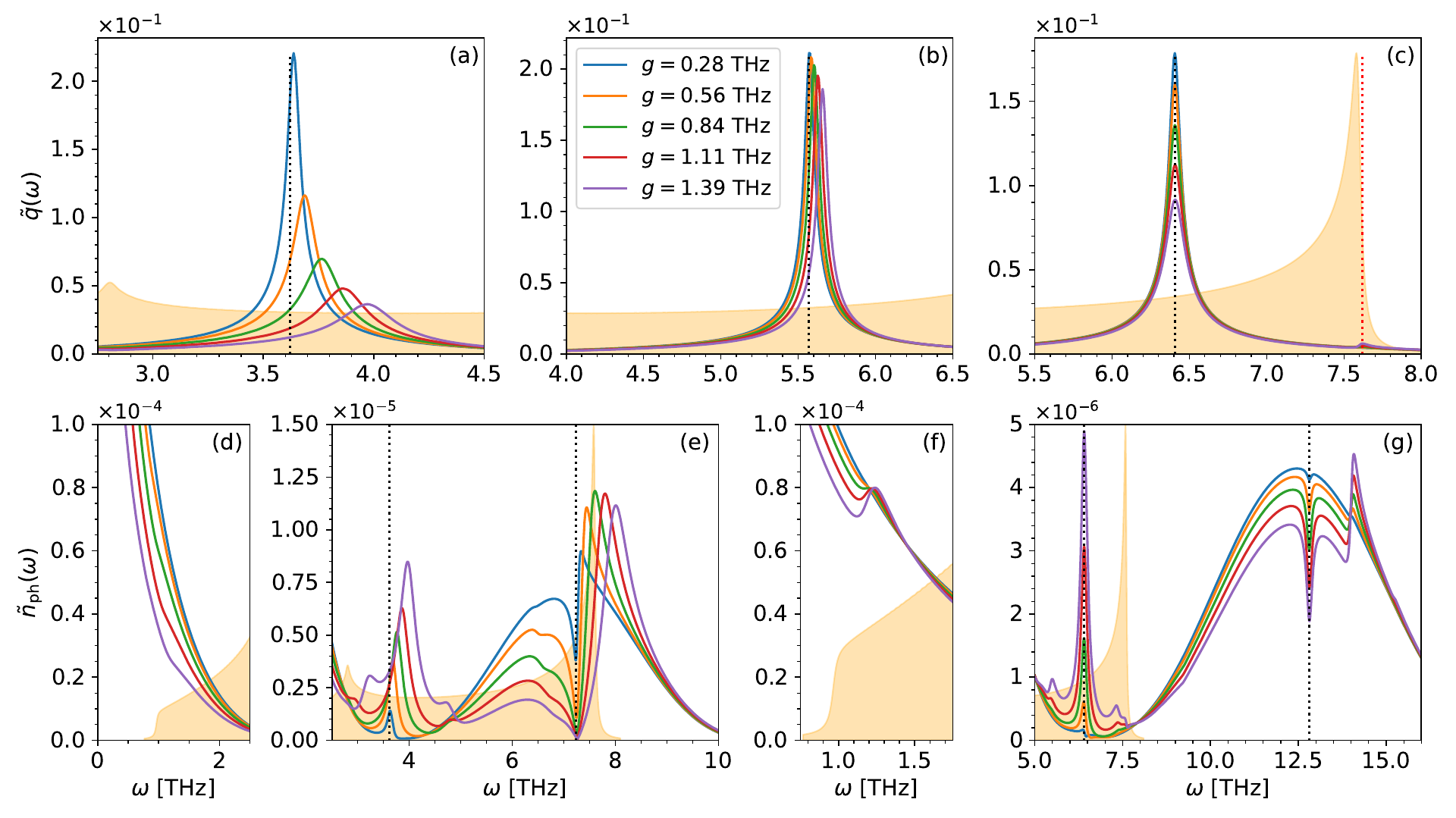}
	\caption{Absolute values of the Fourier transforms of the phonon displacement, \(\tilde{q} (\omega)\) (a-c), and phonon number, \(\tilde{n}_{\text{ph}}(\omega)\) (d-g), for three phonon frequencies above the saddle point [3.62 THz (a,d,e)], at the band center [5.57 THz (b)], and in the upper mid-band [6.41 THz (c,f,g)].}
	\label{fig:2d_ft_mid}
\end{figure*}

Moving to a phonon frequency near the saddle point, in Fig.~\ref{fig:2d_ft_low}(c) we find a clear $g$-induced splitting of the phononic peak, but now the concept of just two mutually repelling hybrids requires generalization. In this case the largest peak in $\tilde{q}(\omega)$, which is that moving upwards with increasing $g$, is repelled by two evident features appearing just below the saddle point and below the lower band edge. Of course the complete picture is that all the bitriplon states are hybridizing with the introduced phonon, altering their energy and contribution as discussed in Secs.~\ref{sec:pba} and \ref{sec:phen}, which here is manifest as both a saddle-point splitting and a band-edge feature. An example where the separation of the two hybrid states expected near the saddle point is governed by $|\omega_0 - \omega_{\rm sp}|$, rather than by $g$, is shown in Fig.~\ref{fig:2d_ft_mid}(a), where one composite state clearly moves upwards, but here the generalized hybridization effect with all lower-energy bitriplons is so uniform that there is no discernible saddle-point or band-edge feature (hence our cutting of the frequency axis). Turning to \(\tilde{n}_{\rm ph}(\omega)\), we find in Fig.~\ref{fig:2d_ft_low}(f) that the zero-frequency component is particularly dominant, masking any possibility of discerning difference-frequency signals. The sum-frequency signals in this more general situation, where Fig.~\ref{fig:2d_ft_low}(c) contains three distinct hybrids (around 0.5, 2.5, and 3.5 THz at large $g$), may now arise from several different combinations, and appear in this case as three broadened peaks.

The response of the system to driven phonons placed in the wide mid-band region between the saddle point and the upper band edge is dominated by the combination of weak hybridization effects with distant peaks in the bitriplon density of states and the small coupling coefficients $y'_k$ identified above. In the 2D system it is not possible to set all $y'_k$ to zero simultaneously, but in Fig.~\ref{fig:2d_ft_mid}(b) we show that a driven phonon placed close to the band center ($\omega_0 = 5.57$ THz $\simeq 2J$) has a very small net hybridization. Even at a coupling as strong as $g = 0.5 J$, the height of the phonon peak in $\tilde{q}(\omega)$ is reduced by only 10\%, remaining in essence the response of a weakly damped harmonic oscillator and confirming the extremely slow decay observed in the time domain in Fig.~\ref{fig:2d_t_q_g}(c). Thus the spin system is nearly ``transparent'' to magnetophononic driving at this frequency. 

Nevertheless, the peak position in Fig.~\ref{fig:2d_ft_mid}(b) does rise slowly with $g$, indicating a weak hybridization that is stronger with bitriplon states in the lower half of the band than with those in the upper. In Fig.~\ref{fig:2d_ft_mid}(c) we show that these repulsion effects cancel at $\omega_0 = 6.41$ THz, leaving the position of the phononic hybrid constant with $g$; here the extent of hybridization can be gauged from the fact that the peak height is halved at $g = 0.5 J$, and the hybrid state at the upper band edge has become visible on the same scale. 

At bilinear order, \(\tilde{n}_{\rm ph}(\omega)\) shows no detectable \(\omega_{\rm df}\) feature for lower mid-band phonons [Fig.~\ref{fig:2d_ft_mid}(d)], but upper mid-band phonons begin to display one at \(\omega_{\rm df} = \omega_{\rm ube} - \omega_0\) [Fig.~\ref{fig:2d_ft_mid}(f)]. The sum-frequency signals also differ, with multiple peaks appearing for lower mid-band phonons [Fig.~\ref{fig:2d_ft_mid}(e)] that contain some signatures of saddle-point and lower band-edge hybrids that are in essence indiscernible in $\tilde{q}(\omega)$. Although cursory inspection of Fig.~\ref{fig:2d_ft_mid}(g) suggests a standard three-peak form emerging for upper mid-band phonons, the $2\omega_{\rm ube}$ peak at 15.2 THz is barely discernible and the clear peak is given by $\omega_0 + \omega_{\rm ube}$; the fact that $2\omega_0$ is marked by a dip in the center of a very broad peak, despite the presence of a dominant sharp peak in $\tilde{q}(\omega)$, reinforces the message that matrix elements can be quite different in \(\tilde{n}_{\rm ph}(\omega)\) and mutual repulsion effects remain relevant. 

\begin{figure*}[t]
	\includegraphics[width=0.98\linewidth]{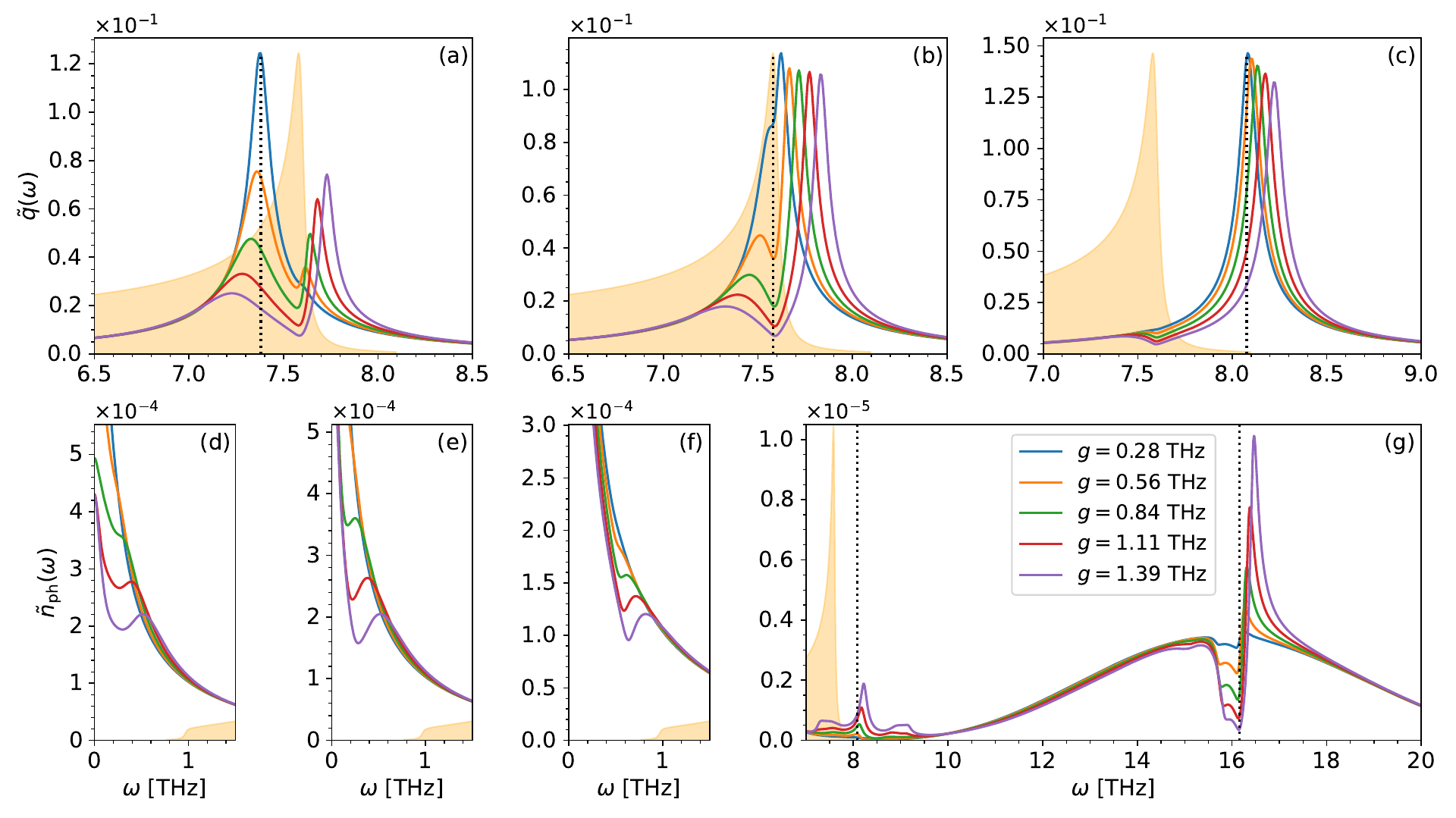}
	\caption{Absolute values of the Fourier transforms of the phonon displacement, \(\tilde{q} (\omega)\) (a-c), and phonon number, \(\tilde{n}_{\text{ph}}(\omega)\) (d-g), for three phonon frequencies below [7.38 THz (a,d)], at [7.72 THz (b,e)], and above [8.08 THz (c,f,g)] the upper band edge.}
	\label{fig:2d_ft_up}
\end{figure*}

Driven phonons placed around the upper band edge, where the density of states is dominated by a near-divergent spike (Fig.~\ref{fig:dos_2d}), provide an excellent mirror of the 1D physics investigated in Secs.~\ref{sec:mm} to \ref{sec:phen}. As Fig.~\ref{fig:2d_ft_up} makes clear, the hybridization of a phonon with the many bitriplon states available in a narrow energy range just below the upper band edge creates two repelling composite states in $\tilde{q}(\omega)$ [Figs.~\ref{fig:2d_ft_up}(a-c)], robust and systematically varying $\omega_{\rm df}$ peaks in \(\tilde{n}_{\rm ph}(\omega)\) [Figs.~\ref{fig:2d_ft_up}(d-f)], and three mostly standard sum-frequency peaks [Fig.~\ref{fig:2d_ft_up}(g)], all at fully realistic values of $g$. The phenomenology is the same as at the lower band edge in 1D (Figs.~\ref{fig:ft_q_w0_145} and \ref{fig:ft_nph_nt_w0_145}), with one phonon-bitriplon mode repelled outside the band while an inner collection of modes is summed to create a broad resonance that shifts downwards (i.e. deeper into the band) with increasing $g$. The only difference is quantitative, in that the $\omega_{\rm out}$ peak in Figs.~\ref{fig:2d_ft_up}(b,c) decays only weakly with $g$ while the $\omega_{\rm in}$ peak is weak and broad; this is again a consequence of the $y'_k$ coefficients, which become small due to the high energies appearing in the denominator. Nonetheless, these frequencies are evidently the best for finding clear signatures in the magnetophononic response of a material with the magnetic excitation spectrum of CuGeO$_3$. 

To summarize, all of the phenomenology observed in the 1D case can be found in a 2D spin system. The clearest signals will be observed in systems with strong peaks in the density of magnetic states, and by targeting phonons that ensure the visibility of difference-frequency features, which means placing them at frequencies that are not too small in order to avoid being masked by the large zero-frequency component. Quantitative differences between 1D magnetophononics and systems in higher dimensions are readily understood and offer wider possibilities for controlling the nature of the response function. 

Ultrafast experiments using magnetophononic driving to investigate spin-phonon coupling in CuGeO$_3$ will require three types of input. Here we have focused on a representative model of the spin spectrum in order to explore the frequency-dependence of the response. The second type of input concerns the precise frequencies of IR phonons available for coherent driving in THz experiments. As noted above, there are 60 phonon modes in the dimerized phase of CuGeO$_3$, of which 30 are IR-active and an unknown number of the 30 Raman-active modes may also contribute by nonlinear phononic mixing. Early investigations of CuGeO$_3$ already revealed the presence of IR modes over a frequency range that spans the bitriplon spectrum \cite{popov95,brade02}, and selected examples were illustrated in Ref.~\cite{yarmo23}. 

According to a recent systematic study \cite{spitz25}, the phonon spectrum at the $\Gamma$ point in CuGeO$_3$ has no optic modes below the bitriplon gap and only one (at 1.6 THz) below the saddle point. Above this, however, are 8 (IR and Raman) phonons in a narrow energy range (2.8--3.5 THz) just above the saddle point, 9 phonons in a mid-band range (4.9--6.2 THz), and 11 phonons in a range (7.6--9.0 THz) at and above the upper band edge. From the standpoint of frequency-selection highlighted in this section, CuGeO$_3$ therefore offers excellent possibilities for observing all the phenomena revealed in our minimal model. The third step would then be to model the eigenvectors of selected phonon modes \cite{spitz25} and their specific effects on the parameters $J$, $\lambda$, and $\mu$ in Eq.~\eqref{eq:disp_2d}. We close our considerations by drawing attention once again to the sum-frequency response [Figs.~\ref{fig:2d_ft_mid}(g) and \ref{fig:2d_ft_up}(g)], which was not considered previously \cite{yarmo23} but should be both readily detectable and more readily visible than the difference-frequency response. 

\section{Discussion and conclusion}
\label{sec:dc}

Experimental techniques based on pumping and probing with ultrafast coherent light have produced a revolution in the phenomena that can be observed and controlled in condensed matter \cite{zhang17b,salen19,oka19,torre21}. In magnetic materials, modulating the spin interactions by using the lattice excitations allows a valuable degree of frequency-selection \cite{fechn18,juras20,disa20,afana21}, and this magnetophononic route is a key mechanism in systems with no ordered moment \cite{giorg23}. Here we have considered pulsed driving in a minimal magnetophononic model consisting of one driven phonon modulating the spin interactions in a gapped quantum magnet, for which we use an alternating spin chain \cite{yarmo21,yarmo23}. Because an ultrashort pulse contains a very broad range of frequencies, even a one-pulse protocol can allow simultaneous driving and detection over much of the excitation spectrum.
	
We observe mutually repelling composite hybrid excitations accompanied by spectral features at their sum and difference frequencies. The composite phonon-bitriplon excitations are the dominant modes in an equilibrium system with spin-phonon coupling, and are particularly marked when the density of magnetic states has strong peaks. By contrast, the sum and difference features are nonequilibrium collective modes that are observed only due to the intrinsically transient nature of any pulsed driving. The difference-frequency mode in particular has a ready physical interpretation as a slow beat between the two fundamental excitations, and its frequency has a systematic dependence on the spin-phonon coupling, $g$. Directly after the pulse, it manifests as an oscillation of energy between the lattice and spin sectors that can reach a very large extent, exactly analogous to the behavior of two classical coupled pendula, where one or other can stop completely. Whereas the slow mode can be lost in the zero-frequency peak of a Fourier transform, the sum-frequency mode is a robust fingerprint with little detection confusion. 

In investigating the effect of the pulse intensity, we have found transient shifts of the excitation frequencies that are the pulsed analog of the spin-band engineering found \cite{yarmo23} with continuous driving. These shifts are also evident in the sum- and difference-frequency excitations. Despite its intrinsic interest, this single-pulse ``self-band-engineering'' is not a very targeted effect, and its deeper investigation or application would require at minimum a separation of pump and probe pulses, or other ``multi-dimensional'' spectroscopic methods \cite{liu25} involving two or more pulses. More generally, we remark also that very strong pulses raise the possibility of having to consider a multiplicity of phonons, some not necessarily with very strong coupling coefficients, that would be excited by the same short driving pulse.
	
We have shown by the worked example of the quasi-1D spin-Peierls material CuGeO$_3$ how these features should become observable in a THz pump-probe experiment. CuGeO$_3$ has the special properties that the gap is very small compared to the band center, and the density of magnetic states has no significant peaks anywhere other than the upper band edge. Although this does not impact the appearance of discernible composite excitations at any driving frequencies, it does complicate the observation of sum- and difference-frequency features. We also find a more general form of hybridization effect, where three or more mutually repelling peaks can appear in the spectrum and compete in determining the net frequency shifts. Driven phonons close to the band center can exhibit very low coupling coefficients even when the density of states and the spin-phonon coupling are large, leading to anomalously slow phonon decay (``phonon transparency''). One more special property of CuGeO$_3$ is to lose its dimerization at the spin-Peierls transition ($T_{\rm sp} = 14.2$ K), thereby requiring that experiments be performed at the lowest temperatures. 

More generally, these results underline the need to focus ultrafast studies of quantum magnets on systems with appreciable structure and contrast in their densities of magnetic excitations. In two and higher dimensions, this suggests the importance of highly frustrated magnetic materials with extended flat bands in their magnetic excitation spectrum. We have restricted our present study to the level of noninteracting triplons, which ensures the computationally straightforward solution of diagonal (bi)triplon matrices; although generalizing this part of the framework is not a conceptually challenging task, we do not anticipate that realistic triplon-triplon interactions would have a strong effect on our results. We also restricted our considerations to the low-temperature limit, which for a material with the magnetic interactions of CuGeO$_3$ ($J,J' \gg T$) is readily achieved in experiment. Clearly the effect of magnetic correlations is lost when $T \gg J,J'$, but one may envision narrow-band materials ($J' \ll J$) in a temperature regime $J' < T < J$, where collective triplon excitations become thermally incoherent but local triplet and bitriplet excitations remain, which form a band of width $T$ centered at $2J$ that still renormalizes the phonon dynamics due to the strong $g$. Finally, to focus on the frequency domain we have also restricted our diagnostic tools to quantities readily accessible within our methodology (Sec.~\ref{sec:mm}), namely the phonon and triplon numbers, leaving for future work the more experimentally oriented task of modelling the appearance of the same features in quantities including polarization rotation, circular dichroism, or nonlinear susceptibilties. 

Here we have also focused on the $S = 1/2$ Heisenberg Hamiltonian, which is a paradigm in quantum magnetism but nevertheless only a small region in the broad landscape of coupled spin-lattice systems. The Heisenberg model is special in that its lack of spin anisotropy forbids a direct coupling to the electric field of light, meaning that one photon cannot create one magnon (except through the light magnetic field, which is normally very weak). Further, its generic lattice coupling, specified in Eq.~\eqref{eq:hsp}, constrains one phonon to interact with two triplons, meaning that one phonon also cannot create one magnon. Relaxing these constraints in models with spin anisotropy, which are generic to materials with appreciable spin-orbit coupling, would lead to quite different types of spin-phonon hybridization in momentum space, although standard light-based probes would still investigate the zero-momentum response. 

We remark again in closing that we have worked in the regime of ``one-phonon magnetophononics,'' where the phonon frequencies have direct overlap with the spin (bitriplon) excitation spectrum and magnetophononic phenomena are created by targeting a single driven phonon mode. While this situation is exemplified well in CuGeO$_3$, quantum magnetic materials come with no guaranteed magnetic energy scales, which can vary from below 0.1 to above 100 meV. The harder optic phonons in inorganic materials can cover the upper end of this energy range, but accessing the lower end requires nonlinear mixing of the driving phonons. This type of ``two-phonon magnetophononics'' has been realized in the $S = 1/2$ Heisenberg antiferromagnet SrCu$_2$(BO$_3$)$_2$, where a characteristic magnetic excitation at 0.89 THz could be populated only by the difference-frequency component obtained from two driven phonons around 4 THz \cite{giorg23}. As for the bitriplons in our present model, the other feature required of the magnetic excitation in this case was $S = 0$ nature, which is a property of the lowest-lying component of the two-triplon bound state in SrCu$_2$(BO$_3$)$_2$. This example illustrates the wide horizons available in using nonequilibrium pump-probe methods to detect and control complex phononic, magnetic,  and hybridized composite states in quantum magnetic materials.

\begin{acknowledgments}
We thank D.~Bossini, N.~Colonna, K.~Deltenre, F.~Giorgianni, S.~Nikitin, Ch.~R\"uegg, and especially L.~Spitz for helpful discussions. We are grateful to Swiss National Science Foundation (SNF) for financial support through project 200021\_207438 and to the German Research Foundation (DFG) for financial support through projects UH 90-14/1 and UH 90-14/2, as well as project B8 of ICRC 160.
\end{acknowledgments}

\section*{Data Availability}

The data that support the findings of this article are openly available \cite{datadump}.

\bibliography{pmbib}

\begin{thebibliography}{54}%
\makeatletter
\providecommand \@ifxundefined [1]{%
 \@ifx{#1\undefined}
}%
\providecommand \@ifnum [1]{%
 \ifnum #1\expandafter \@firstoftwo
 \else \expandafter \@secondoftwo
 \fi
}%
\providecommand \@ifx [1]{%
 \ifx #1\expandafter \@firstoftwo
 \else \expandafter \@secondoftwo
 \fi
}%
\providecommand \natexlab [1]{#1}%
\providecommand \enquote  [1]{``#1''}%
\providecommand \bibnamefont  [1]{#1}%
\providecommand \bibfnamefont [1]{#1}%
\providecommand \citenamefont [1]{#1}%
\providecommand \href@noop [0]{\@secondoftwo}%
\providecommand \href [0]{\begingroup \@sanitize@url \@href}%
\providecommand \@href[1]{\@@startlink{#1}\@@href}%
\providecommand \@@href[1]{\endgroup#1\@@endlink}%
\providecommand \@sanitize@url [0]{\catcode `\\12\catcode `\$12\catcode
  `\&12\catcode `\#12\catcode `\^12\catcode `\_12\catcode `\%12\relax}%
\providecommand \@@startlink[1]{}%
\providecommand \@@endlink[0]{}%
\providecommand \url  [0]{\begingroup\@sanitize@url \@url }%
\providecommand \@url [1]{\endgroup\@href {#1}{\urlprefix }}%
\providecommand \urlprefix  [0]{URL }%
\providecommand \Eprint [0]{\href }%
\providecommand \doibase [0]{https://doi.org/}%
\providecommand \selectlanguage [0]{\@gobble}%
\providecommand \bibinfo  [0]{\@secondoftwo}%
\providecommand \bibfield  [0]{\@secondoftwo}%
\providecommand \translation [1]{[#1]}%
\providecommand \BibitemOpen [0]{}%
\providecommand \bibitemStop [0]{}%
\providecommand \bibitemNoStop [0]{.\EOS\space}%
\providecommand \EOS [0]{\spacefactor3000\relax}%
\providecommand \BibitemShut  [1]{\csname bibitem#1\endcsname}%
\let\auto@bib@innerbib\@empty
\bibitem [{\citenamefont {Kampfrath}\ \emph {et~al.}(2013)\citenamefont
  {Kampfrath}, \citenamefont {Tanaka},\ and\ \citenamefont {Nelson}}]{kampf13}%
  \BibitemOpen
  \bibfield  {author} {\bibinfo {author} {\bibfnamefont {T.}~\bibnamefont
  {Kampfrath}}, \bibinfo {author} {\bibfnamefont {K.}~\bibnamefont {Tanaka}},\
  and\ \bibinfo {author} {\bibfnamefont {K.~A.}\ \bibnamefont {Nelson}},\
  }\bibfield  {title} {\bibinfo {title} {Resonant and nonresonant control over
  matter and light by intense terahertz transients},\ }\href
  {https://doi.org/10.1038/nphoton.2013.184} {\bibfield  {journal} {\bibinfo
  {journal} {Nat. Photonics}\ }\textbf {\bibinfo {volume} {7}},\ \bibinfo
  {pages} {680} (\bibinfo {year} {2013})}\BibitemShut {NoStop}%
\bibitem [{\citenamefont {Zhang}\ \emph {et~al.}(2017)\citenamefont {Zhang},
  \citenamefont {Shkurinov},\ and\ \citenamefont {Zhang}}]{zhang17b}%
  \BibitemOpen
  \bibfield  {author} {\bibinfo {author} {\bibfnamefont {X.~C.}\ \bibnamefont
  {Zhang}}, \bibinfo {author} {\bibfnamefont {A.}~\bibnamefont {Shkurinov}},\
  and\ \bibinfo {author} {\bibfnamefont {Y.}~\bibnamefont {Zhang}},\ }\bibfield
   {title} {\bibinfo {title} {Extreme terahertz science},\ }\href
  {https://doi.org/10.1038/nphoton.2016.249} {\bibfield  {journal} {\bibinfo
  {journal} {Nat. Photonics}\ }\textbf {\bibinfo {volume} {11}},\ \bibinfo
  {pages} {16} (\bibinfo {year} {2017})}\BibitemShut {NoStop}%
\bibitem [{\citenamefont {Sal\'en}\ \emph {et~al.}(2019)\citenamefont
  {Sal\'en}, \citenamefont {Basini}, \citenamefont {Bonetti}, \citenamefont
  {Hebling}, \citenamefont {Krasilnikov}, \citenamefont {Nikitin},
  \citenamefont {Shamuilov}, \citenamefont {Tibai}, \citenamefont
  {Zhaunerchyk},\ and\ \citenamefont {Goryashko}}]{salen19}%
  \BibitemOpen
  \bibfield  {author} {\bibinfo {author} {\bibfnamefont {P.}~\bibnamefont
  {Sal\'en}}, \bibinfo {author} {\bibfnamefont {M.}~\bibnamefont {Basini}},
  \bibinfo {author} {\bibfnamefont {S.}~\bibnamefont {Bonetti}}, \bibinfo
  {author} {\bibfnamefont {J.}~\bibnamefont {Hebling}}, \bibinfo {author}
  {\bibfnamefont {M.}~\bibnamefont {Krasilnikov}}, \bibinfo {author}
  {\bibfnamefont {A.~Y.}\ \bibnamefont {Nikitin}}, \bibinfo {author}
  {\bibfnamefont {G.}~\bibnamefont {Shamuilov}}, \bibinfo {author}
  {\bibfnamefont {Z.}~\bibnamefont {Tibai}}, \bibinfo {author} {\bibfnamefont
  {V.}~\bibnamefont {Zhaunerchyk}},\ and\ \bibinfo {author} {\bibfnamefont
  {V.}~\bibnamefont {Goryashko}},\ }\bibfield  {title} {\bibinfo {title}
  {{Matter manipulation with extreme terahertz light: Progress in the enabling
  {THz} technology}},\ }\href
  {https://doi.org/https://doi.org/10.1016/j.physrep.2019.09.002} {\bibfield
  {journal} {\bibinfo  {journal} {Phys. Rep.}\ }\textbf {\bibinfo {volume}
  {836-837}},\ \bibinfo {pages} {1} (\bibinfo {year} {2019})}\BibitemShut
  {NoStop}%
\bibitem [{\citenamefont {Cavalleri}(2018)}]{caval18}%
  \BibitemOpen
  \bibfield  {author} {\bibinfo {author} {\bibfnamefont {A.}~\bibnamefont
  {Cavalleri}},\ }\bibfield  {title} {\bibinfo {title} {Photo-induced
  superconductivity},\ }\href {https://doi.org/10.1080/00107514.2017.1406623}
  {\bibfield  {journal} {\bibinfo  {journal} {Contemp. Phys.}\ }\textbf
  {\bibinfo {volume} {59}},\ \bibinfo {pages} {31} (\bibinfo {year}
  {2018})}\BibitemShut {NoStop}%
\bibitem [{\citenamefont {Murakami}\ \emph {et~al.}(2023)\citenamefont
  {Murakami}, \citenamefont {Golež}, \citenamefont {Eckstein},\ and\
  \citenamefont {Werner}}]{murak23}%
  \BibitemOpen
  \bibfield  {author} {\bibinfo {author} {\bibfnamefont {Y.}~\bibnamefont
  {Murakami}}, \bibinfo {author} {\bibfnamefont {D.}~\bibnamefont {Golež}},
  \bibinfo {author} {\bibfnamefont {M.}~\bibnamefont {Eckstein}},\ and\
  \bibinfo {author} {\bibfnamefont {P.}~\bibnamefont {Werner}},\ }\href@noop {}
  {\bibinfo {title} {{Photo-induced nonequilibrium states in Mott insulators}}}
  (\bibinfo {year} {2023}),\ \Eprint {https://arxiv.org/abs/2310.05201}
  {arXiv:2310.05201} \BibitemShut {NoStop}%
\bibitem [{\citenamefont {Oka}\ and\ \citenamefont {Kitamura}(2019)}]{oka19}%
  \BibitemOpen
  \bibfield  {author} {\bibinfo {author} {\bibfnamefont {T.}~\bibnamefont
  {Oka}}\ and\ \bibinfo {author} {\bibfnamefont {S.}~\bibnamefont {Kitamura}},\
  }\bibfield  {title} {\bibinfo {title} {{Floquet Engineering of Quantum
  Materials}},\ }\href@noop {} {\bibfield  {journal} {\bibinfo  {journal}
  {Annu. Rev. Condens. Matter Phys.}\ }\textbf {\bibinfo {volume} {10}},\
  \bibinfo {pages} {387} (\bibinfo {year} {2019})}\BibitemShut {NoStop}%
\bibitem [{\citenamefont {Disa}\ \emph {et~al.}(2021)\citenamefont {Disa},
  \citenamefont {Nova}, \citenamefont {Liu},\ and\ \citenamefont
  {Cavalleri}}]{disa21}%
  \BibitemOpen
  \bibfield  {author} {\bibinfo {author} {\bibfnamefont {A.~S.}\ \bibnamefont
  {Disa}}, \bibinfo {author} {\bibfnamefont {T.~F.}\ \bibnamefont {Nova}},
  \bibinfo {author} {\bibfnamefont {B.}~\bibnamefont {Liu}},\ and\ \bibinfo
  {author} {\bibfnamefont {A.}~\bibnamefont {Cavalleri}},\ }\bibfield  {title}
  {\bibinfo {title} {Engineering crystal structures with light},\ }\href
  {https://doi.org/10.1038/s41567-021-01366-1} {\bibfield  {journal} {\bibinfo
  {journal} {Nat. Phys.}\ }\textbf {\bibinfo {volume} {17}},\ \bibinfo {pages}
  {1092} (\bibinfo {year} {2021})}\BibitemShut {NoStop}%
\bibitem [{\citenamefont {de~la Torre}\ \emph {et~al.}(2021)\citenamefont
  {de~la Torre}, \citenamefont {Kennes}, \citenamefont {Claassen},
  \citenamefont {Gerber}, \citenamefont {McIver},\ and\ \citenamefont
  {Sentef}}]{torre21}%
  \BibitemOpen
  \bibfield  {author} {\bibinfo {author} {\bibfnamefont {A.}~\bibnamefont
  {de~la Torre}}, \bibinfo {author} {\bibfnamefont {D.~M.}\ \bibnamefont
  {Kennes}}, \bibinfo {author} {\bibfnamefont {M.}~\bibnamefont {Claassen}},
  \bibinfo {author} {\bibfnamefont {S.}~\bibnamefont {Gerber}}, \bibinfo
  {author} {\bibfnamefont {J.~W.}\ \bibnamefont {McIver}},\ and\ \bibinfo
  {author} {\bibfnamefont {M.}~\bibnamefont {Sentef}},\ }\bibfield  {title}
  {\bibinfo {title} {{Nonthermal pathways to ultrafast control in quantum
  materials}},\ }\href {https://doi.org/10.1103/RevModPhys.93.041002}
  {\bibfield  {journal} {\bibinfo  {journal} {Rev. Mod. Phys.}\ }\textbf
  {\bibinfo {volume} {93}},\ \bibinfo {pages} {041002} (\bibinfo {year}
  {2021})}\BibitemShut {NoStop}%
\bibitem [{\citenamefont {Kampfrath}\ \emph {et~al.}(2011)\citenamefont
  {Kampfrath}, \citenamefont {Sell}, \citenamefont {Klatt}, \citenamefont
  {Pashkin}, \citenamefont {Mährlein}, \citenamefont {Dekorsy}, \citenamefont
  {Wolf}, \citenamefont {Fiebig}, \citenamefont {Leitenstorfer},\ and\
  \citenamefont {Huber}}]{kampf11}%
  \BibitemOpen
  \bibfield  {author} {\bibinfo {author} {\bibfnamefont {T.}~\bibnamefont
  {Kampfrath}}, \bibinfo {author} {\bibfnamefont {A.}~\bibnamefont {Sell}},
  \bibinfo {author} {\bibfnamefont {G.}~\bibnamefont {Klatt}}, \bibinfo
  {author} {\bibfnamefont {A.}~\bibnamefont {Pashkin}}, \bibinfo {author}
  {\bibfnamefont {S.}~\bibnamefont {Mährlein}}, \bibinfo {author}
  {\bibfnamefont {T.}~\bibnamefont {Dekorsy}}, \bibinfo {author} {\bibfnamefont
  {M.}~\bibnamefont {Wolf}}, \bibinfo {author} {\bibfnamefont {M.}~\bibnamefont
  {Fiebig}}, \bibinfo {author} {\bibfnamefont {A.}~\bibnamefont
  {Leitenstorfer}},\ and\ \bibinfo {author} {\bibfnamefont {R.}~\bibnamefont
  {Huber}},\ }\bibfield  {title} {\bibinfo {title} {Coherent terahertz control
  of antiferromagnetic spin waves},\ }\href
  {https://doi.org/10.1038/nphoton.2010.259} {\bibfield  {journal} {\bibinfo
  {journal} {Nat. Photonics}\ }\textbf {\bibinfo {volume} {5}},\ \bibinfo
  {pages} {31} (\bibinfo {year} {2011})}\BibitemShut {NoStop}%
\bibitem [{\citenamefont {Kubacka}\ \emph {et~al.}(2014)\citenamefont
  {Kubacka}, \citenamefont {Johnson}, \citenamefont {Hoffmann}, \citenamefont
  {Vicario}, \citenamefont {de~Jong}, \citenamefont {Beaud}, \citenamefont
  {Gr{\"u}bel}, \citenamefont {Huang}, \citenamefont {Huber}, \citenamefont
  {Patthey}, \citenamefont {Chuang}, \citenamefont {Turner}, \citenamefont
  {Dakovski}, \citenamefont {Lee}, \citenamefont {Minitti}, \citenamefont
  {Schlotter}, \citenamefont {Moore}, \citenamefont {Hauri}, \citenamefont
  {Koohpayeh}, \citenamefont {Scagnoli}, \citenamefont {Ingold}, \citenamefont
  {Johnson},\ and\ \citenamefont {Staub}}]{kubac14}%
  \BibitemOpen
  \bibfield  {author} {\bibinfo {author} {\bibfnamefont {T.}~\bibnamefont
  {Kubacka}}, \bibinfo {author} {\bibfnamefont {J.~A.}\ \bibnamefont
  {Johnson}}, \bibinfo {author} {\bibfnamefont {M.~C.}\ \bibnamefont
  {Hoffmann}}, \bibinfo {author} {\bibfnamefont {C.}~\bibnamefont {Vicario}},
  \bibinfo {author} {\bibfnamefont {S.}~\bibnamefont {de~Jong}}, \bibinfo
  {author} {\bibfnamefont {P.}~\bibnamefont {Beaud}}, \bibinfo {author}
  {\bibfnamefont {S.}~\bibnamefont {Gr{\"u}bel}}, \bibinfo {author}
  {\bibfnamefont {S.-W.}\ \bibnamefont {Huang}}, \bibinfo {author}
  {\bibfnamefont {L.}~\bibnamefont {Huber}}, \bibinfo {author} {\bibfnamefont
  {L.}~\bibnamefont {Patthey}}, \bibinfo {author} {\bibfnamefont {Y.-D.}\
  \bibnamefont {Chuang}}, \bibinfo {author} {\bibfnamefont {J.~J.}\
  \bibnamefont {Turner}}, \bibinfo {author} {\bibfnamefont {G.~L.}\
  \bibnamefont {Dakovski}}, \bibinfo {author} {\bibfnamefont {W.-S.}\
  \bibnamefont {Lee}}, \bibinfo {author} {\bibfnamefont {M.~P.}\ \bibnamefont
  {Minitti}}, \bibinfo {author} {\bibfnamefont {W.}~\bibnamefont {Schlotter}},
  \bibinfo {author} {\bibfnamefont {R.~G.}\ \bibnamefont {Moore}}, \bibinfo
  {author} {\bibfnamefont {C.~P.}\ \bibnamefont {Hauri}}, \bibinfo {author}
  {\bibfnamefont {S.~M.}\ \bibnamefont {Koohpayeh}}, \bibinfo {author}
  {\bibfnamefont {V.}~\bibnamefont {Scagnoli}}, \bibinfo {author}
  {\bibfnamefont {G.}~\bibnamefont {Ingold}}, \bibinfo {author} {\bibfnamefont
  {S.~L.}\ \bibnamefont {Johnson}},\ and\ \bibinfo {author} {\bibfnamefont
  {U.}~\bibnamefont {Staub}},\ }\bibfield  {title} {\bibinfo {title}
  {Large-amplitude spin dynamics driven by a {THz} pulse in resonance with an
  electromagnon},\ }\href {https://doi.org/10.1126/science.1242862} {\bibfield
  {journal} {\bibinfo  {journal} {Science}\ }\textbf {\bibinfo {volume}
  {343}},\ \bibinfo {pages} {1333} (\bibinfo {year} {2014})}\BibitemShut
  {NoStop}%
\bibitem [{\citenamefont {Baierl}\ \emph {et~al.}(2016)\citenamefont {Baierl},
  \citenamefont {Hohenleutner}, \citenamefont {Kampfrath}, \citenamefont
  {Zvezdin}, \citenamefont {Kimel}, \citenamefont {Huber},\ and\ \citenamefont
  {Mikhaylovskiy}}]{baier16}%
  \BibitemOpen
  \bibfield  {author} {\bibinfo {author} {\bibfnamefont {S.}~\bibnamefont
  {Baierl}}, \bibinfo {author} {\bibfnamefont {M.}~\bibnamefont
  {Hohenleutner}}, \bibinfo {author} {\bibfnamefont {T.}~\bibnamefont
  {Kampfrath}}, \bibinfo {author} {\bibfnamefont {A.~K.}\ \bibnamefont
  {Zvezdin}}, \bibinfo {author} {\bibfnamefont {A.~V.}\ \bibnamefont {Kimel}},
  \bibinfo {author} {\bibfnamefont {R.}~\bibnamefont {Huber}},\ and\ \bibinfo
  {author} {\bibfnamefont {R.~V.}\ \bibnamefont {Mikhaylovskiy}},\ }\bibfield
  {title} {\bibinfo {title} {{Nonlinear spin control by terahertz-driven
  anisotropy fields}},\ }\href {https://doi.org/10.1038/nphoton.2016.181}
  {\bibfield  {journal} {\bibinfo  {journal} {Nat. Photonics}\ }\textbf
  {\bibinfo {volume} {10}},\ \bibinfo {pages} {715} (\bibinfo {year}
  {2016})}\BibitemShut {NoStop}%
\bibitem [{\citenamefont {Lu}\ \emph {et~al.}(2017)\citenamefont {Lu},
  \citenamefont {Li}, \citenamefont {Hwang}, \citenamefont {Ofori-Okai},
  \citenamefont {Kurihara}, \citenamefont {Suemoto},\ and\ \citenamefont
  {Nelson}}]{lu17}%
  \BibitemOpen
  \bibfield  {author} {\bibinfo {author} {\bibfnamefont {J.}~\bibnamefont
  {Lu}}, \bibinfo {author} {\bibfnamefont {X.}~\bibnamefont {Li}}, \bibinfo
  {author} {\bibfnamefont {H.~Y.}\ \bibnamefont {Hwang}}, \bibinfo {author}
  {\bibfnamefont {B.~K.}\ \bibnamefont {Ofori-Okai}}, \bibinfo {author}
  {\bibfnamefont {T.}~\bibnamefont {Kurihara}}, \bibinfo {author}
  {\bibfnamefont {T.}~\bibnamefont {Suemoto}},\ and\ \bibinfo {author}
  {\bibfnamefont {K.~A.}\ \bibnamefont {Nelson}},\ }\bibfield  {title}
  {\bibinfo {title} {{Coherent Two-Dimensional Terahertz Magnetic Resonance
  Spectroscopy of Collective Spin Waves}},\ }\href
  {https://doi.org/10.1103/PhysRevLett.118.207204} {\bibfield  {journal}
  {\bibinfo  {journal} {Phys. Rev. Lett.}\ }\textbf {\bibinfo {volume} {118}},\
  \bibinfo {pages} {207204} (\bibinfo {year} {2017})}\BibitemShut {NoStop}%
\bibitem [{\citenamefont {Mashkovich}\ \emph {et~al.}(2019)\citenamefont
  {Mashkovich}, \citenamefont {Grishunin}, \citenamefont {Mikhaylovskiy},
  \citenamefont {Zvezdin}, \citenamefont {Pisarev}, \citenamefont {Strugatsky},
  \citenamefont {Christianen}, \citenamefont {Rasing},\ and\ \citenamefont
  {Kimel}}]{mashk19}%
  \BibitemOpen
  \bibfield  {author} {\bibinfo {author} {\bibfnamefont {E.}~\bibnamefont
  {Mashkovich}}, \bibinfo {author} {\bibfnamefont {K.}~\bibnamefont
  {Grishunin}}, \bibinfo {author} {\bibfnamefont {R.}~\bibnamefont
  {Mikhaylovskiy}}, \bibinfo {author} {\bibfnamefont {A.}~\bibnamefont
  {Zvezdin}}, \bibinfo {author} {\bibfnamefont {R.}~\bibnamefont {Pisarev}},
  \bibinfo {author} {\bibfnamefont {M.}~\bibnamefont {Strugatsky}}, \bibinfo
  {author} {\bibfnamefont {P.}~\bibnamefont {Christianen}}, \bibinfo {author}
  {\bibfnamefont {T.}~\bibnamefont {Rasing}},\ and\ \bibinfo {author}
  {\bibfnamefont {A.}~\bibnamefont {Kimel}},\ }\bibfield  {title} {\bibinfo
  {title} {{Terahertz Optomagnetism: Nonlinear THz Excitation of GHz Spin Waves
  in Antiferromagnetic FeBO$_3$}},\ }\href
  {https://doi.org/10.1103/PhysRevLett.123.157202} {\bibfield  {journal}
  {\bibinfo  {journal} {Phys. Rev. Lett.}\ }\textbf {\bibinfo {volume} {123}},\
  \bibinfo {pages} {157202} (\bibinfo {year} {2019})}\BibitemShut {NoStop}%
\bibitem [{\citenamefont {Behovits}\ \emph {et~al.}(2023)\citenamefont
  {Behovits}, \citenamefont {Chekhov}, \citenamefont {Bodnar}, \citenamefont
  {Gueckstock}, \citenamefont {Reimers}, \citenamefont {Lytvynenko},
  \citenamefont {Skourski}, \citenamefont {Wolf}, \citenamefont {Seifert},
  \citenamefont {Gomonay}, \citenamefont {Kl\"aui}, \citenamefont {Jourdan},\
  and\ \citenamefont {T.Kampfrath}}]{behov23}%
  \BibitemOpen
  \bibfield  {author} {\bibinfo {author} {\bibfnamefont {Y.}~\bibnamefont
  {Behovits}}, \bibinfo {author} {\bibfnamefont {A.~L.}\ \bibnamefont
  {Chekhov}}, \bibinfo {author} {\bibfnamefont {S.~Y.}\ \bibnamefont {Bodnar}},
  \bibinfo {author} {\bibfnamefont {O.}~\bibnamefont {Gueckstock}}, \bibinfo
  {author} {\bibfnamefont {S.}~\bibnamefont {Reimers}}, \bibinfo {author}
  {\bibfnamefont {Y.}~\bibnamefont {Lytvynenko}}, \bibinfo {author}
  {\bibfnamefont {Y.}~\bibnamefont {Skourski}}, \bibinfo {author}
  {\bibfnamefont {M.}~\bibnamefont {Wolf}}, \bibinfo {author} {\bibfnamefont
  {T.~S.}\ \bibnamefont {Seifert}}, \bibinfo {author} {\bibfnamefont
  {O.}~\bibnamefont {Gomonay}}, \bibinfo {author} {\bibfnamefont
  {M.}~\bibnamefont {Kl\"aui}}, \bibinfo {author} {\bibfnamefont
  {M.}~\bibnamefont {Jourdan}},\ and\ \bibinfo {author} {\bibnamefont
  {T.Kampfrath}},\ }\bibfield  {title} {\bibinfo {title} {Terahertz {N\'eel
  spin-orbit torques drive nonlinear magnon dynamics in antiferromagnetic
  Mn$_2$Au}},\ }\href {https://doi.org/10.1038/s41467-023-41569-z} {\bibfield
  {journal} {\bibinfo  {journal} {Nat. Commun.}\ }\textbf {\bibinfo {volume}
  {14}},\ \bibinfo {pages} {6038} (\bibinfo {year} {2023})}\BibitemShut
  {NoStop}%
\bibitem [{\citenamefont {Khudoyberdiev}\ and\ \citenamefont
  {Uhrig}(2024)}]{khudo24}%
  \BibitemOpen
  \bibfield  {author} {\bibinfo {author} {\bibfnamefont {A.}~\bibnamefont
  {Khudoyberdiev}}\ and\ \bibinfo {author} {\bibfnamefont {G.~S.}\ \bibnamefont
  {Uhrig}},\ }\bibfield  {title} {\bibinfo {title} {Switching of {magnetization
  in quantum antiferromagnets with time-dependent control fields}},\ }\href
  {https://doi.org/10.1103/PhysRevB.109.174419} {\bibfield  {journal} {\bibinfo
   {journal} {Phys. Rev. B}\ }\textbf {\bibinfo {volume} {109}},\ \bibinfo
  {pages} {174419} (\bibinfo {year} {2024})}\BibitemShut {NoStop}%
\bibitem [{\citenamefont {Walowski}\ and\ \citenamefont
  {Münzenberg}(2016)}]{walow16}%
  \BibitemOpen
  \bibfield  {author} {\bibinfo {author} {\bibfnamefont {J.}~\bibnamefont
  {Walowski}}\ and\ \bibinfo {author} {\bibfnamefont {M.}~\bibnamefont
  {Münzenberg}},\ }\bibfield  {title} {\bibinfo {title} {{Perspective:
  Ultrafast magnetism and THz spintronics}},\ }\href
  {https://doi.org/10.1063/1.4958846} {\bibfield  {journal} {\bibinfo
  {journal} {J. Appl. Phys.}\ }\textbf {\bibinfo {volume} {120}},\ \bibinfo
  {pages} {140901} (\bibinfo {year} {2016})}\BibitemShut {NoStop}%
\bibitem [{\citenamefont {N\v{e}mec}\ \emph {et~al.}(2018)\citenamefont
  {N\v{e}mec}, \citenamefont {Fiebig}, \citenamefont {Kampfrath},\ and\
  \citenamefont {Kimel}}]{nemec18}%
  \BibitemOpen
  \bibfield  {author} {\bibinfo {author} {\bibfnamefont {P.}~\bibnamefont
  {N\v{e}mec}}, \bibinfo {author} {\bibfnamefont {M.}~\bibnamefont {Fiebig}},
  \bibinfo {author} {\bibfnamefont {T.}~\bibnamefont {Kampfrath}},\ and\
  \bibinfo {author} {\bibfnamefont {A.~V.}\ \bibnamefont {Kimel}},\ }\bibfield
  {title} {\bibinfo {title} {Antiferromagnetic opto-spintronics},\ }\href
  {https://doi.org/10.1038/s41567-018-0051x} {\bibfield  {journal} {\bibinfo
  {journal} {Nat. Phys.}\ }\textbf {\bibinfo {volume} {14}},\ \bibinfo {pages}
  {229} (\bibinfo {year} {2018})}\BibitemShut {NoStop}%
\bibitem [{\citenamefont {Pirro}\ \emph {et~al.}(2021)\citenamefont {Pirro},
  \citenamefont {Vasyuchka}, \citenamefont {Serga},\ and\ \citenamefont
  {Hillebrands}}]{pirro21}%
  \BibitemOpen
  \bibfield  {author} {\bibinfo {author} {\bibfnamefont {P.}~\bibnamefont
  {Pirro}}, \bibinfo {author} {\bibfnamefont {V.~I.}\ \bibnamefont
  {Vasyuchka}}, \bibinfo {author} {\bibfnamefont {A.~A.}\ \bibnamefont
  {Serga}},\ and\ \bibinfo {author} {\bibfnamefont {B.}~\bibnamefont
  {Hillebrands}},\ }\bibfield  {title} {\bibinfo {title} {{Advances in coherent
  magnonics}},\ }\href {https://doi.org/10.1038/s41578-021-00332-w} {\bibfield
  {journal} {\bibinfo  {journal} {Nat. Rev. Mater.}\ }\textbf {\bibinfo
  {volume} {6}},\ \bibinfo {pages} {1114} (\bibinfo {year} {2021})}\BibitemShut
  {NoStop}%
\bibitem [{\citenamefont {Broholm}\ \emph {et~al.}(2020)\citenamefont
  {Broholm}, \citenamefont {Cava}, \citenamefont {Kivelson}, \citenamefont
  {Nocera}, \citenamefont {Norman},\ and\ \citenamefont {Senthil}}]{broho20}%
  \BibitemOpen
  \bibfield  {author} {\bibinfo {author} {\bibfnamefont {C.}~\bibnamefont
  {Broholm}}, \bibinfo {author} {\bibfnamefont {R.~J.}\ \bibnamefont {Cava}},
  \bibinfo {author} {\bibfnamefont {S.~A.}\ \bibnamefont {Kivelson}}, \bibinfo
  {author} {\bibfnamefont {D.~G.}\ \bibnamefont {Nocera}}, \bibinfo {author}
  {\bibfnamefont {M.~R.}\ \bibnamefont {Norman}},\ and\ \bibinfo {author}
  {\bibfnamefont {T.}~\bibnamefont {Senthil}},\ }\bibfield  {title} {\bibinfo
  {title} {Quantum spin liquids},\ }\href
  {https://doi.org/10.1126/science.aay0668} {\bibfield  {journal} {\bibinfo
  {journal} {Science}\ }\textbf {\bibinfo {volume} {367}},\ \bibinfo {pages}
  {eaay0668} (\bibinfo {year} {2020})}\BibitemShut {NoStop}%
\bibitem [{\citenamefont {Nova}\ \emph {et~al.}(2017)\citenamefont {Nova},
  \citenamefont {Cartella}, \citenamefont {Cantaluppi}, \citenamefont
  {F\"orst}, \citenamefont {Bossini}, \citenamefont {Mikhaylovskiy},
  \citenamefont {Kimel}, \citenamefont {Merlin},\ and\ \citenamefont
  {Cavalleri}}]{nova17}%
  \BibitemOpen
  \bibfield  {author} {\bibinfo {author} {\bibfnamefont {T.~F.}\ \bibnamefont
  {Nova}}, \bibinfo {author} {\bibfnamefont {A.}~\bibnamefont {Cartella}},
  \bibinfo {author} {\bibfnamefont {A.}~\bibnamefont {Cantaluppi}}, \bibinfo
  {author} {\bibfnamefont {M.}~\bibnamefont {F\"orst}}, \bibinfo {author}
  {\bibfnamefont {D.}~\bibnamefont {Bossini}}, \bibinfo {author} {\bibfnamefont
  {R.~V.}\ \bibnamefont {Mikhaylovskiy}}, \bibinfo {author} {\bibfnamefont
  {A.~V.}\ \bibnamefont {Kimel}}, \bibinfo {author} {\bibfnamefont
  {R.}~\bibnamefont {Merlin}},\ and\ \bibinfo {author} {\bibfnamefont
  {A.}~\bibnamefont {Cavalleri}},\ }\bibfield  {title} {\bibinfo {title} {An
  effective magnetic field from optically driven phonons},\ }\href
  {https://doi.org/10.1038/nphys3925} {\bibfield  {journal} {\bibinfo
  {journal} {Nat. Phys.}\ }\textbf {\bibinfo {volume} {13}},\ \bibinfo {pages}
  {132} (\bibinfo {year} {2017})}\BibitemShut {NoStop}%
\bibitem [{\citenamefont {Kozina}\ \emph {et~al.}(2019)\citenamefont {Kozina},
  \citenamefont {Fechner}, \citenamefont {Marsik}, \citenamefont {van Driel},
  \citenamefont {Glownia}, \citenamefont {Bernhard}, \citenamefont {Radovic},
  \citenamefont {Zhu}, \citenamefont {Bonetti}, \citenamefont {Staub},\ and\
  \citenamefont {Hoffmann}}]{kozin19}%
  \BibitemOpen
  \bibfield  {author} {\bibinfo {author} {\bibfnamefont {M.}~\bibnamefont
  {Kozina}}, \bibinfo {author} {\bibfnamefont {M.}~\bibnamefont {Fechner}},
  \bibinfo {author} {\bibfnamefont {P.}~\bibnamefont {Marsik}}, \bibinfo
  {author} {\bibfnamefont {T.}~\bibnamefont {van Driel}}, \bibinfo {author}
  {\bibfnamefont {J.~M.}\ \bibnamefont {Glownia}}, \bibinfo {author}
  {\bibfnamefont {C.}~\bibnamefont {Bernhard}}, \bibinfo {author}
  {\bibfnamefont {M.}~\bibnamefont {Radovic}}, \bibinfo {author} {\bibfnamefont
  {D.}~\bibnamefont {Zhu}}, \bibinfo {author} {\bibfnamefont {S.}~\bibnamefont
  {Bonetti}}, \bibinfo {author} {\bibfnamefont {U.}~\bibnamefont {Staub}},\
  and\ \bibinfo {author} {\bibfnamefont {M.~C.}\ \bibnamefont {Hoffmann}},\
  }\bibfield  {title} {\bibinfo {title} {Terahertz-driven phonon upconversion
  in {SrTiO$_3$}},\ }\href {https://doi.org/10.1038/s41567-018-0408-1}
  {\bibfield  {journal} {\bibinfo  {journal} {Nat. Phys.}\ }\textbf {\bibinfo
  {volume} {15}},\ \bibinfo {pages} {387} (\bibinfo {year} {2019})}\BibitemShut
  {NoStop}%
\bibitem [{\citenamefont {Fechner}\ \emph {et~al.}(2018)\citenamefont
  {Fechner}, \citenamefont {Sukhov}, \citenamefont {Chotorlishvili},
  \citenamefont {Kenel}, \citenamefont {Berakdar},\ and\ \citenamefont
  {Spaldin}}]{fechn18}%
  \BibitemOpen
  \bibfield  {author} {\bibinfo {author} {\bibfnamefont {M.}~\bibnamefont
  {Fechner}}, \bibinfo {author} {\bibfnamefont {A.}~\bibnamefont {Sukhov}},
  \bibinfo {author} {\bibfnamefont {L.}~\bibnamefont {Chotorlishvili}},
  \bibinfo {author} {\bibfnamefont {C.}~\bibnamefont {Kenel}}, \bibinfo
  {author} {\bibfnamefont {J.}~\bibnamefont {Berakdar}},\ and\ \bibinfo
  {author} {\bibfnamefont {N.~A.}\ \bibnamefont {Spaldin}},\ }\bibfield
  {title} {\bibinfo {title} {{Magnetophononics: Ultrafast spin control through
  the lattice}},\ }\href {https://doi.org/10.1103/PhysRevMaterials.2.064401}
  {\bibfield  {journal} {\bibinfo  {journal} {Phys. Rev. Mater.}\ }\textbf
  {\bibinfo {volume} {2}},\ \bibinfo {pages} {064401} (\bibinfo {year}
  {2018})}\BibitemShut {NoStop}%
\bibitem [{\citenamefont {Juraschek}\ \emph {et~al.}(2020)\citenamefont
  {Juraschek}, \citenamefont {Narang},\ and\ \citenamefont
  {Spaldin}}]{juras20}%
  \BibitemOpen
  \bibfield  {author} {\bibinfo {author} {\bibfnamefont {D.~M.}\ \bibnamefont
  {Juraschek}}, \bibinfo {author} {\bibfnamefont {P.}~\bibnamefont {Narang}},\
  and\ \bibinfo {author} {\bibfnamefont {N.~A.}\ \bibnamefont {Spaldin}},\
  }\bibfield  {title} {\bibinfo {title} {{Phono-magnetic analogs to
  opto-magnetic effects}},\ }\href
  {https://doi.org/10.1103/PhysRevResearch.2.043035} {\bibfield  {journal}
  {\bibinfo  {journal} {Phys. Rev. Research}\ }\textbf {\bibinfo {volume}
  {2}},\ \bibinfo {pages} {043035} (\bibinfo {year} {2020})}\BibitemShut
  {NoStop}%
\bibitem [{\citenamefont {Yarmohammadi}\ \emph {et~al.}(2023)\citenamefont
  {Yarmohammadi}, \citenamefont {Krebs}, \citenamefont {Uhrig},\ and\
  \citenamefont {Normand}}]{yarmo23}%
  \BibitemOpen
  \bibfield  {author} {\bibinfo {author} {\bibfnamefont {M.}~\bibnamefont
  {Yarmohammadi}}, \bibinfo {author} {\bibfnamefont {M.}~\bibnamefont {Krebs}},
  \bibinfo {author} {\bibfnamefont {G.~S.}\ \bibnamefont {Uhrig}},\ and\
  \bibinfo {author} {\bibfnamefont {B.}~\bibnamefont {Normand}},\ }\bibfield
  {title} {\bibinfo {title} {{Strong-coupling magnetophononics: Self-blocking,
  phonon-bitriplons, and spin-band engineering}},\ }\href
  {https://doi.org/10.1103/PhysRevB.107.174415} {\bibfield  {journal} {\bibinfo
   {journal} {Phys. Rev. B}\ }\textbf {\bibinfo {volume} {107}},\ \bibinfo
  {pages} {174415} (\bibinfo {year} {2023})}\BibitemShut {NoStop}%
\bibitem [{\citenamefont {Giorgianni}\ \emph {et~al.}(2023)\citenamefont
  {Giorgianni}, \citenamefont {Wehinger}, \citenamefont {Allenspach},
  \citenamefont {Colonna}, \citenamefont {Vicario}, \citenamefont {Puphal},
  \citenamefont {Pomjakushina}, \citenamefont {Normand},\ and\ \citenamefont
  {Rüegg}}]{giorg23}%
  \BibitemOpen
  \bibfield  {author} {\bibinfo {author} {\bibfnamefont {F.}~\bibnamefont
  {Giorgianni}}, \bibinfo {author} {\bibfnamefont {B.}~\bibnamefont
  {Wehinger}}, \bibinfo {author} {\bibfnamefont {S.}~\bibnamefont
  {Allenspach}}, \bibinfo {author} {\bibfnamefont {N.}~\bibnamefont {Colonna}},
  \bibinfo {author} {\bibfnamefont {C.}~\bibnamefont {Vicario}}, \bibinfo
  {author} {\bibfnamefont {P.}~\bibnamefont {Puphal}}, \bibinfo {author}
  {\bibfnamefont {E.}~\bibnamefont {Pomjakushina}}, \bibinfo {author}
  {\bibfnamefont {B.}~\bibnamefont {Normand}},\ and\ \bibinfo {author}
  {\bibfnamefont {C.}~\bibnamefont {Rüegg}},\ }\bibfield  {title} {\bibinfo
  {title} {{Ultrafast frustration breaking and magnetophononic driving of
  singlet excitations in a quantum magnet}},\ }\href
  {https://doi.org/10.1103/PhysRevB.107.184440} {\bibfield  {journal} {\bibinfo
   {journal} {Phys. Rev. B}\ }\textbf {\bibinfo {volume} {107}},\ \bibinfo
  {pages} {184440} (\bibinfo {year} {2023})}\BibitemShut {NoStop}%
\bibitem [{\citenamefont {Disa}\ \emph {et~al.}(2020)\citenamefont {Disa},
  \citenamefont {Fechner}, \citenamefont {Nova}, \citenamefont {Liu},
  \citenamefont {F\"orst}, \citenamefont {Prabhakaran}, \citenamefont
  {Radaelli},\ and\ \citenamefont {Cavalleri}}]{disa20}%
  \BibitemOpen
  \bibfield  {author} {\bibinfo {author} {\bibfnamefont {A.~S.}\ \bibnamefont
  {Disa}}, \bibinfo {author} {\bibfnamefont {M.}~\bibnamefont {Fechner}},
  \bibinfo {author} {\bibfnamefont {T.~F.}\ \bibnamefont {Nova}}, \bibinfo
  {author} {\bibfnamefont {B.}~\bibnamefont {Liu}}, \bibinfo {author}
  {\bibfnamefont {M.}~\bibnamefont {F\"orst}}, \bibinfo {author} {\bibfnamefont
  {D.}~\bibnamefont {Prabhakaran}}, \bibinfo {author} {\bibfnamefont {P.~G.}\
  \bibnamefont {Radaelli}},\ and\ \bibinfo {author} {\bibfnamefont
  {A.}~\bibnamefont {Cavalleri}},\ }\bibfield  {title} {\bibinfo {title}
  {Polarizing an antiferromagnet by optical engineering of the crystal field},\
  }\href {https://doi.org/10.1038/s41567-020-0936-3} {\bibfield  {journal}
  {\bibinfo  {journal} {Nat. Phys.}\ }\textbf {\bibinfo {volume} {16}},\
  \bibinfo {pages} {937} (\bibinfo {year} {2020})}\BibitemShut {NoStop}%
\bibitem [{\citenamefont {Afanasiev}\ \emph {et~al.}(2021)\citenamefont
  {Afanasiev}, \citenamefont {Hortensius}, \citenamefont {Ivanov},
  \citenamefont {Sasani}, \citenamefont {Bousquet}, \citenamefont {Blanter},
  \citenamefont {Mikhaylovskiy}, \citenamefont {Kimel},\ and\ \citenamefont
  {Caviglia}}]{afana21}%
  \BibitemOpen
  \bibfield  {author} {\bibinfo {author} {\bibfnamefont {D.}~\bibnamefont
  {Afanasiev}}, \bibinfo {author} {\bibfnamefont {J.~R.}\ \bibnamefont
  {Hortensius}}, \bibinfo {author} {\bibfnamefont {B.~A.}\ \bibnamefont
  {Ivanov}}, \bibinfo {author} {\bibfnamefont {A.}~\bibnamefont {Sasani}},
  \bibinfo {author} {\bibfnamefont {E.}~\bibnamefont {Bousquet}}, \bibinfo
  {author} {\bibfnamefont {Y.~M.}\ \bibnamefont {Blanter}}, \bibinfo {author}
  {\bibfnamefont {R.~V.}\ \bibnamefont {Mikhaylovskiy}}, \bibinfo {author}
  {\bibfnamefont {A.~V.}\ \bibnamefont {Kimel}},\ and\ \bibinfo {author}
  {\bibfnamefont {A.~D.}\ \bibnamefont {Caviglia}},\ }\bibfield  {title}
  {\bibinfo {title} {Ultrafast control of magnetic interactions via
  light-driven phonons},\ }\href {https://doi.org/10.1038/s41563-021-00922-7}
  {\bibfield  {journal} {\bibinfo  {journal} {Nat. Mater.}\ }\textbf {\bibinfo
  {volume} {20}},\ \bibinfo {pages} {607} (\bibinfo {year} {2021})}\BibitemShut
  {NoStop}%
\bibitem [{\citenamefont {Mashkovich}\ \emph {et~al.}(2021)\citenamefont
  {Mashkovich}, \citenamefont {Grishunin}, \citenamefont {Dubrovin},
  \citenamefont {Zvezdin}, \citenamefont {Pisarev},\ and\ \citenamefont
  {Kimel}}]{mashk21}%
  \BibitemOpen
  \bibfield  {author} {\bibinfo {author} {\bibfnamefont {E.~A.}\ \bibnamefont
  {Mashkovich}}, \bibinfo {author} {\bibfnamefont {K.~A.}\ \bibnamefont
  {Grishunin}}, \bibinfo {author} {\bibfnamefont {R.~M.}\ \bibnamefont
  {Dubrovin}}, \bibinfo {author} {\bibfnamefont {A.~K.}\ \bibnamefont
  {Zvezdin}}, \bibinfo {author} {\bibfnamefont {R.~V.}\ \bibnamefont
  {Pisarev}},\ and\ \bibinfo {author} {\bibfnamefont {A.~V.}\ \bibnamefont
  {Kimel}},\ }\bibfield  {title} {\bibinfo {title} {{Terahertz light–driven
  coupling of antiferromagnetic spins to lattice}},\ }\href
  {https://doi.org/10.1126/science.abk1121} {\bibfield  {journal} {\bibinfo
  {journal} {Science}\ }\textbf {\bibinfo {volume} {373}},\ \bibinfo {pages}
  {1608} (\bibinfo {year} {2021})}\BibitemShut {NoStop}%
\bibitem [{\citenamefont {Bossini}\ \emph {et~al.}(2021)\citenamefont
  {Bossini}, \citenamefont {Conte}, \citenamefont {Terschanski}, \citenamefont
  {Springholz}, \citenamefont {Bonanni}, \citenamefont {Deltenre},
  \citenamefont {Anders}, \citenamefont {Uhrig}, \citenamefont {Cerullo},\ and\
  \citenamefont {Cinchetti}}]{bossi21}%
  \BibitemOpen
  \bibfield  {author} {\bibinfo {author} {\bibfnamefont {D.}~\bibnamefont
  {Bossini}}, \bibinfo {author} {\bibfnamefont {S.~D.}\ \bibnamefont {Conte}},
  \bibinfo {author} {\bibfnamefont {M.}~\bibnamefont {Terschanski}}, \bibinfo
  {author} {\bibfnamefont {G.}~\bibnamefont {Springholz}}, \bibinfo {author}
  {\bibfnamefont {A.}~\bibnamefont {Bonanni}}, \bibinfo {author} {\bibfnamefont
  {K.}~\bibnamefont {Deltenre}}, \bibinfo {author} {\bibfnamefont
  {F.}~\bibnamefont {Anders}}, \bibinfo {author} {\bibfnamefont {G.~S.}\
  \bibnamefont {Uhrig}}, \bibinfo {author} {\bibfnamefont {G.}~\bibnamefont
  {Cerullo}},\ and\ \bibinfo {author} {\bibfnamefont {M.}~\bibnamefont
  {Cinchetti}},\ }\bibfield  {title} {\bibinfo {title} {Femtosecond phononic
  coupling to both spins and charges in a room-temperature antiferromagnetic
  semiconductor},\ }\href {https://doi.org/10.1103/PhysRevB.104.224424}
  {\bibfield  {journal} {\bibinfo  {journal} {Phys. Rev. B}\ }\textbf {\bibinfo
  {volume} {104}},\ \bibinfo {pages} {224424} (\bibinfo {year}
  {2021})}\BibitemShut {NoStop}%
\bibitem [{\citenamefont {Yarmohammadi}\ \emph {et~al.}(2021)\citenamefont
  {Yarmohammadi}, \citenamefont {Meyer}, \citenamefont {Fauseweh},
  \citenamefont {Normand},\ and\ \citenamefont {Uhrig}}]{yarmo21}%
  \BibitemOpen
  \bibfield  {author} {\bibinfo {author} {\bibfnamefont {M.}~\bibnamefont
  {Yarmohammadi}}, \bibinfo {author} {\bibfnamefont {C.}~\bibnamefont {Meyer}},
  \bibinfo {author} {\bibfnamefont {B.}~\bibnamefont {Fauseweh}}, \bibinfo
  {author} {\bibfnamefont {B.}~\bibnamefont {Normand}},\ and\ \bibinfo {author}
  {\bibfnamefont {G.~S.}\ \bibnamefont {Uhrig}},\ }\bibfield  {title} {\bibinfo
  {title} {{Dynamical properties of a driven dissipative dimerized $S = 1/2$
  chain}},\ }\href {https://doi.org/10.1103/PhysRevB.103.045132} {\bibfield
  {journal} {\bibinfo  {journal} {Phys. Rev. B}\ }\textbf {\bibinfo {volume}
  {103}},\ \bibinfo {pages} {045132} (\bibinfo {year} {2021})}\BibitemShut
  {NoStop}%
\bibitem [{\citenamefont {Hase}\ \emph {et~al.}(1993)\citenamefont {Hase},
  \citenamefont {Terasaki},\ and\ \citenamefont {Uchinokura}}]{hase93}%
  \BibitemOpen
  \bibfield  {author} {\bibinfo {author} {\bibfnamefont {M.}~\bibnamefont
  {Hase}}, \bibinfo {author} {\bibfnamefont {I.}~\bibnamefont {Terasaki}},\
  and\ \bibinfo {author} {\bibfnamefont {K.}~\bibnamefont {Uchinokura}},\
  }\bibfield  {title} {\bibinfo {title} {{Observation of the spin-Peierls
  transition in linear Cu$^{2+}$ (spin-1/2) chains in an inorganic compound
  CuGeO$_3$}},\ }\href {https://doi.org/10.1103/PhysRevLett.70.3651} {\bibfield
   {journal} {\bibinfo  {journal} {Phys. Rev. Lett.}\ }\textbf {\bibinfo
  {volume} {70}},\ \bibinfo {pages} {3651} (\bibinfo {year}
  {1993})}\BibitemShut {NoStop}%
\bibitem [{\citenamefont {Popovi\'c}\ \emph {et~al.}(1995)\citenamefont
  {Popovi\'c}, \citenamefont {Devi\'c}, \citenamefont {Popov}, \citenamefont
  {Dhalenne},\ and\ \citenamefont {Revcolevschi}}]{popov95}%
  \BibitemOpen
  \bibfield  {author} {\bibinfo {author} {\bibfnamefont {Z.~V.}\ \bibnamefont
  {Popovi\'c}}, \bibinfo {author} {\bibfnamefont {S.~D.}\ \bibnamefont
  {Devi\'c}}, \bibinfo {author} {\bibfnamefont {V.~N.}\ \bibnamefont {Popov}},
  \bibinfo {author} {\bibfnamefont {G.}~\bibnamefont {Dhalenne}},\ and\
  \bibinfo {author} {\bibfnamefont {A.}~\bibnamefont {Revcolevschi}},\
  }\bibfield  {title} {\bibinfo {title} {{Phonons in CuGeO$_3$ studied using
  polarized far-infrared and Raman-scattering spectroscopies}},\ }\href
  {https://doi.org/10.1103/PhysRevB.52.4185} {\bibfield  {journal} {\bibinfo
  {journal} {Phys. Rev. B}\ }\textbf {\bibinfo {volume} {52}},\ \bibinfo
  {pages} {4185} (\bibinfo {year} {1995})}\BibitemShut {NoStop}%
\bibitem [{\citenamefont {Braden}\ \emph {et~al.}(1998)\citenamefont {Braden},
  \citenamefont {Hennion}, \citenamefont {Reichardt}, \citenamefont
  {Dhalenne},\ and\ \citenamefont {Revcolevschi}}]{brade98a}%
  \BibitemOpen
  \bibfield  {author} {\bibinfo {author} {\bibfnamefont {M.}~\bibnamefont
  {Braden}}, \bibinfo {author} {\bibfnamefont {B.}~\bibnamefont {Hennion}},
  \bibinfo {author} {\bibfnamefont {W.}~\bibnamefont {Reichardt}}, \bibinfo
  {author} {\bibfnamefont {G.}~\bibnamefont {Dhalenne}},\ and\ \bibinfo
  {author} {\bibfnamefont {A.}~\bibnamefont {Revcolevschi}},\ }\bibfield
  {title} {\bibinfo {title} {{Spin-Phonon Coupling in CuGeO$_3$}},\ }\href
  {https://doi.org/10.1103/PhysRevLett.80.3634} {\bibfield  {journal} {\bibinfo
   {journal} {Phys. Rev. Lett.}\ }\textbf {\bibinfo {volume} {80}},\ \bibinfo
  {pages} {3634} (\bibinfo {year} {1998})}\BibitemShut {NoStop}%
\bibitem [{\citenamefont {Braden}\ \emph {et~al.}(2002)\citenamefont {Braden},
  \citenamefont {Reichardt}, \citenamefont {Hennion}, \citenamefont
  {Dhalenne},\ and\ \citenamefont {Revcolevschi}}]{brade02}%
  \BibitemOpen
  \bibfield  {author} {\bibinfo {author} {\bibfnamefont {M.}~\bibnamefont
  {Braden}}, \bibinfo {author} {\bibfnamefont {W.}~\bibnamefont {Reichardt}},
  \bibinfo {author} {\bibfnamefont {B.}~\bibnamefont {Hennion}}, \bibinfo
  {author} {\bibfnamefont {G.}~\bibnamefont {Dhalenne}},\ and\ \bibinfo
  {author} {\bibfnamefont {A.}~\bibnamefont {Revcolevschi}},\ }\bibfield
  {title} {\bibinfo {title} {{Lattice dynamics of CuGeO$_3$:~Inelastic neutron
  scattering and model calculations}},\ }\href
  {https://doi.org/10.1103/PhysRevB.66.214417} {\bibfield  {journal} {\bibinfo
  {journal} {Phys. Rev. B}\ }\textbf {\bibinfo {volume} {66}},\ \bibinfo
  {pages} {214417} (\bibinfo {year} {2002})}\BibitemShut {NoStop}%
\bibitem [{\citenamefont {Regnault}\ \emph {et~al.}(1996)\citenamefont
  {Regnault}, \citenamefont {A\"\i{}n}, \citenamefont {Hennion}, \citenamefont
  {Dhalenne},\ and\ \citenamefont {Revcolevschi}}]{regna96a}%
  \BibitemOpen
  \bibfield  {author} {\bibinfo {author} {\bibfnamefont {L.~P.}\ \bibnamefont
  {Regnault}}, \bibinfo {author} {\bibfnamefont {M.}~\bibnamefont {A\"\i{}n}},
  \bibinfo {author} {\bibfnamefont {B.}~\bibnamefont {Hennion}}, \bibinfo
  {author} {\bibfnamefont {G.}~\bibnamefont {Dhalenne}},\ and\ \bibinfo
  {author} {\bibfnamefont {A.}~\bibnamefont {Revcolevschi}},\ }\bibfield
  {title} {\bibinfo {title} {{Inelastic neutron scattering investigation of the
  spin-Peierls system CuGeO$_3$}},\ }\href
  {https://doi.org/10.1103/PhysRevB.53.5579} {\bibfield  {journal} {\bibinfo
  {journal} {Phys. Rev. B}\ }\textbf {\bibinfo {volume} {53}},\ \bibinfo
  {pages} {5579} (\bibinfo {year} {1996})}\BibitemShut {NoStop}%
\bibitem [{\citenamefont {Uhrig}(1997)}]{uhrig97a}%
  \BibitemOpen
  \bibfield  {author} {\bibinfo {author} {\bibfnamefont {G.~S.}\ \bibnamefont
  {Uhrig}},\ }\bibfield  {title} {\bibinfo {title} {{Symmetry and Dimension of
  the Dispersion of Inorganic Spin-Peierls Systems}},\ }\href
  {https://doi.org/10.1103/PhysRevLett.79.163} {\bibfield  {journal} {\bibinfo
  {journal} {Phys. Rev. Lett.}\ }\textbf {\bibinfo {volume} {79}},\ \bibinfo
  {pages} {163} (\bibinfo {year} {1997})}\BibitemShut {NoStop}%
\bibitem [{\citenamefont {Sachdev}\ and\ \citenamefont
  {Bhatt}(1990)}]{sachd90}%
  \BibitemOpen
  \bibfield  {author} {\bibinfo {author} {\bibfnamefont {S.}~\bibnamefont
  {Sachdev}}\ and\ \bibinfo {author} {\bibfnamefont {R.~N.}\ \bibnamefont
  {Bhatt}},\ }\bibfield  {title} {\bibinfo {title} {Bond-operator
  representation of quantum spins: Mean-field theory of frustrated quantum
  {Heisenberg} antiferromagnets},\ }\href
  {https://doi.org/10.1103/PhysRevB.41.9323} {\bibfield  {journal} {\bibinfo
  {journal} {Phys. Rev. B}\ }\textbf {\bibinfo {volume} {41}},\ \bibinfo
  {pages} {9323} (\bibinfo {year} {1990})}\BibitemShut {NoStop}%
\bibitem [{\citenamefont {Gopalan}\ \emph {et~al.}(1994)\citenamefont
  {Gopalan}, \citenamefont {Rice},\ and\ \citenamefont {Sigrist}}]{gopal94}%
  \BibitemOpen
  \bibfield  {author} {\bibinfo {author} {\bibfnamefont {S.}~\bibnamefont
  {Gopalan}}, \bibinfo {author} {\bibfnamefont {T.~M.}\ \bibnamefont {Rice}},\
  and\ \bibinfo {author} {\bibfnamefont {M.}~\bibnamefont {Sigrist}},\
  }\bibfield  {title} {\bibinfo {title} {Spin ladders with spin gaps: A
  description of a class of cuprates},\ }\href
  {https://doi.org/10.1103/PhysRevB.49.8901} {\bibfield  {journal} {\bibinfo
  {journal} {Phys. Rev. B}\ }\textbf {\bibinfo {volume} {49}},\ \bibinfo
  {pages} {8901} (\bibinfo {year} {1994})}\BibitemShut {NoStop}%
\bibitem [{\citenamefont {Normand}\ and\ \citenamefont
  {R\"uegg}(2011)}]{norma11}%
  \BibitemOpen
  \bibfield  {author} {\bibinfo {author} {\bibfnamefont {B.}~\bibnamefont
  {Normand}}\ and\ \bibinfo {author} {\bibfnamefont {C.}~\bibnamefont
  {R\"uegg}},\ }\bibfield  {title} {\bibinfo {title} {Complete bond-operator
  theory of the two-chain spin ladder},\ }\href
  {https://doi.org/10.1103/PhysRevB.83.054415} {\bibfield  {journal} {\bibinfo
  {journal} {Phys. Rev. B}\ }\textbf {\bibinfo {volume} {83}},\ \bibinfo
  {pages} {054415} (\bibinfo {year} {2011})}\BibitemShut {NoStop}%
\bibitem [{\citenamefont {Breuer}\ and\ \citenamefont
  {Petruccione}(2007)}]{breue07}%
  \BibitemOpen
  \bibfield  {author} {\bibinfo {author} {\bibfnamefont {H.-P.}\ \bibnamefont
  {Breuer}}\ and\ \bibinfo {author} {\bibfnamefont {F.}~\bibnamefont
  {Petruccione}},\ }\href
  {https://doi.org/10.1093/acprof:oso/9780199213900.001.0001} {\emph {\bibinfo
  {title} {{The Theory of Open Quantum Systems}}}}\ (\bibinfo  {publisher}
  {Oxford University Press},\ \bibinfo {address} {Oxford},\ \bibinfo {year}
  {2007})\BibitemShut {NoStop}%
\bibitem [{\citenamefont {Weiss}(2021)}]{weiss21}%
  \BibitemOpen
  \bibfield  {author} {\bibinfo {author} {\bibfnamefont {U.}~\bibnamefont
  {Weiss}},\ }\href {https://doi.org/10.1142/12402} {\emph {\bibinfo {title}
  {{Quantum Dissipative Systems}}}},\ \bibinfo {edition} {5th}\ ed.\ (\bibinfo
  {publisher} {World Scientific},\ \bibinfo {address} {Singapore},\ \bibinfo
  {year} {2021})\BibitemShut {NoStop}%
\bibitem [{\citenamefont {Lindblad}(1976)}]{lindb76}%
  \BibitemOpen
  \bibfield  {author} {\bibinfo {author} {\bibfnamefont {G.}~\bibnamefont
  {Lindblad}},\ }\bibfield  {title} {\bibinfo {title} {{On the Generators of
  Quantum Dynamical Semigroups}},\ }\href {https://doi.org/10.1007/BF01608499}
  {\bibfield  {journal} {\bibinfo  {journal} {Comm. Math. Phys.}\ }\textbf
  {\bibinfo {volume} {48}},\ \bibinfo {pages} {119} (\bibinfo {year}
  {1976})}\BibitemShut {NoStop}%
\bibitem [{\citenamefont {Giavazzi}\ \emph {et~al.}(2022)\citenamefont
  {Giavazzi}, \citenamefont {di~Maiolo},\ and\ \citenamefont
  {Painelli}}]{giava22}%
  \BibitemOpen
  \bibfield  {author} {\bibinfo {author} {\bibfnamefont {D.}~\bibnamefont
  {Giavazzi}}, \bibinfo {author} {\bibfnamefont {F.}~\bibnamefont
  {di~Maiolo}},\ and\ \bibinfo {author} {\bibfnamefont {A.}~\bibnamefont
  {Painelli}},\ }\bibfield  {title} {\bibinfo {title} {{The fate of molecular
  excited states: modeling donor–acceptor dyes}},\ }\href
  {https://doi.org/10.1039/d1cp05971h} {\bibfield  {journal} {\bibinfo
  {journal} {Phys. Chem. Chem. Phys.}\ }\textbf {\bibinfo {volume} {24}},\
  \bibinfo {pages} {5555} (\bibinfo {year} {2022})}\BibitemShut {NoStop}%
\bibitem [{\citenamefont {Bezanson}\ \emph {et~al.}(2017)\citenamefont
  {Bezanson}, \citenamefont {Edelman}, \citenamefont {Karpinski},\ and\
  \citenamefont {Shah}}]{bezanson17}%
  \BibitemOpen
  \bibfield  {author} {\bibinfo {author} {\bibfnamefont {J.}~\bibnamefont
  {Bezanson}}, \bibinfo {author} {\bibfnamefont {A.}~\bibnamefont {Edelman}},
  \bibinfo {author} {\bibfnamefont {S.}~\bibnamefont {Karpinski}},\ and\
  \bibinfo {author} {\bibfnamefont {V.~B.}\ \bibnamefont {Shah}},\ }\bibfield
  {title} {\bibinfo {title} {Julia: A fresh approach to numerical computing},\
  }\href {https://doi.org/10.1137/141000671} {\bibfield  {journal} {\bibinfo
  {journal} {SIAM Rev.}\ }\textbf {\bibinfo {volume} {59}},\ \bibinfo {pages}
  {65} (\bibinfo {year} {2017})}\BibitemShut {NoStop}%
\bibitem [{\citenamefont {Virtanen}\ \emph {et~al.}(2020)\citenamefont
  {Virtanen}, \citenamefont {Gommers}, \citenamefont {Oliphant}, \citenamefont
  {Haberland}, \citenamefont {Reddy}, \citenamefont {Cournapeau}, \citenamefont
  {Burovski}, \citenamefont {Peterson}, \citenamefont {Weckesser},
  \citenamefont {Bright}, \citenamefont {{van der Walt}}, \citenamefont
  {Brett}, \citenamefont {Wilson}, \citenamefont {Millman}, \citenamefont
  {Mayorov}, \citenamefont {Nelson}, \citenamefont {Jones}, \citenamefont
  {Kern}, \citenamefont {Larson}, \citenamefont {Carey}, \citenamefont {Polat},
  \citenamefont {Feng}, \citenamefont {Moore}, \citenamefont {{VanderPlas}},
  \citenamefont {Laxalde}, \citenamefont {Perktold}, \citenamefont {Cimrman},
  \citenamefont {Henriksen}, \citenamefont {Quintero}, \citenamefont {Harris},
  \citenamefont {Archibald}, \citenamefont {Ribeiro}, \citenamefont
  {Pedregosa}, \citenamefont {{van Mulbregt}},\ and\ \citenamefont {{SciPy 1.0
  Contributors}}}]{virtanen20}%
  \BibitemOpen
  \bibfield  {author} {\bibinfo {author} {\bibfnamefont {P.}~\bibnamefont
  {Virtanen}}, \bibinfo {author} {\bibfnamefont {R.}~\bibnamefont {Gommers}},
  \bibinfo {author} {\bibfnamefont {T.~E.}\ \bibnamefont {Oliphant}}, \bibinfo
  {author} {\bibfnamefont {M.}~\bibnamefont {Haberland}}, \bibinfo {author}
  {\bibfnamefont {T.}~\bibnamefont {Reddy}}, \bibinfo {author} {\bibfnamefont
  {D.}~\bibnamefont {Cournapeau}}, \bibinfo {author} {\bibfnamefont
  {E.}~\bibnamefont {Burovski}}, \bibinfo {author} {\bibfnamefont
  {P.}~\bibnamefont {Peterson}}, \bibinfo {author} {\bibfnamefont
  {W.}~\bibnamefont {Weckesser}}, \bibinfo {author} {\bibfnamefont
  {J.}~\bibnamefont {Bright}}, \bibinfo {author} {\bibfnamefont {S.~J.}\
  \bibnamefont {{van der Walt}}}, \bibinfo {author} {\bibfnamefont
  {M.}~\bibnamefont {Brett}}, \bibinfo {author} {\bibfnamefont
  {J.}~\bibnamefont {Wilson}}, \bibinfo {author} {\bibfnamefont {K.~J.}\
  \bibnamefont {Millman}}, \bibinfo {author} {\bibfnamefont {N.}~\bibnamefont
  {Mayorov}}, \bibinfo {author} {\bibfnamefont {A.~R.~J.}\ \bibnamefont
  {Nelson}}, \bibinfo {author} {\bibfnamefont {E.}~\bibnamefont {Jones}},
  \bibinfo {author} {\bibfnamefont {R.}~\bibnamefont {Kern}}, \bibinfo {author}
  {\bibfnamefont {E.}~\bibnamefont {Larson}}, \bibinfo {author} {\bibfnamefont
  {C.~J.}\ \bibnamefont {Carey}}, \bibinfo {author} {\bibfnamefont
  {{\.I}.}~\bibnamefont {Polat}}, \bibinfo {author} {\bibfnamefont
  {Y.}~\bibnamefont {Feng}}, \bibinfo {author} {\bibfnamefont {E.~W.}\
  \bibnamefont {Moore}}, \bibinfo {author} {\bibfnamefont {J.}~\bibnamefont
  {{VanderPlas}}}, \bibinfo {author} {\bibfnamefont {D.}~\bibnamefont
  {Laxalde}}, \bibinfo {author} {\bibfnamefont {J.}~\bibnamefont {Perktold}},
  \bibinfo {author} {\bibfnamefont {R.}~\bibnamefont {Cimrman}}, \bibinfo
  {author} {\bibfnamefont {I.}~\bibnamefont {Henriksen}}, \bibinfo {author}
  {\bibfnamefont {E.~A.}\ \bibnamefont {Quintero}}, \bibinfo {author}
  {\bibfnamefont {C.~R.}\ \bibnamefont {Harris}}, \bibinfo {author}
  {\bibfnamefont {A.~M.}\ \bibnamefont {Archibald}}, \bibinfo {author}
  {\bibfnamefont {A.~H.}\ \bibnamefont {Ribeiro}}, \bibinfo {author}
  {\bibfnamefont {F.}~\bibnamefont {Pedregosa}}, \bibinfo {author}
  {\bibfnamefont {P.}~\bibnamefont {{van Mulbregt}}},\ and\ \bibinfo {author}
  {\bibnamefont {{SciPy 1.0 Contributors}}},\ }\bibfield  {title} {\bibinfo
  {title} {{{SciPy} 1.0: fundamental algorithms for scientific computing in
  Python}},\ }\href {https://doi.org/10.1038/s41592-019-0686-2} {\bibfield
  {journal} {\bibinfo  {journal} {Nat. Methods}\ }\textbf {\bibinfo {volume}
  {17}},\ \bibinfo {pages} {261} (\bibinfo {year} {2020})}\BibitemShut
  {NoStop}%
\bibitem [{\citenamefont {Gros}\ and\ \citenamefont {Werner}(1998)}]{gros98}%
  \BibitemOpen
  \bibfield  {author} {\bibinfo {author} {\bibfnamefont {C.}~\bibnamefont
  {Gros}}\ and\ \bibinfo {author} {\bibfnamefont {R.}~\bibnamefont {Werner}},\
  }\bibfield  {title} {\bibinfo {title} {{Dynamics of the Peierls-active phonon
  modes in CuGeO$_3$}},\ }\href {https://doi.org/10.1103/PhysRevB.58.R14677}
  {\bibfield  {journal} {\bibinfo  {journal} {Phys. Rev. B}\ }\textbf {\bibinfo
  {volume} {58}},\ \bibinfo {pages} {R14677} (\bibinfo {year}
  {1998})}\BibitemShut {NoStop}%
\bibitem [{\citenamefont {Takehana}\ \emph {et~al.}(2000)\citenamefont
  {Takehana}, \citenamefont {Takamasu}, \citenamefont {Hase}, \citenamefont
  {Kido},\ and\ \citenamefont {Uchinokura}}]{takeh00}%
  \BibitemOpen
  \bibfield  {author} {\bibinfo {author} {\bibfnamefont {K.}~\bibnamefont
  {Takehana}}, \bibinfo {author} {\bibfnamefont {T.}~\bibnamefont {Takamasu}},
  \bibinfo {author} {\bibfnamefont {M.}~\bibnamefont {Hase}}, \bibinfo {author}
  {\bibfnamefont {G.}~\bibnamefont {Kido}},\ and\ \bibinfo {author}
  {\bibfnamefont {K.}~\bibnamefont {Uchinokura}},\ }\bibfield  {title}
  {\bibinfo {title} {{Far-infrared spectroscopy in the spin-Peierls compound
  CuGeO$_3$ under high magnetic fields}},\ }\href
  {https://doi.org/10.1103/PhysRevB.62.5191} {\bibfield  {journal} {\bibinfo
  {journal} {Phys. Rev. B}\ }\textbf {\bibinfo {volume} {62}},\ \bibinfo
  {pages} {5191} (\bibinfo {year} {2000})}\BibitemShut {NoStop}%
\bibitem [{\citenamefont {Spitz}\ \emph {et~al.}(2025)\citenamefont {Spitz},
  \citenamefont {Razpopov}, \citenamefont {Biswas}, \citenamefont {Lane},
  \citenamefont {Nikitin}, \citenamefont {Iida}, \citenamefont {Kajimoto},
  \citenamefont {Fujita}, \citenamefont {Arai}, \citenamefont {Mourigal},
  \citenamefont {{{Ch.} R\"uegg}}, \citenamefont {Valentí},\ and\
  \citenamefont {Normand}}]{spitz25}%
  \BibitemOpen
  \bibfield  {author} {\bibinfo {author} {\bibfnamefont {L.}~\bibnamefont
  {Spitz}}, \bibinfo {author} {\bibfnamefont {A.}~\bibnamefont {Razpopov}},
  \bibinfo {author} {\bibfnamefont {S.}~\bibnamefont {Biswas}}, \bibinfo
  {author} {\bibfnamefont {H.}~\bibnamefont {Lane}}, \bibinfo {author}
  {\bibfnamefont {S.~E.}\ \bibnamefont {Nikitin}}, \bibinfo {author}
  {\bibfnamefont {K.}~\bibnamefont {Iida}}, \bibinfo {author} {\bibfnamefont
  {R.}~\bibnamefont {Kajimoto}}, \bibinfo {author} {\bibfnamefont
  {M.}~\bibnamefont {Fujita}}, \bibinfo {author} {\bibfnamefont
  {M.}~\bibnamefont {Arai}}, \bibinfo {author} {\bibfnamefont {M.}~\bibnamefont
  {Mourigal}}, \bibinfo {author} {\bibnamefont {{{Ch.} R\"uegg}}}, \bibinfo
  {author} {\bibfnamefont {R.}~\bibnamefont {Valentí}},\ and\ \bibinfo
  {author} {\bibfnamefont {B.}~\bibnamefont {Normand}},\ }\href@noop {}
  {\bibinfo {title} {{Phonon spectrum in the spin-Peierls phase of CuGeO$_3$}}}
  (\bibinfo {year} {2025}),\ \Eprint {https://arxiv.org/abs/2507.12409}
  {arXiv:2507.12409} \BibitemShut {NoStop}%
\bibitem [{\citenamefont {Knetter}\ and\ \citenamefont
  {Uhrig}(2001)}]{knett01}%
  \BibitemOpen
  \bibfield  {author} {\bibinfo {author} {\bibfnamefont {C.}~\bibnamefont
  {Knetter}}\ and\ \bibinfo {author} {\bibfnamefont {G.~S.}\ \bibnamefont
  {Uhrig}},\ }\bibfield  {title} {\bibinfo {title} {{Triplet dispersion in
  CuGeO$_3$: Perturbative analysis}},\ }\href
  {https://doi.org/10.1103/PhysRevB.63.094401} {\bibfield  {journal} {\bibinfo
  {journal} {Phys. Rev. B}\ }\textbf {\bibinfo {volume} {63}},\ \bibinfo
  {pages} {094401} (\bibinfo {year} {2001})}\BibitemShut {NoStop}%
\bibitem [{\citenamefont {Werner}\ \emph {et~al.}(1999)\citenamefont {Werner},
  \citenamefont {Gros},\ and\ \citenamefont {Braden}}]{werne99}%
  \BibitemOpen
  \bibfield  {author} {\bibinfo {author} {\bibfnamefont {R.}~\bibnamefont
  {Werner}}, \bibinfo {author} {\bibfnamefont {C.}~\bibnamefont {Gros}},\ and\
  \bibinfo {author} {\bibfnamefont {M.}~\bibnamefont {Braden}},\ }\bibfield
  {title} {\bibinfo {title} {{Microscopic spin-phonon coupling constants in
  CuGeO$_3$}},\ }\href {https://doi.org/10.1103/PhysRevB.59.14356} {\bibfield
  {journal} {\bibinfo  {journal} {Phys. Rev. B}\ }\textbf {\bibinfo {volume}
  {59}},\ \bibinfo {pages} {14356} (\bibinfo {year} {1999})}\BibitemShut
  {NoStop}%
\bibitem [{\citenamefont {Feldkemper}\ and\ \citenamefont
  {Weber}(2002)}]{feldk02}%
  \BibitemOpen
  \bibfield  {author} {\bibinfo {author} {\bibfnamefont {S.}~\bibnamefont
  {Feldkemper}}\ and\ \bibinfo {author} {\bibfnamefont {W.}~\bibnamefont
  {Weber}},\ }\bibfield  {title} {\bibinfo {title} {{Superexchange via cluster
  states: Calculation of spin-phonon coupling constants for CuGeO$_3$}},\
  }\href {https://doi.org/10.1103/PhysRevB.62.3816} {\bibfield  {journal}
  {\bibinfo  {journal} {Phys. Rev. B}\ }\textbf {\bibinfo {volume} {62}},\
  \bibinfo {pages} {3816} (\bibinfo {year} {2002})}\BibitemShut {NoStop}%
\bibitem [{\citenamefont {Rackauckas}\ and\ \citenamefont
  {Nie}(2017)}]{rackauckas17}%
  \BibitemOpen
  \bibfield  {author} {\bibinfo {author} {\bibfnamefont {C.}~\bibnamefont
  {Rackauckas}}\ and\ \bibinfo {author} {\bibfnamefont {Q.}~\bibnamefont
  {Nie}},\ }\bibfield  {title} {\bibinfo {title} {{DifferentialEquations.jl--A
  Performant and Feature-Rich Ecosystem for Solving Differential Equations in
  Julia}},\ }\href@noop {} {\bibfield  {journal} {\bibinfo  {journal} {J. Open
  Res. Softw.}\ }\textbf {\bibinfo {volume} {5}},\ \bibinfo {pages} {15}
  (\bibinfo {year} {2017})}\BibitemShut {NoStop}%
\bibitem [{\citenamefont {Liu}(2025)}]{liu25}%
  \BibitemOpen
  \bibfield  {author} {\bibinfo {author} {\bibfnamefont {A.}~\bibnamefont
  {Liu}},\ }\bibfield  {title} {\bibinfo {title} {{Multidimensional terahertz
  probes of quantum materials}},\ }\href
  {https://doi.org/10.1038/s41535-025-00741-y} {\bibfield  {journal} {\bibinfo
  {journal} {npj Quant. Mater.}\ }\textbf {\bibinfo {volume} {10}},\ \bibinfo
  {pages} {18} (\bibinfo {year} {2025})}\BibitemShut {NoStop}%
\bibitem [{dat()}]{datadump}%
  \BibitemOpen
  \href@noop {} {}\bibinfo {note} {All of the data shown in the figures of this
  article may be found in the repository at
  https://doi.org/10.5281/zenodo.16269926.}\BibitemShut {Stop}%
\end{thebibliography}%
	
\end{document}